\documentclass[12pt]{article}
\usepackage{graphicx} 
\usepackage[numbers,square]{natbib}

\usepackage[left=2cm,right=2cm,top=2.5cm,bottom=2.5cm]{geometry}

\usepackage{graphicx}
\usepackage{subcaption}
\usepackage{physics}
\usepackage{bigints}
\usepackage{braket}
\usepackage{tikz-feynman}
\usepackage{float}
\usepackage{mathtools}
\usepackage{xcolor}
\usepackage{epsfig}
\usepackage{cancel}
\usepackage{caption}
\usepackage{feynmf} 
\usepackage{amssymb}
\usepackage{amsfonts}
\usepackage{epsf}
\usepackage{rotating}
\usepackage{graphicx}
\usepackage{amsmath}
\usepackage{fancyhdr}
\usepackage{graphics}
\usepackage{color}
\usepackage{frontespizio}

\usepackage{type1ec}
\usepackage[T1]{fontenc}
\usepackage{lettrine}
\usepackage{bbold}
\usepackage{calligra}
\usepackage{tikz}
\usepackage{mathrsfs}
\usepackage{curve2e}
\usepackage{setspace}
\usepackage{indentfirst}
\usepackage{emptypage}
\usepackage[babel]{csquotes} 
\usepackage[font=small,labelfont=bf,labelsep=quad]{caption} 
\usepackage{graphicx} 
\usepackage{listings} 
\usetikzlibrary{patterns}
\usepackage{relsize}
\usetikzlibrary{intersections,positioning}
\usetikzlibrary{decorations.pathmorphing,decorations.markings,arrows,positioning}
\usepackage{braket}
\usepackage{mathrsfs}
\usepackage{stackengine}
\usepackage{calc}
\newlength\shlength
\newcommand\xshlongvec[2][0]{\setlength\shlength{#1pt}%
	\stackengine{-5.6pt}{$#2$}{\smash{$\kern\shlength%
			\stackengine{7.55pt}{$\mathchar"017E$}%
			{\rule{\widthof{$#2$}}{.57pt}\kern.4pt}{O}{r}{F}{F}{L}\kern-\shlength$}}%
	{O}{c}{F}{T}{S}}

\usepackage{hyperref}
\hypersetup{
	colorlinks,
	citecolor=black,
	filecolor=black,
	linkcolor=black,
	urlcolor=black
}

\IfFileExists{dsfont.sty}
{\usepackage{dsfont}
	\let\mathbb=\mathds
	\newcommand{\id}{\mathds{1}}}
{\typeout{Package dsfont.sty was not found, using alternative macros.}
	\let\mathds=\mathbb
	\newcommand{\id}{\mbox{1 \kern-.59em {\rm l}}}}

\usepackage{slashed}
\usepackage{units}
\usepackage{setspace}
\let\a=\alpha   \let\b=\beta      \let\d=\delta
 \let\z=\zeta        
      \let\l=\lambda  \let\m=\mu
\let\n=\nu                 \let\r=\rho
\let\s=\sigma  \let\t=\tau      
\let\c=\chi         

\let\d=\delta
\let\s=\sigma
\renewcommand{\a}{\alpha}

\renewcommand{\th}{\theta}

\renewcommand{\r}{\rho}
\renewcommand{\t}{\tau}

\newcommand{\Y}{\Upsilon}

\newcommand{\Z}{\mathds{Z}}

\def\nbox#1#2{\vcenter{\hrule \hbox{\vrule height#2in
			\kern#1in \vrule} \hrule}}
\def\sq{\,\raise.5pt\hbox{$\nbox{.09}{.09}$}\,}
\def\sqb{\,\raise.5pt\hbox{$\overline{\nbox{.09}{.09}}$}\,}

\newcommand{\bea}{\begin{eqnarray}}
\newcommand{\eea}{\end{eqnarray}}
\newcommand{\be}{\begin{equation}}
\newcommand{\ee}{\end{equation}}

\newcommand{\bes}{\begin{subequations}}
	\newcommand{\ees}{\end{subequations}}

\newcommand{\brak}[1]{\left({#1}\right)}
\numberwithin{equation}{section}

\usepackage{accents}

\makeatletter
\newcommand{\xLine}[2][]{\ext@arrow 0359\Rightarrowfill@{#1}{#2}}
\makeatother
\xdefinecolor{darkgreen}{RGB}{102, 204, 70}
\xdefinecolor{darkblue}{RGB}{0, 0, 153}

\usepackage{comment}
\title{331 Review}
\author{Dario MELLE}
\date{February 2024}

\begin{document}
\vspace{1.5cm}
\begin{center}
{\bf  \Large The  $SU(3)_C\times SU(3)_L\times U(1)_X$  (331) Model: \\ } 
\vspace{0.2cm}
{\bf \Large Addressing the Fermion Families Problem\\ within Horizontal Anomalies Cancellation \\}
\vspace{0.2cm}
{\bf  \Large }

{\Large \bf  }
\vspace{0.1cm}
{\Large \bf }

 \vspace{0.3cm}
\vspace{1cm}
{\bf Claudio Corian\`o and Dario Melle \\}

\vspace{1cm}
{\it Dipartimento di Matematica e Fisica, Universit\`{a} del Salento \\
and INFN Sezione di Lecce, Via Arnesano 73100 Lecce, Italy\\
National Center for HPC, Big Data and Quantum Computing\\}
\vspace{0.5cm}

\begin{abstract}
    One of the most important and unanswered problems in particle physics is the origin of the three generations of quarks and leptons. The standard Model does not provide any hint regarding its sequential charge assignments, which remain a fundamental mystery of Nature. One possible solution of the puzzle is to look for charge assignments, in a given gauge theory, that are inter-generational, by employing the cancellation of the gravitational and gauge anomalies horizontally. The 331 model, based on an $SU(3)_C\times SU(3)_L \times U(1)_X$ does it in an economic way, and defines a possible extension of the Standard Model, where the number of families has to be necessarily three. We review the model in Frampton's formulation, that predicts the existence of bileptons. Another characteristics of the model is to unify the $SU(3)_C\times SU(2)_L \times U(1)_X$ into the 331 symmetry at a scale which is in the TeV range, and can be tested at the LHC. Expressions of the scalar mass eigenstates and of the renormalization group equations of the model are also presented.

\end{abstract}
\end{center}
\newpage

\section{Introduction}
In the quest to unveil new physics governing fundamental interactions at the Large Hadron Collider (LHC), resolving several crucial questions remains a challenge within the Standard Model (SM). These include the gauge hierarchy problem in the Higgs sector and the origin of light neutrino masses.\\
Addressing these issues often requires theories involving larger gauge groups and a broader spectrum of particles. Grand Unified Theories (GUTs) offer promising avenues, but their high energy scales (around $10^{12}$ to $10^{15}$ GeV) far exceed the electroweak scale probed by the LHC.\\
Bridging the gap between the GUT scale and the TeV scale, where the LHC operates, to identify signatures of symmetry breaking presents a significant challenge due to the increased complexity of these extended theories.\\
However, specific scenarios exist where evidence for enlarged gauge symmetries might be discovered or excluded at the LHC scale, suggesting alternative exploration paths.\\
One such example is the 331 model $(SU(3)_c \times SU(3)_L \times U(1)_X)$, where the constraint of real gauge couplings significantly restricts the parameter space for potential signal searches.
This model was proposed as a potential extension to the SM in order to address certain theoretical and experimental shortcomings, as well as to provide explanations for phenomena not accounted for within the SM.\\
The 331 model introduces a new gauge group, $SU(3)_L$, which is isomorphic to the color gauge group $SU(3)_C$. This implies that the strong force acting between quarks within hadrons is now governed by the $SU(3)_L$ symmetry, in addition to the color symmetry. \\
The fermion content of the 331 model differs from the SM due to the extended gauge symmetry. Typically, in the 331 model, the quarks and leptons are organized into multiplets that transform under the representations of the $SU(3)_L$ gauge group. For example, quarks and leptons may be arranged in triplets or antitriplets of $SU(3)_L$, depending on their electric charge and other quantum numbers.
One notable feature of the 331 model is the presence of new gauge bosons called "bileptons." These are bosons carrying both lepton number and electric charge, with charges $Q=\pm 2$ and $L=\pm 2$. Bileptons arise due to the extended gauge symmetry and can have significant implications for various phenomena, including neutrino masses and decays of heavy particles.\\
Similar to the SM, the 331 model also involves spontaneous symmetry breaking, where the gauge symmetries are broken at some energy scale. This results in the generation of particle masses and the emergence of the familiar gauge bosons such as the $W^\pm$ and $Z^0$ bosons.\\
The 331 model offers potential explanations for various phenomena beyond the scope of the SM, including neutrino masses and mixing, and the unification of fundamental forces at high energies.\\
Overall, the 331 model represents an intriguing extension of the SM, offering new avenues for exploring fundamental physics beyond the established framework. However, it remains subject to experimental scrutiny and theoretical refinement to fully ascertain its validity and implications for our understanding of the fundamental forces and particles in nature.\\
In the model, the constraint of real gauge couplings significantly restricts the parameter space for potential signal searches. 
This property establishes the vacuum expectation values (vevs) of the Higgs bosons, responsible for symmetry breaking from the 331 scale to the electroweak scale, around the TeV region.
The model under consideration incorporates the presence of bileptons, denoted as gauge bosons $(Y^{--}, Y^{++})$, possessing a charge $Q=\pm 2$ and lepton number $L=\pm 2$. Consequently, we dub this framework the "bilepton model." Within the array of 331 models, the existence of bileptons within the spectrum arises only through specific embeddings of the $U(1)_X$ symmetry, as well as the charge $(Q)$ and hypercharge $(Y)$ generators within the local gauge structure.\\
An additional noteworthy aspect of this model is its departure from the conventions of the Standard Model or typical chiral models seen thus far. In contrast to merely extending the SM spectrum and symmetries, the determination of the (chiral) fermion generations hinges on the interfamily cancellation of gauge anomalies. Remarkably, gauge anomalies cancel across distinct fermion families, thereby pinpointing the number of generations to be 3. From this vantage point, the model emerges as distinctly singular. Furthermore, in the framework outlined by Frampton, which we adopt henceforth, the treatment of the third fermion family is asymmetrical in comparison to the initial two families.\\
Our work is organized as follows: We will first discuss the general structure of the model, starting with the anomaly constraints that crucially characterize the charge assignments of the spectrum. We then move on to characterize the gauge boson spectrum, turning afterwards to the Higgs sector. The structure of the potential is thoroughly examined, both in its triplet and sextet contributions. We discuss the energy bound present in the model, induced by the structure of the gauge coupling relation coming from the embedding of the Standard Model into the 331. This bound on the energy scale at which the model is characterized by real (as opposed to complex) values of the couplings is one of the salient features of this theory. A second important feature is the presence of a Landau pole in the renormalization group equations (RGEs) of the gauge couplings, which we briefly illustrate numerically. The results are based on an analysis of the RGEs, which we derive and present in Appendix B. A list of results for the mass eigenstates of the Higgs sector is contained in Appendix A. Notice that an important feature of the model is the identification of the electric charge operator in terms of the diagonal generators of the fundamental gauge symmetry. In general this is given by 
\begin{equation}
    Q=T^3+ \beta\,T^8+X \mathbb{1}\ 
\end{equation}
where $\beta$ is a parameter of the model, with $T^3$ and $T^8$ generators of $SU(3)_L$. Our discussion will be focused on the choice $\beta=\sqrt{3}$, as in Frampton's original formulation. This choice induces the presence of bileptons, which are doubly charged gauge bosons carrying lepton number $L=\pm 2$. We will refer to this version of the model as to the minimal one.

\section{The particle content of the minimal 331 Model}\label{331}
One of the most important questions that arise in particle physics is why there are only three families of quarks and leptons. There are many observables in particle physics that depend on the family number and they all agree to constrain this number to three. In the Standard Model there is no mechanism that prohibits the existence of more families than those observed. To answer this question P. H. Frampton, F. Pisano and V. Pleitez \cite {frampton_chiral_1992} \cite {pisano_su3ensuremathbigotimesu1_1992} have proposed a new model that extends the Standard Model and could provide an elegant answer to this fundamental question. Specifically it is called the 331 Model, and it is built on the gauge group
\begin{equation}
    SU(3)_C\times SU(3)_L\times U(1)_X\ ,
\end{equation}
which enlarge the $SU(2)_L$ symmetry of the Standard Model. Three exotic quarks must be added to the particle content of the Standard Model to allow for $ SU(3)_L $ symmetry in the quark sector. The 331 Model democratically treats leptons in each of the three families, in fact color singlets are $ SU (3)_L $ anti-triplets
\begin{equation}
    \begin{pmatrix}
    e\\-\n_e\\ e^c
    \end{pmatrix}_L,
    \begin{pmatrix}
    \m\\-\n_\m\\ \m^c
    \end{pmatrix}_L,
    \begin{pmatrix}
    \t\\-\n_\t\\ \t^c
    \end{pmatrix}_L\rightarrow \brak{\textbf{1},\bar{\textbf{3}},0}
\end{equation}
each with $X=0$. Where with $e^c\ ,\m^c$ and $\t^c$ we are denoting the left-handed Weyl spinors of the relative charge conjugate field. On the contrary, in the quark sector the three families are treated differently, the first two generations are in triplets of $ SU (3)_L $ with the corresponding left-handed field of exotic quark
\begin{equation}
    \begin{pmatrix}
    u\\d\\ D
    \end{pmatrix}_L,
    \begin{pmatrix}
    c\\s\\ S
    \end{pmatrix}_L\rightarrow \brak{\textbf{3},\textbf{3},-\frac{1}{3}}\ ,
\end{equation}
both with $X=-\frac{1}{3}$. On the other hand, the third generation in the quark sector is embedded in anti-triplets of $SU(3)_L$
\begin{equation}
    \begin{pmatrix}
    b\\-t\\ T
    \end{pmatrix}_L
    \rightarrow \brak{\textbf{3},\bar{\textbf{3}},\frac{2}{3}}\ ,
\end{equation}
with $X=\frac{2}{3}$. To each left handed field there is a corresponding right handed field singlet under $SU(3)_L$ 
\begin{equation}
\begin{matrix}
    (u^c)_L\\
    (c^c)_L\\ 
    (t^c)_L
\end{matrix} \rightarrow \brak{\textbf{1},\textbf{1},-\frac{2}{3}}\ ,
\end{equation}

\begin{equation}
\begin{matrix}
     (d^c)_L\\
     (s^c)_L\\
     (b^c)_L
\end{matrix} \rightarrow \brak{\textbf{1},\textbf{1},\frac{1}{3}}\ ,
\end{equation}

\begin{equation}
\begin{matrix}
    (D^c)_L\\(S^c)_L
\end{matrix} \rightarrow \brak{\textbf{1},\textbf{1},\frac{4}{3}}\ ,
\end{equation}

\begin{equation}
\begin{matrix}
    (T^c)_L
\end{matrix} \rightarrow \brak{\textbf{1},\textbf{1},-\frac{5}{3}}\ ,
\end{equation}
The $U(1)_X$ charges are respectively $-\frac{2}{3}$, $\frac{1}{3}$ and $\frac{4}{3}$ for $u^c$, $d^c$, $D^c$ and the corresponding field of the second generation. The $X$ quantum numbers of the third family instead are respectively  $\frac{1}{3}$, $-\frac{2}{3}$ and $-\frac{5}{3}$ for $b^c$, $t^c$, $T^c$. 
Three scalar triplets under $SU(3)_L$ are necessary to ensure the spontaneous symmetry breaking
\begin{equation}
    \r=\begin{pmatrix}
        \r^{++}\\ \r^+\\ \r^0
    \end{pmatrix}
    \eta=\begin{pmatrix}
        \eta^{+}_1\\ \eta^0\\ \eta^-_2
    \end{pmatrix}
    \chi=\begin{pmatrix}
        \chi^{0}\\ \chi^-\\ \chi^{--}
    \end{pmatrix}
\end{equation}
respectively with $X$ charge $X=1,0,-1$, and a scalar sextet 
\begin{equation}
    \s=\begin{pmatrix}
      \s^{++}_1&\frac{\s_1^+}{\sqrt{2}}&\frac{\s_1^0}{\sqrt{2}}\\
      \frac{\s_1^+}{\sqrt{2}}&\s_2^0&\frac{\s^-_2}{\sqrt{2}}\\ 
      \frac{\s_1^0}{\sqrt{2}}&\frac{\s^-_2}{\sqrt{2}}&\s_2^{--}\\
    \end{pmatrix}\ ,
\end{equation}
in order to generate physical masses for leptons as we will see in Section \ref{331lep}.
All quantum numbers under $SU(3)_C\times SU(3)_L\times U(1)_X$ in the 331 Model can be found in Table \ref{Tab:minimal331_qn}. This unconventional assignment of quantum numbers in the model ensures that gauge anomalies are not canceled vertically for each family, as in the Standard Model. It is necessary to add the contribution of each quark in the triangle anomaly to obtain total cancellation. This is one of the most important and attractive feature of the model because it provides a possible explanation for the number of generations. This could provide a first step towards understanding the flavor puzzle and perhaps serve as a guide for new models inspired by it.

\begin{table}
\centering
    \begin{tabular}{cccc}
    \hline
    \hline
    \rule[-4mm]{0mm}{1cm}
     & $SU(3)_C$ & $SU(3)_L$ & $U(1)_X$\\
    \hline\\
    $u^c\ \ \ \ c^c$  & $\bar{\textbf{3}}$ & $\textbf{1}$ & $-\frac{2}{3}$\\\\
    
    $t^c$  & $\bar{\textbf{3}}$ & $\textbf{1}$ & $\frac{1}{3}$\\\\

    $d^c\ \ \ \ s^c$ & $\bar{\textbf{3}}$ & $\textbf{1}$ & $\frac{1}{3}$\\\\
    
    $b^c$ & $\bar{\textbf{3}}$ & $\textbf{1}$ & $-\frac{2}{3}$\\\\
    
    $D^c\ \ \ \ S^c$ & $\bar{\textbf{3}}$ & $\textbf{1}$ & $\frac{4}{3}$\\\\
    
    $T^c$ & $\bar{\textbf{3}}$ & $\textbf{1}$ & $-\frac{5}{3}$\\\\

    $ \begin{pmatrix}
    u\\d\\ D
    \end{pmatrix}_L\ \ \ \ \begin{pmatrix}
    c\\s\\ S
    \end{pmatrix}_L\ \ \ \ $ & $\textbf{3}$ & $\textbf{3}$ & $-\frac{1}{3}$\\\\
    
    $\begin{pmatrix}
    b\\t\\ T
    \end{pmatrix}_L$& $\textbf{3}$ & $\bar{\textbf{3}}$ & $\frac{2}{3}$\\\\\\


    $\begin{pmatrix}
    e^-\\\n_e\\ e^+
    \end{pmatrix}_L\ \ \ \ \begin{pmatrix}
    \m^-\\\n_\m\\ \m^+
    \end{pmatrix}_L\ \ \ \ \begin{pmatrix}
    \t^-\\\n_\t\\ \t^+
    \end{pmatrix}_L$ & $\textbf{1}$ & $\bar{\textbf{3}}$ & $0$\\\\
\\
    $X_\mu$ & $\textbf{1}$ & $\textbf{1}$ & $0$\\\\

    $W^a_\mu$ & $\textbf{1}$ & $\textbf{8}$ & $0$\\\\

    $G^b_\mu$ & $\textbf{8}$ & $\textbf{1}$ & $0$\\\\\\

    $\r$ & $\textbf{1}$ & $\textbf{3}$ & $1$\\\\
    
    $\eta$ & $\textbf{1}$ & $\textbf{3}$ & $0$\\\\
    
    $\chi$ & $\textbf{1}$ & $\textbf{3}$ & $-1$\\\\
    
    $\s$ & $\textbf{1}$ & $\textbf{6}$ & $0$\\\\
    \hline
    \hline
    \end{tabular}
    \caption{Quantum numbers of the particle spectrum of the minimal 331 Model}
    \label{Tab:minimal331_qn}
\end{table}
\section{Cross-Family Anomaly cancellation and the Flavor question}
In this section, we will analyze the non-trivial cancellation of anomalies in the 331 Model and discuss why this method of eliminating them could represent an initial step towards addressing the flavor question.

There are six types of anomalies that occur in the 331 Model
\begin{align}
    SU(3)_C^3\ ,&\ \ \ \ \ \ \ \ \ \ \ \ SU(3)_C^2U(1)_X\ ,\ \ \ \ \ \ \ \ \ \ \ \ SU(3)_L^3\ ,\nonumber\\ SU(3)_L^2U&(1)_X\ ,\ \ \ \ \ \ \ \ \ \ \ \ U(1)_X^3\ ,\ \ \ \ \ \ \ \ \ \ \ \  \textrm{grav}^2\,U(1)_X,\ 
\end{align}
where $\textrm{grav}^2\,U(1)_X$ is the mixed chiral anomaly with two gravitons and a chiral $U(1)_X$ gauge current, i.e. the gravitational chiral anomaly. \\
Each vertex collects a factor 
\begin{equation}
2tr(T^a_R\{T_R^b,T_R^c\})=A(R)\, d^{abc}\ ,\ \ \ \ \ \ \ \textnormal{with  }A(\textrm{fund})=1
\end{equation}
 from the gauge current group generators, where $A(R)$ is the representation dependent anomaly coefficient, with $A(\textrm{fund})$ in the fundamental representation of $SU(3)$, and $d^{abc}$ is the totally symmetric invariant $SU(3)$ tensor.

The anomaly involving only gluons is obviously zero as in the Standard Model.  Another anomaly arises from the mixed $SU(3)_C^2\, U(1)_X$ vertex, which imposes the following constraint on the $X$ charges  
\begin{equation}
    SU(3)_C^2U(1)_X\rightarrow  3X_{Q_i}+X_{u^c_i}+X_{d^c_i}+X_{J^c_i}=0.\ 
\end{equation}
Here, the index $i$ ranges over families, but there is no summation over it, $Q_i$ stands for quark triplets, while $J_i^c$ denotes the complex conjugate of the exotic quark fields. Indeed, for this reason, the anomaly cancellation in this case occurs vertically between families, as in the Standard Model.\\
The same cannot be said for another anomaly, which involves only the gauge group $SU(3)_L$. In the Standard Model, the $SU(2)_L$ group possesses a vanishing anomaly coefficient, rendering it unnecessary. However, in the case of $SU(3)$ or, in general, $SU(N)$ with $N>2$, it exhibits a non-zero anomaly coefficient. In the 331 Model, such $SU(3)_L^3$ anomaly cancels due to the equal number of fermions in the $\textbf{3}$ and $\bar{\textbf{3}}$ representations of $SU(3)_L$.\\
In the minimal 331 Model, the $SU(3)_L^2U(1)_X$ anomaly can only be canceled by accounting for contributions from all three families. This type of anomaly cancellation is referred to as a horizontal cancellation. If we consider anomaly cancellation on a generation-by-generation basis, it does not vanish, and requires the summation over different families. The relative constraint is
\begin{equation}
    SU(3)_L^2\, U(1)_X\rightarrow \sum_{i=1}^3 X_{Q_i}=0\ .
\end{equation}
The same motivations lead to the cancellation of the cubic anomaly
\begin{equation}
    U(1)_X^3\rightarrow \sum_{i=1}^3 \brak{3X_{Q_i}^3+X_{u^c_i}^3+X_{d^c_i}^3+X_{J^c_i}^3}=0\, ,
\end{equation}
with the relative constraint.

\begin{table}
    \centering
    \begin{tabular}{cccc}
    \hline
    \hline
    \rule[-4mm]{0mm}{1cm}
     Anomaly &&\\
    \hline\\
    $SU(3)_C^2U(1)_X $&&$3X_Q+X_{u^c_i}+X_{d^c_i}+X_{J^c_i}=0$\\\\

    $SU(3)_L^3 $&&$\textnormal{Equal number of $\textbf{3}_L$ and $\bar{\textbf{3}}_L$ representations} $\\\\
    
    $ SU(3)_L^2U(1)_X $&& $\sum_{i=1}^3 X_{Q_i}=0$\\\\
    
    $ U(1)_X^3$&&$\sum_{i=1}^3 \brak{3X_{Q_i}^3+X_{u^c_i}^3+X_{d^c_i}^3+X_{J^c_i}^3}=0 $\\\\
    
    $  grav^2U(1)_X$&&$ 3X_{Q_i}+X_{u^c_i}+X_{d^c_i}+X_{J^c_i}=0$\\\\
    \hline
    \hline
    \end{tabular}
    \caption{Anomaly cancellation constraints on the fermion charges in the minimal 331 Model}
    \label{Tab:331_anomaly}
\end{table}
The last anomaly that needs to be checked is the gravitational anomaly, but it is not difficult to show that it leads to the same constraint as for $SU(3)^2_C\,U(1)_X$
\begin{equation}
    \textrm{grav}^2\, U(1)_X\rightarrow 3X_{Q_i}+X_{u^c_i}+X_{d^c_i}+X_{J^c_i}=0\ .
\end{equation}
A summary of anomalies and relative constraints can be found in Table \ref{Tab:331_anomaly}.\\
The non-trivial cancellation of anomalies in the model is arguably one of its most intriguing and distinctive features. The horizontal approach, involving all three generations of quarks and leptons, inherently constrains the number of families to three. Unlike the Standard Model, which lacks a mechanism to limit the number of fermion generations, the 331 model provides such a constraint. \\
Notice that in the Standard Model the number of generations is fixed esperimentally by the annihilation process $e^+e^-\rightarrow \text{hadrons}$, which is sensitive to the number of families. Experimental constraints based on these observables restrict the number to 3, neverthless, the model lacks a theoretical guiding principle for predicting such a number. The horizontal approach to anomaly cancellation of the 331, on the other end, as already mentioned, can serve as a guiding principle for investigating other beyond the Standard Model (BSM)  UV-completions, that aim to address the flavor question.

\section{Spontaneous symmetry breaking}
Below the electroweak scale, in the Standard Model, the gauge symmetry is $SU(3)_C \times U(1)_{em}$. Therefore, also in the 331 model, spontaneous symmetry breaking (SSB) must occur, in order to reduce  the $SU(3)_C\times SU(3)_L\times U(1)_X$ gauge symmetry. The breaking can be divided into two stages. Initially, at energy scales greater than $246 \,\textnormal{GeV}$, the gauge symmetry of the 331 model can be broken down to that of the Standard Model, and subsequently to $SU(3)_C \times U(1)_{em}$. This can be represented as
\begin{equation}
    SU(3)_C\times SU(3)_L\times U(1)_X\rightarrow SU(3)_C\times SU(2)_L\times U(1)_{Y}\rightarrow SU(3)_C\times U(1)_{em}\ .
\end{equation}
Achieving this requires a more intricate Higgs sector comprising three scalar triplets and a scalar sextet of $SU(3)_L$. In the following two sections, we will analyze the pattern of SSB and explore how it predicts the existence of bileptons, specifically massive double-charged gauge bosons which carry lepton number of $L=\pm 2$.

\subsection{The breaking $SU(3)_C\times SU(3)_L\times U(1)_X\rightarrow SU(3)_C\times SU(2)_L\times U(1)_{Y}$}
The spontaneous symmetry breaking to the Standard Model gauge group can be accomplished by means of a vacuum expectation value of a scalar triplet belonging to $SU(3)_L$, denoted as $\r$,
\begin{equation}
    \r=\begin{pmatrix}
        \r^{++}\\ \r^+\\ \r^0
    \end{pmatrix}
\end{equation}
which carries charge under $U(1)_X$, namely $X=1$
\begin{equation}
    \expval{\r}=\begin{pmatrix}
         0\\0\\v_\r
    \end{pmatrix}\ .
\end{equation}
The $SU(3)_L\times U(1)_X$ covariant derivative can be written as follow
\begin{align}\label{covdev}
D_\m=&\ \partial_\m -ig_1XX_\m-ig_2\frac{\l^a}{2}W^a_\m=\nonumber\\
=&\ \brak{\partial_\m-i\sqrt{\frac{2}{3}}g_1X X_\m}
\begin{pmatrix}
  1&0&0\\
  0&1&0\\
  0&0&1
\end{pmatrix}-ig_2
\begin{pmatrix}
      \frac{   {W^3_\m}   }{2}+\frac{   {W^8_\m}   }{2 \sqrt{3}} & \frac{   {W^1_\m}   }{2}-\frac{i    {W^2_\m}   }{2} & \frac{   {W^4_\m}   }{2}-\frac{i    {W^5_\m}   }{2} \\
 \frac{   {W^1_\m}   }{2}+\frac{i    {W^2_\m}   }{2} &    -\frac{   {W^3_\m}   }{2}+\frac{   {W^8_\m}   }{2 \sqrt{3}} & \frac{   {W^6_\m}   }{2}-\frac{i    {W^7_\m}   }{2} \\
 \frac{   {W^4_\m}   }{2}+\frac{i    {W^5_\m}   }{2} & \frac{   {W^6_\m}   }{2}+\frac{i    {W^7_\m}   }{2} &    -\frac{   {W^8_\m}   }{\sqrt{3}} 
\end{pmatrix}\ ,
\end{align}
where $X$ is the charge under $U(1)_X$ of the fermion, $X_\mu$ is the corresponding gauge boson and 
$W_\mu^a$ are the generators of $SU(3)_L$. 
$\l^a$ are the Gell-Mann matrices normalized as $\Tr(\l^a \l^b)=2\d^{ab}$. Once the Higgs triplet $\rho$ acquires the vacuum expectation value, its kinetic term gives
\begin{align}\label{d}
\brak{D_\m\begin{pmatrix}
        0\\0\\v_\r
    \end{pmatrix}}^\dagger\brak{D_\m\begin{pmatrix}
        0\\0\\v_\r
    \end{pmatrix}}=-\frac{1}{6} v_\r^2 \biggl(&-4 \sqrt{2} g_1g_2 X_\m W^{\m 8} +4g_1^2 X^\m X_\m+3 g_2^2Y^{++\m} Y^{--}_\m\nonumber \\&+3 g_2^2 V^{\m+}V_\m^- + 2g_2^2 W^8_\m W^{\m 8}\biggl)\ .
\end{align}
From equation \eqref{d} it is immediate to observe that the mass terms obtained are expressed in a basis that is not completely diagonal. This implies that in order to obtain the mass eigenstates of the bosons it is necessary to perform an orthogonal rotation on the corresponding states. Before doing this, we mention that the $ W^\pm $ bosons, given by
\begin{equation}
    W^\pm_\m=\frac{1}{\sqrt{2}}\brak{W^1_\m\mp W^2_\m}\ ,
\end{equation}
remain massless. This is due to the fact the residual symmetry $SU(2)_L$ remains unbroken at this stage.  We recognise also two correctly diagonalized kinetic energy contributions, given in terms of the charge operator eigenstates. As we will see, the charge operator in the 331 Model is embedded as $Q=\frac{1}{2}\l^3+\frac{\sqrt{3}}{2}\l^8+X\mathbb{1}$. We have  
\begin{equation}
    Y^{\pm\pm}_\m=\frac{1}{\sqrt{2}}\brak{W^4_\m\mp i W^5_\m}\ ,
\end{equation}
for the bileptons and 
\begin{equation}
    V^{\pm}_\m=\frac{1}{\sqrt{2}}\brak{W^6_\m\mp i W^7_\m}\ ,
\end{equation}
for the exotic charged gauge bosons. The mass matrix that needs to be diagonalized in the $\{X,W^8\}$ bases is the following
\begin{equation}
    \begin{pmatrix}
          \frac{g_2^2v_\r^2}{3}&-\frac{\sqrt{2}}{3}g_1g_2v_\r^2\\
          -\frac{\sqrt{2}}{3}g_1g_2v_\r^2&\frac{2}{3}g_1^2v_\r^2
    \end{pmatrix}\ .
\end{equation}
The diagonalization can be easily achieved through the orthogonal transformations
\begin{equation}\label{cb1}
    Z'_\m=\frac{1}{\sqrt{g_2^2+2g_1^2}}\brak{g_2W^8_\m+\sqrt{2}g_1X_\m}\ ,
\end{equation}
\begin{equation}\label{cb2}
    B_\m=\frac{1}{\sqrt{g_2^2+2g_1^2}}\brak{\sqrt{2}g_1W^8_\m-g_2X_\m}\ ,
\end{equation}
which can be thought as a rotation from the basis $\{X,W^8\}$ to $\{B,Z'\}$, with an angle 
\begin{equation}
    \sin{\th_{331}}=\frac{g_2}{\sqrt{g_2^2+\frac{g_1^2}{2}}}\ ,
\end{equation}
that gives
\begin{equation}\label{m}
    \begin{pmatrix}
         0&0\\
          0&\frac{1}{3} v_\r^2 \left(g_2^2+2 g_1^2\right)
    \end{pmatrix}\ .
\end{equation}
From the matrix in equation \eqref{m}, we can read off the squared masses of the mass eigenstates $\{B,Z'\}$
\begin{equation}
    M_B^2=0\ \ \ \ \ \ \ \ \ \ M_{Z'}^2=\frac{1}{3} v_\r^2 \left(g_2^2+2 g_1^2\right)\ .
\end{equation}
It is clear that we obtain a massless boson related to the $U(1)_Y$ symmetry of the Standard Model. It is not difficult to identify the embedding of the $Y$ charge operator in the 331 Model, namely
\begin{equation}
    \frac{Y}{2}=\sqrt{3}T^8+X\mathbb{1}\ , 
\end{equation}
where $T^8$ is the eighth generator of $SU(3)$. The matching condition between the $U(1)_X$ coupling and the Standard Model hypercharge can be easily computed, leading to the relation
\begin{equation}
    \frac{1}{g_Y^2}=\frac{6}{g_1^2}+\frac{3}{g_2^2}\ .
\end{equation}

\subsection{The breaking $SU(3)_C\times SU(2)_L\times U(1)_{Y}\rightarrow SU(3)_C\otimes U(1)_{em}$}
Once the gauge symmetry has been decomposed into that of the Standard Model, another symmetry breaking is necessary to end up with the residual $SU(3)_C\otimes U(1)_{em}$ gauge symmetry. To realize the correct breaking scheme, we require two Higgs triplets, $\eta$ and $\chi$, which acquire the vacuum expectation values

\begin{equation}
    \expval{\eta}=\begin{pmatrix}
         0\\ \frac{v_\eta}{\sqrt{2}}\\0
    \end{pmatrix} \ \ \ \ \ \ \textnormal{with\ } X=0\ ,
\end{equation}
and
\begin{equation}
    \expval{\chi}=\begin{pmatrix}
         \frac{v_\chi}{\sqrt{2}}\\0\\0
    \end{pmatrix} \ \ \ \ \ \ \textnormal{with\ } X=-1\ ,
\end{equation}
and a sextet of $SU(3)_L$ 
\begin{equation}
    \expval{\s}=\begin{pmatrix}
    0&0&\frac{v_\s}{2}\\
    0&0&0\\
    \frac{v_\s}{2}&0&0
    \end{pmatrix} \ \ \ \ \ \ \textnormal{with\ } X=0\ .
\end{equation}
After the first symmetry breaking, the covariant derivative can be written in terms of the mass eigenstate fields, namely in the basis $\{B,Z'\}$. Inserting the inverse of the equations \eqref{cb1} and\eqref{cb2} into \eqref{covdev} gives
\begin{equation}
    D_\m=
\begin{pmatrix}
     \partial_\m-ig_2\frac{ {W^3_\m}   }{2}+K_1 & -ig_2\brak{\frac{   {W^1_\m}   }{2}-\frac{i    {W^2_\m}   }{2}} & -ig_2\brak{\frac{   {W^4_\m}   }{2}-\frac{i    {W^5_\m}   }{2}} \\
 -ig_2\brak{\frac{   {W^1_\m}   }{2}+\frac{i    {W^2_\m}   }{2}} &\partial_\m    -ig_2\frac{   {W^3_\m}   }{2}+K_1 & -ig_2\brak{\frac{   {W^6_\m}   }{2}-\frac{i    {W^7_\m}   }{2}} \\
 -ig_2\brak{\frac{   {W^4_\m}   }{2}+\frac{i    {W^5_\m}   }{2}} &-ig_2\brak{ \frac{   {W^6_\m}   }{2}+\frac{i    {W^7_\m}   }{2}} & \partial_\m   +K_2 \\
\end{pmatrix}
\end{equation}
where $K_1$ and $K_2$ are given by 
\begin{equation}
    K_1=\frac{i \left(\sqrt{2} B_\m g_2 {g_1} (2 X-1)-Z'_\m \left(g_2^2+4 {g_1}^2 X\right)\right)}{2 \sqrt{3}}\ ,
\end{equation}
and
\begin{equation}
    K_2=\frac{i \left(\sqrt{2} B_\m g_2 {g_1} (X+1)+Z'_\m \left(g^2_2-2 {g_1}^2 X\right)\right)}{\sqrt{3}}\ .
\end{equation}
Once all the scalar field acquire a vacuum expectation value, the gauge fields  $ W^\pm $ and $ Z $ become massive too, while $ Y^{\pm \pm} $, $ V^\pm $ and $ Z'$ get more involved mass terms.
The squared masses of $ W^\pm $, $ V^\pm $ and $ Y^{\pm \pm} $ are given by 
\begin{equation}
    M^2_W= \frac{g_2^2 v_\eta^2}{4}+\frac{g_2^2 v_\chi^2}{4}+\frac{g_2^2 v_\sigma^2}{4} \ ,
\end{equation}

\begin{equation}
    M^2_V= \frac{g_2^2 v_\rho^2}{4}+\frac{g_2^2 v_\eta^2}{4}+\frac{g_2^2 v_\sigma^2}{4}\ ,
\end{equation}

\begin{equation}
    M^2_Y= \frac{g_2^2 v_\rho^2}{4}+\frac{g_2^2 v_\chi^2}{4}+g_2^2 v_\sigma^2 \ ,
\end{equation}
where we recall that  
\begin{equation}
    W^{\pm}_\m=\frac{1}{\sqrt{2}}\brak{W^1_\m\mp i W^2_\m}\ ,\ \ \ \ \ \ V^{\pm}_\m=\frac{1}{\sqrt{2}}\brak{W^6_\m\mp i W^7_\m}\ ,\ \ \ \ \ \ Y^{\pm\pm}_\m=\frac{1}{\sqrt{2}}\brak{W^4_\m\mp i W^5_\m}\ .
\end{equation}
On the other side, the neutral gauge bosons gain also non-diagonal mass terms, that in the $\{W^3,W^8,X\}$ basis are given by the following matrix
\begin{equation}
    \left(
\begin{array}{ccc}
 g_1^2 v_\r^2+g_1^2 v_\chi^2 & -\frac{g_1 g_2 v_\r^2}{\sqrt{3}}-\frac{g_1 g_2 v_\chi^2}{2 \sqrt{3}} & -\frac{1}{2} g_1 g_2 v_\chi^2 \\
 -\frac{g_1 g_2 v_\r^2}{\sqrt{3}}-\frac{g_1 g_2 v_\chi^2}{2 \sqrt{3}} & \frac{g_2^2 v_\r^2}{3}+\frac{g_2^2 v_\eta^2}{12}+\frac{g_2^2 v_\chi^2}{12}+\frac{g_2^2 v_\sigma^2}{12} & -\frac{g_2^2 v_\eta^2}{4 \sqrt{3}}+\frac{g_2^2 v_\chi^2}{4 \sqrt{3}}-\frac{g_2^2 v_\sigma^2}{4 \sqrt{3}} \\
 -\frac{1}{2} g_1 g_2 v_\chi^2 & -\frac{g_2^2 v_\eta^2}{4 \sqrt{3}}+\frac{g_2^2 v_\chi^2}{4 \sqrt{3}}-\frac{g_2^2 v_\sigma^2}{4 \sqrt{3}} & \frac{g_2^2 v_\eta^2}{4}+\frac{g_2^2 v_\chi^2}{4}+\frac{g_2^2 v_\sigma^2}{4} \\
\end{array}
\right)
\end{equation}
In a first step, the two neutral gauge bosons $W^8$ and $X$ mix, giving rise to the two bosons $B$ and $Z$. The mixing angle is denoted by $\theta_{331}$ and is given by 
\begin{equation}\label{th331}
    \sin{\th_{331}}=\frac{g_2}{\sqrt{g_2^2+\frac{g_1^2}{2}}}\ ,
\end{equation}
which was the rotation discussed in the previous Section.
Then we can proceed in complete analogy with the Standard Model, where $B$ mixes with $W^3$ through the Weinberg angle $\theta_W$, which in the minimal 331 model takes the form
\begin{equation}\label{thw}
    \sin{\th_{W}}=\frac{g_Y}{\sqrt{g^2+g_Y^2}}\ .
\end{equation}
Using \eqref{thw} is straightforward to show that we can express the $\th_{331}$ angle in terms of the Weinberg angle, namely
\begin{equation}
    \cos{\th_{331}}=\sqrt{3}\tan{\th_W}\ ,
\end{equation}
therefore the photon field $A$ and the two massive neutral gauge bosons $Z$ and $Z'$ are identified 
as
\begin{equation}\label{aa}
    A=\sin{\th_W}W_3+\cos{\th_W}\brak{\sqrt{3}\tan{\th_W} W_8+\sqrt{1-3\tan^2{\th_W}}X} \ ,
\end{equation}
\begin{equation}\label{bb}
    Z=\cos{\th_W}W_3-\sin{\th_W}\brak{\sqrt{3}\tan{\th_W} W_8+\sqrt{1-3\tan^2{\th_W}}X}\ ,
\end{equation}
\begin{equation}\label{cc}
    Z'=-\sqrt{1-3\tan^2{\th_W}} W_8+\sqrt{3}\tan{\th_W}X\ .
\end{equation}
The Weinberg angle also relate the coupling of $SU(3)_L$ which by matching is equal to the $SU(2)_L$ of the Standard Model, to the $U(1)_X$ coupling, through
\begin{equation}
\label{boundd}
    \frac{g^2_1}{g^2_2}=\frac{6\sin^2{\th_W}}{1-4\sin^2{\th_W}}\ ,
\end{equation}
which can be obtained from equations \eqref{th331} and \eqref{thw}.
From equations \eqref{aa} \eqref{bb} and \eqref{cc} it is evident that there is a residual mixing between the massive gauge boson, which is given by the matrix
\begin{equation}
    \left(
\begin{array}{cc}
 C_{ZZ} & C_{ZZ'} \\
 C_{ZZ'}& C_{Z'Z'} \\
\end{array}
\right)
\end{equation}
with entries 
\begin{equation}
   C_{ZZ}= g_2^2(2 \theta_W) \left(v_\r^2+v_\eta^2+v_\s^2\right)\ ,
\end{equation}
\begin{equation}
    C_{Z'Z'}=\frac{g_2^2\sin^2\th_W \left(\csc ^2(\theta_W) \left(v_\r^2+v_\eta^2+4 v_\chi^2+v_\s^2\right)+9 \sec ^2(\theta_W) \left(v_\r^2+v_\eta^2+v_\s^2\right)-4 \left(v_\r^2+4 v_\eta^2+v_\chi^2+4 v_\s^2\right)\right)}{24 \cos (2 \theta_W)-12} \ ,
\end{equation}
\begin{equation}
    C_{ZZ'}= \frac{g_2^2 \sec ^3(\theta_W) \left(-\cos (2 \theta_W) \left(v_\r^2+2 \left(v_\eta^2+v_\s^2\right)\right)+2 v_\r^2+v_\eta^2+v_\s^2\right)}{4 \sqrt{12-9 \sec ^2(\theta_W)}}\ .
\end{equation}
Therefore a further rotation is needed 
\begin{equation}
    \begin{pmatrix}
        Z_1\\Z_2
    \end{pmatrix}=\begin{pmatrix}
        \cos{\th_Z}&-\sin{\th_Z}\\
        \sin{\th_Z}&\cos{\th_Z}
    \end{pmatrix} \begin{pmatrix}
        Z\\Z'
    \end{pmatrix}
\end{equation}
in order to obtain the masses of propagating gauge bosons, namely
\begin{align}
    M_Z^2=&\frac{1}{6} (3 g_1^2 (v_\rho^2+v_\chi^2)-(9 g_1^4 (v_\rho^2+v_\chi^2)^2+6 g_1^2 g_2^2 (v_\rho^4-v_\rho^2 (v_\eta^2+v_\sigma^2)\nonumber\\&+v_\chi^2 (-v_\eta^2+v_\chi^2-v_\sigma^2))+ g_2^4 (v_\rho^4-v_\rho^2 (v_\eta^2+v_\chi^2+v_\sigma^2)+v_\eta^4+v_\sigma^2 (2 v_\eta^2-v_\chi^2)\nonumber\\ &-v_\eta^2 v_\chi^2+v_\chi^4+v_\sigma^4)^\frac{1}{2})+g_2^2 (v_\rho^2+v_\eta^2+v_\chi^2+v_\sigma^2))\ ,
\end{align}
and
\begin{align}
    M_{Z'}^2=&\frac{1}{6} (3 g_1^2 (v_\rho^2+v_\chi^2)+(9 g_1^4 (v_\rho^2+v_\chi^2)^2+6 g_1^2 g_2^2 (v_\rho^4-v_\rho^2 (v_\eta^2+v_\sigma^2)\nonumber\\&+v_\chi^2 (-v_\eta^2+v_\chi^2-v_\sigma^2))+g_2^4 (v_\rho^4-v_\rho^2 (v_\eta^2+v_\chi^2+v_\sigma^2)+v_\eta^4+v_\sigma^2 (2 v_\eta^2-v_\chi^2)\nonumber\\&-v_\eta^2 v_\chi^2+v_\chi^4+v_\sigma^4))^{\frac{1}{2}}+g_2^2 (v_\rho^2+v_\eta^2+v_\chi^2+v_\sigma^2))\ .
\end{align}

\begin{figure}[]
        \centering
        \begin{subfigure}[t]{0.45\textwidth}
            \centering
        \includegraphics[width=\linewidth,height=50mm, keepaspectratio]{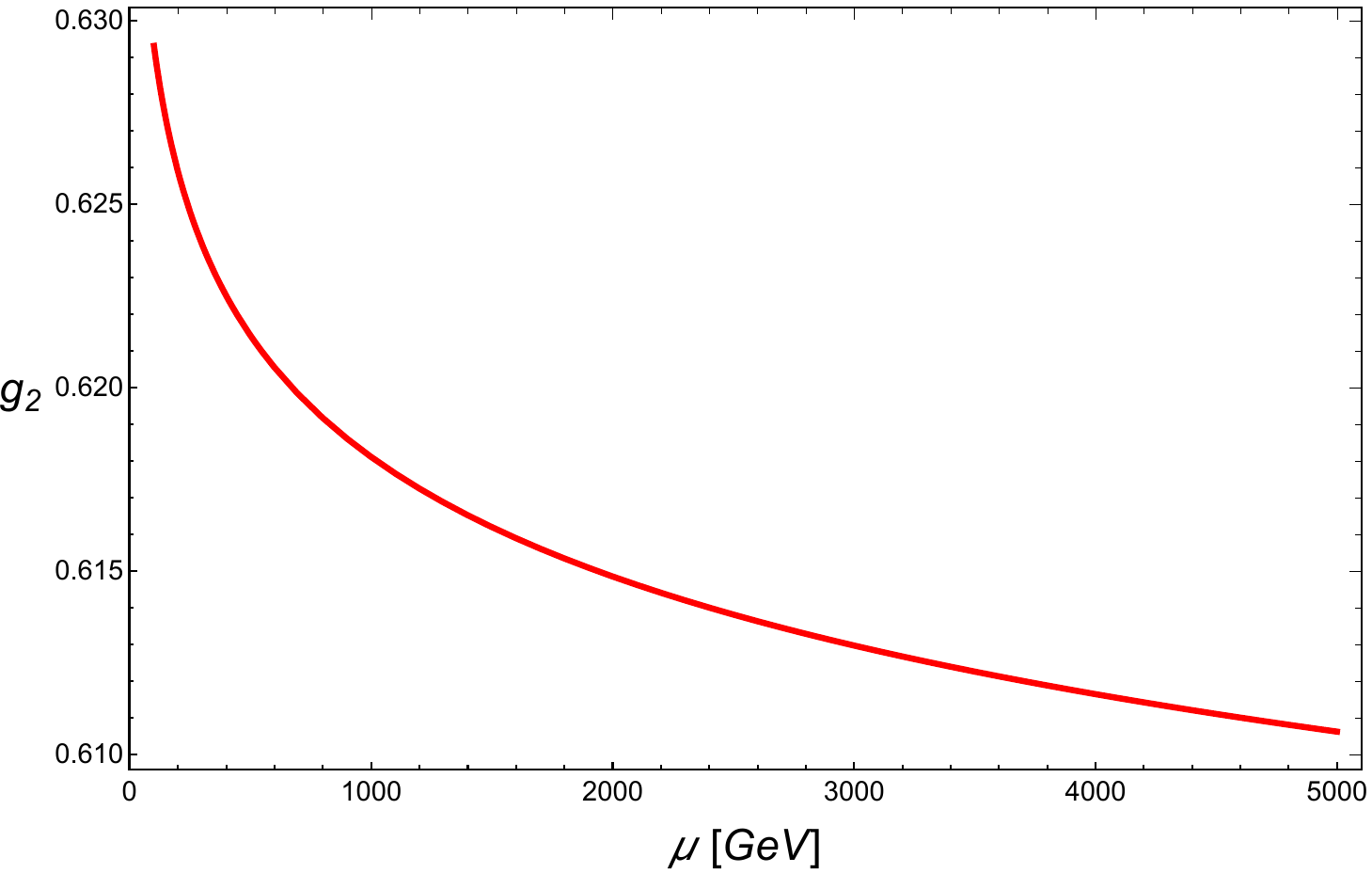}
            \caption{Running of the coupling $g_2$ in the Standard Model.}
            \label{fig:fig-a}
        \end{subfigure}
        \quad
        \begin{subfigure}[t]{0.45\textwidth}
            \centering
        \includegraphics[width=\linewidth,height=50mm, keepaspectratio]{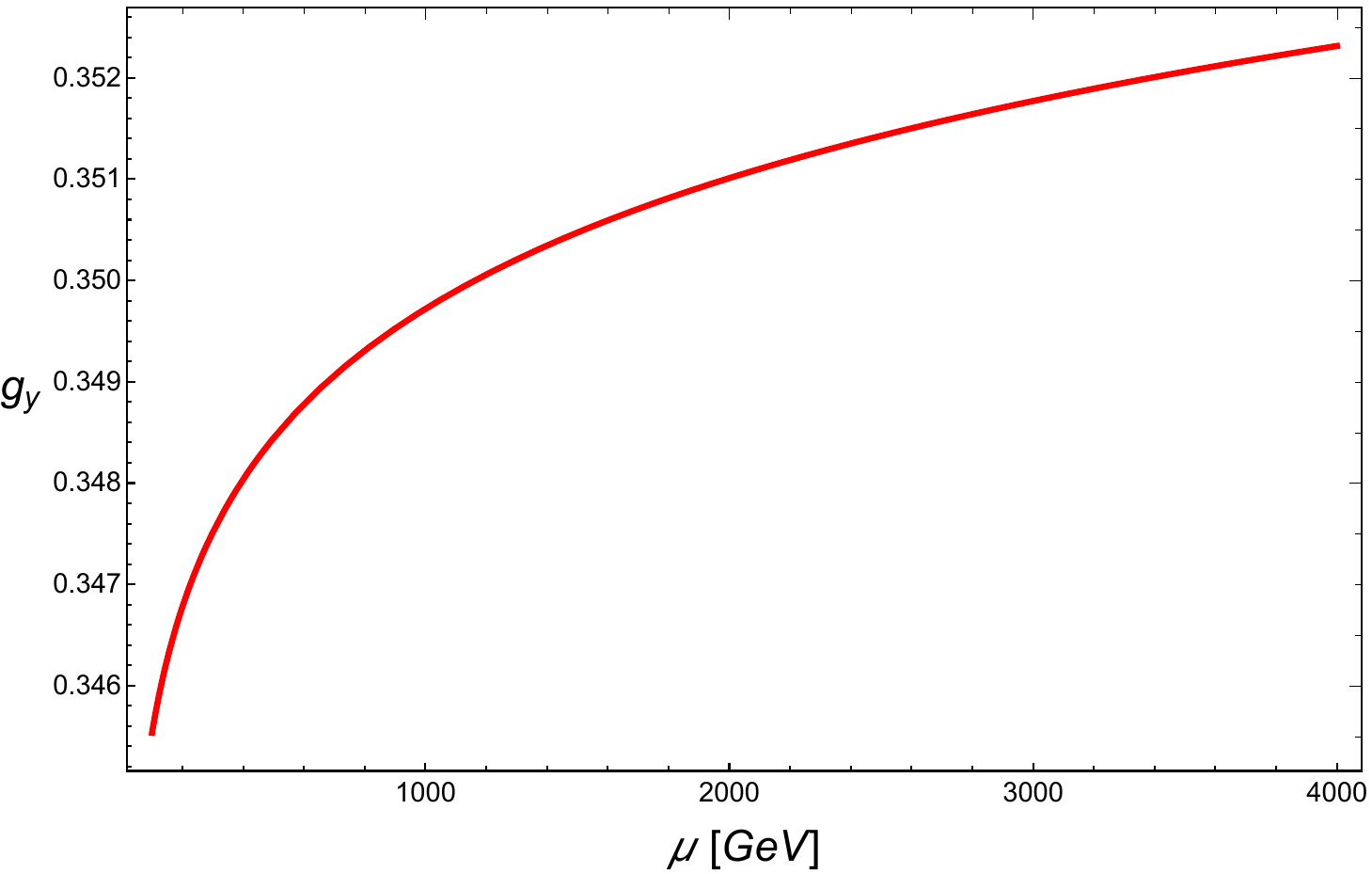}
            \caption{Running of the coupling $g_y$ in the Standard Model.}
            \label{fig:fig-b}
        \end{subfigure}
        \par\bigskip
        \begin{subfigure}[t]{0.45\textwidth}
            \centering
        \includegraphics[width=\linewidth,height=50mm, keepaspectratio]{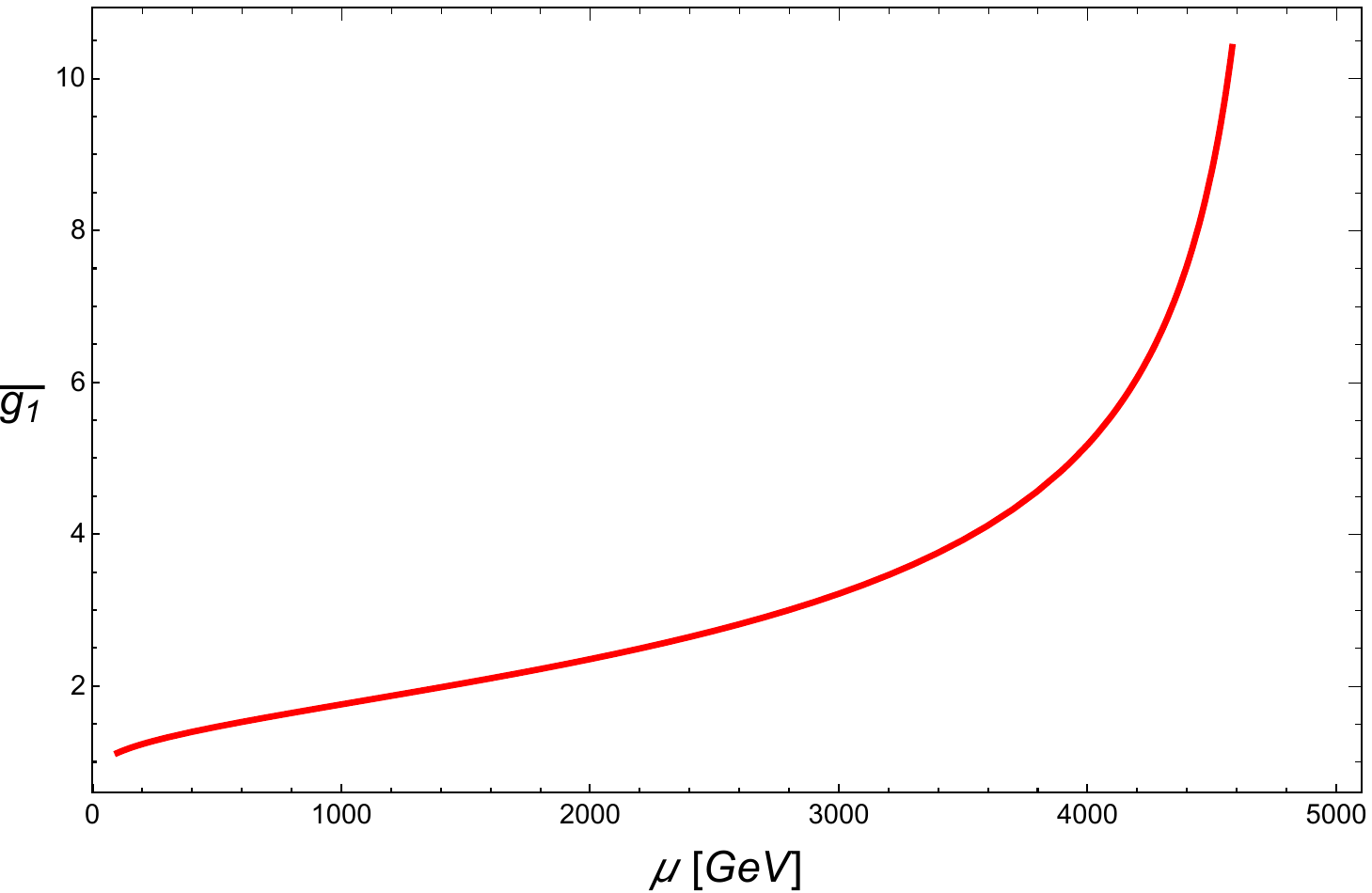}
            \caption{Evolution of the matching condition of the coupling $g_1$ of the 331 Model with the Standard Model coupling constants.}
            \label{fig:fig-c}
        \end{subfigure}
        \quad
        \begin{subfigure}[t]{0.45\textwidth}
            \centering
        \includegraphics[width=\linewidth,height=50mm, keepaspectratio]{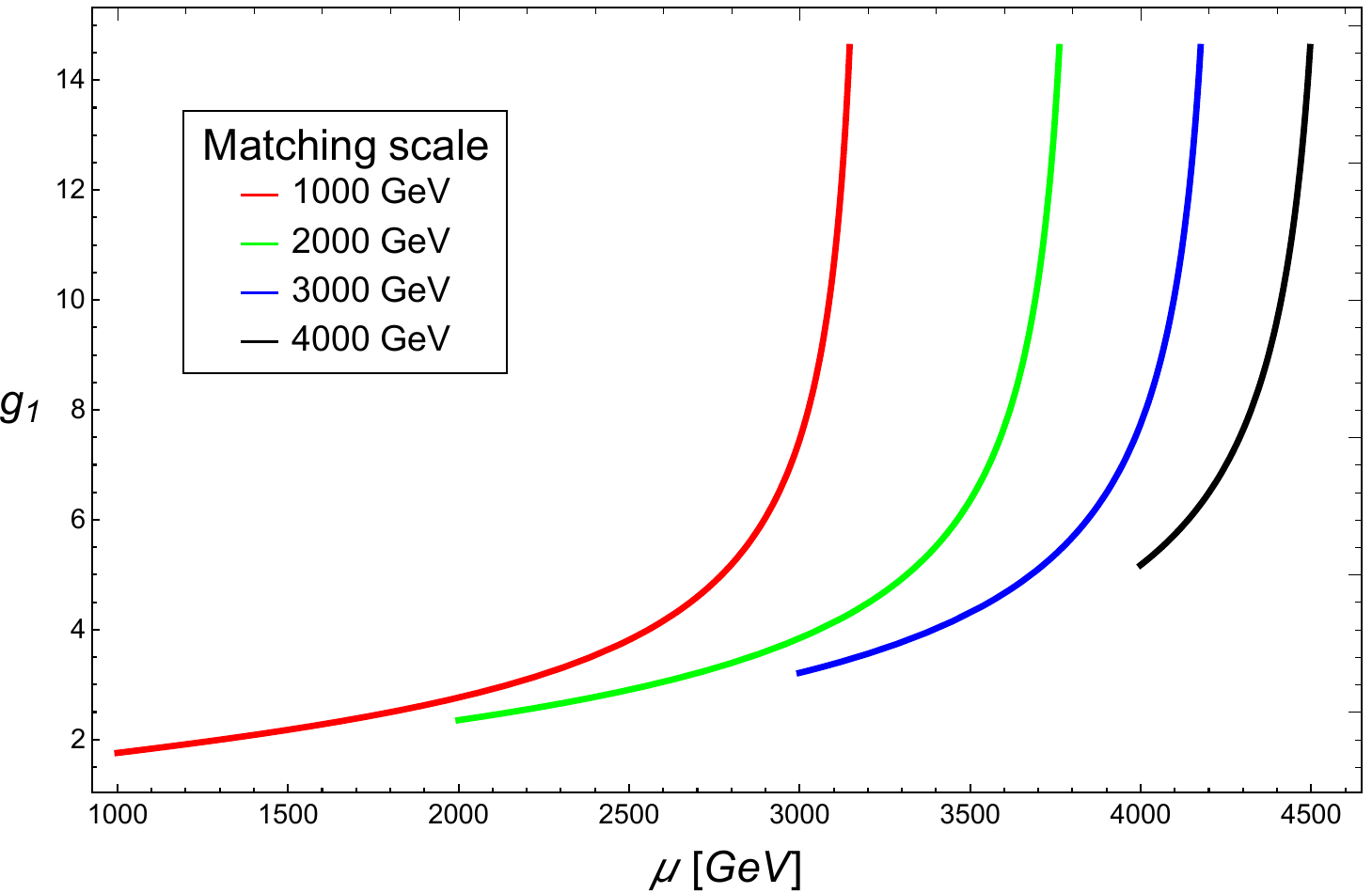}
            \caption{Running of the coupling $g_1$ in the minimal 331 Model.}
            \label{fig:fig-d}
        \end{subfigure}
        \caption{Behaviour of the matching condition between the Standard Model and 331 Model, which make it clear how the Landau pole at the TeV scale emerge from the matching with the Standard Model.}
        \label{fig:my-fig}
\end{figure}

Once the orthogonal rotations have been performed and the Lagrangian has been written in terms of mass eigenstates of bosonic fields, it is possible to extract the values of the couplings. In particular, it possible to derive the expression of the electric charge in terms of the couplings of the minimal 331  
\begin{equation}
    e=\frac{g_1 g_2}{\sqrt{ g^2_2+4 g_1^2}}\ ,
\end{equation}
with the embedding of the charge operator given by
\begin{equation}
    Q=T^3+\sqrt{3}T^8+X \mathbb{1}\ .
\end{equation}
One of the most interesting features of the model is that the embedding of the Standard Model gauge group into the 331 Model gauge group induces a bound on the UV completion of the model \cite{martinez_landau_2007}. From \eqref{boundd}
it is clear that we need to satisfy the following condition
\begin{equation}\label{dis}
    \sin^2\theta_W\leq \frac{1}{4}\ ,
\end{equation}
in order to guarantee that the $g_1$ coupling of the minimal 331 Model is finite.
When $\sin^2\theta_W(\mu)=1/4$, the coupling constant $g_1(\mu)$ diverges, indicating a Landau pole in the renormalization group evolution of the model, which causes the theory to loose its perturbative character even at energy scales lower than $\mu$. The alternative scenario where $g_2$ tends towards zero is disregarded, as $g_2$ coincides with the Standard Model's $SU(2)_L$ coupling, $g_2$, due to the full embedding of $SU(2)_L$ into $SU(3)_L$.  We show the running of the Standard Model couplings $g_2$ in Figure \ref{fig:fig-a} and $g_y$ in Figure \ref{fig:fig-b} and in Figure \ref{fig:fig-c} we show how the matching condition of the coupling $g_1$ evolve in terms of the matching scale. Here we have defined 
\begin{equation}
    \overline{g_1}=\frac{g_2 g_y}{\sqrt{g_2^2-3 g_y^2}}  \ ,
\end{equation}
which express the value of the $U(1)_X$ coupling in terms of the Standard Model. Finally the evolution of $g_1$ in the context of the 331 Model is showed in Figure \ref{fig:fig-d}. some more recent study was performed in \cite{w_barela_3-3-1_2024}

The occurrence of a Landau-like pole in the minimal 331 model is not surprising, as many non 
asymptotically free theories exhibit similar behavior. What distinguishes some of these models is the possibility of encountering this behavior at energies as low as a few TeVs. Consequently, the cutoff scale, $\Lambda_{_{cutoff}}$, cannot be removed by taking $\Lambda_{_{cutoff}} \to\infty$, as in other renormalizable theories. 

From a phenomenological perspective, this result is not overly concerning. The necessity of embedding QED within the electroweak theory at energies of a few hundred GeVs, along with the requirement to account for weak and strong corrections in calculations of physical observables, is already acknowledged. Nevertheless, as a mathematical exercise, studying pure QED at infinitesimal distances proves intriguing. Lattice calculations suggest that chiral symmetry breaking within QED mitigates the Landau pole issue by shifting it above the cutoff scale. Interestingly, the potential existence of the Landau pole or the triviality of the theory arises even at low orders in perturbation theory, suggesting that this phenomenon is not merely a perturbative artifact.\\
The renormalization group offers qualitative insights into the asymptotic behavior of theories at very high energies, even when coupling constants at the relevant scale prohibit the use of perturbation theory. However, it's essential to remember that both QED and the Standard Model are effective, not fundamental, theories. Consequently, effective operators with dimensions higher than $d=4$ must be considered for a realistic continuum limit in lattice calculations. Thus, employing the pure versions of these models remains inconclusive, and the renormalization group may provide valuable insights into this issue within minimal 331 Model.

\section{Higgs Sector}
The inclusion of the sextet representation in the potential enriches the phenomenology of the model and enlarges the number of physical states in the spectrum. In fact we now have, after electroweak symmetry breaking (EWSB) $SU(3)_L\times U(1)_X \to SU(2)_L\times U(1)_Y\to   U(1)_{\rm{em}}$,
five scalar Higgses, three pseudoscalar Higgses, four charged Higgses and three doubly-charged Higgses. The (lepton-number conserving) potential of the model is given by  \cite{tully_scalar_2003}
\begin{align}\label{pot}
    V=&m_1\r^\dagger\r+m_2\eta^\dagger\eta+m_3\chi^\dagger\chi+\l_1(\rho^\dagger\r)^2+\l_2(\eta^\dagger\eta)^2+\l_3(\chi^\dagger\chi)^2\nonumber\\
    &+\l_{12}\r^\dagger\r\eta^\dagger\eta+\l_{13}\r^\dagger\r\chi^\dagger\chi+\l_{23}\chi^\dagger\chi\eta^\dagger\eta+\z_{12}\r^\dagger\eta\eta^\dagger\r+\z_{13}\r^\dagger\chi\chi^\dagger\r+\z_{23}\eta^\dagger\chi\chi^\dagger\eta\nonumber\\
    &+m_4\Tr(\s^\dagger\s)+\l_4(\Tr(\s^\dagger\s))^2+\l_{14}\r^\dagger\r\Tr(\s^\dagger\s)+\l_{24}\eta^\dagger\eta\Tr(\s^\dagger\s)+\l_{34}\chi^\dagger\chi\Tr(\s^\dagger\s)\nonumber\\
    &+\l_{44}\Tr(\s^\dagger\s\s^\dagger\s)+\z_{14}\r^\dagger\s\s^\dagger\rho+\z_{24}\eta^\dagger\s\s^\dagger\eta+\z_{34}\chi^\dagger\s\s^\dagger\chi\nonumber\\
    &+(\sqrt{2}f_{\r\eta\chi}\epsilon^{ijk}\r_i\eta_j\chi_k+\sqrt{2}f_{\r\s\chi}\r^T\s^\dagger\chi\nonumber\\
    &+\xi_{14}\epsilon^{ijk}\r^{*l}\s_{li}\r_j\eta_k+\xi_{24}\epsilon^{ijk}\epsilon^{lmn}\eta_i\eta_l\s_{jm}\s_{kn}+\xi_{34}\epsilon^{ijk}\chi^{*l}\s_{li}\chi_j\eta_k )+\textnormal{h.c.}\ .
\end{align}
The EWSB mechanism will cause a mixing among the Higgs fields \cite{costantini_theoretical_2020}. From Eq. (\ref{pot}) it is possible to obtain the explicit expressions of the mass matrices of the scalar, pseudoscalar, charged and doubly-charged Higgses, by using
standard procedures. In the broken Higgs phase, the minimization conditions
\be\label{mincond}
\frac{\partial V}{\partial v_\phi}=0, \quad \langle \phi^0\rangle=v_\phi, \quad \phi=\rho, \eta, \chi, \sigma
\ee
will define the tree-level vacuum. We remind that we are considering massless neutrinos by choosing the vev of the neutral field $\sigma_2^0$ to be zero. This was the choice in Frampton's original formulation. It can be 
generalized in order to give a small Majorana neutrino mass to the neutrinos \cite{tully_generating_2001}.

The explicit expressions of the minimization conditions are then given by
\begin{align}\label{minpot1}
m_1 v_\rho + \lambda_1 v_\rho^3 + \frac{1}{2}\lambda_{12}v_\rho v_\eta^2-f_{\rho\eta\chi} v_\eta v_\chi+\frac{1}{2}\lambda_{13}v_\rho v_\chi^2 - \frac{1}{\sqrt2}\xi_{14}v_\rho v_\eta v_\sigma + f_{\rho\sigma\chi}v_\chi v_\sigma&\\
+ \frac{1}{2}\lambda_{14}v_\rho v_\sigma^2 + \frac{1}{4}\zeta_{14}v_\rho v_\sigma^2&=0\nonumber\\
m_2 v_\eta + \frac{1}{2}\lambda_{12}v_\rho^2 v_\eta +\lambda_2 v_\eta^3 - f_{\rho\eta\chi} v_\rho v_\chi +\frac{1}{2}\lambda_{23} v_\eta v_\chi^2 - \frac{1}{2\sqrt2}\xi_{14}v_\rho^2 v_\sigma+ \frac{1}{2\sqrt2}v_\chi^2 v_\sigma&\\
+\frac{1}{2}\lambda_{24} v_\eta v_\sigma^2-\xi_{24} v_\eta v_\sigma^2&=0\nonumber\\
m_3 v_\chi + \lambda_3 v_\chi^3 + \frac{1}{2} \lambda_{13} v_\rho^2 v_\chi - f_{\rho\eta\chi} v_\rho v_\eta +\frac{1}{2}\lambda_{23}v_\eta^2 v_\chi +\frac{1}{\sqrt2}\xi_{34}v_\eta v_\chi v_\sigma + f_{\rho\sigma\chi} v_\rho v_\sigma&\\
+\frac{1}{2}\lambda_{34} v_\chi v_\sigma^2 + \frac{1}{4}\zeta_{34} v_\chi v_\sigma^2&=0\nonumber\\
\label{minpot2}
m_4 v_\sigma + \frac{1}{2}\lambda_{14}v_\rho^2 v_\sigma + \lambda_{44} v_\sigma^3 + \frac{1}{2}\lambda_4 v_\sigma^3 + f_{\rho\sigma\chi} v_\rho v_\chi - \frac{1}{2\sqrt2} \xi_{14} v_\rho^2 v_\eta + \frac{1}{2\sqrt2} \xi_{34} v_\eta v_\chi^2&\\
+\frac{1}{2}\lambda_{14}v_\rho^2 v_\sigma + \frac{1}{4}\zeta_{14} v_\rho^2 v_\sigma + \frac{1}{2}\lambda_{24} v_\eta^2 v_\sigma - \xi_{24} v_\eta^2 v_\sigma + \frac{1}{2} \lambda_{34} v_\chi^2 v_\sigma + \frac{1}{4} \zeta_{34} v_\chi^2 v_\sigma&=0
\nonumber
\end{align}
These conditions are inserted into the the tree-level mass matrices of the CP-even and CP-odd Higgs sectors, derived from $M_{ij}=\left.{\partial^2 V}/{\partial \phi_i\partial \phi_j}\right|_{vev}$, where $V$ is the potential in
Eq.~(\ref{pot}). The mass eigenstates are defined as follow
\begin{equation}
   h=R^S \begin{pmatrix}
        {\rm Re}\, \rho^0\\ {\rm Re}\, \eta^0\\ {\rm Re}\, \chi^0 \\ {\rm Re}\, \sigma^0_1\\ \s^{0}_2
    \end{pmatrix}
  Ah=R^P   \begin{pmatrix}
         {\rm Im}\, \rho^0\\ {\rm Im}\, \eta^0\\ {\rm Im}\, \chi^0 \\ {\rm Im}\, \sigma^0_1
    \end{pmatrix}
    H^+=R^C \begin{pmatrix}
         \r^{+}\\ \chi^{+}\\ \eta^{+}_1\\ \eta^{+}_2\\ \s^{+}_1 \\ \s^{+}_2
    \end{pmatrix}
   H^{++}=R^{2C} \begin{pmatrix}
         \r^{++}\\ \chi^{++}\\ \s^{++}_1 \\ \s^{++}_2
    \end{pmatrix}
\end{equation}
where the explicit expressions of the mass matrices
are too cumbersome to be presented here, and are given in Appendix \ref{massmat}. \\
In this case we have 5 scalar Higgs bosons, one of them will be the SM
Higgs of mass about 125 GeV, along with 4 neutral pseudoscalar Higgs bosons, out of which, 2  are the Goldstones of the $Z$ and the $Z^\prime$ massive vector bosons. In addition, there are 6 charged Higgses, 2 of which are the charged Goldstones, and 3 are doubly-charged Higgses, one of which is a Goldstone boson.\\
Hereafter we shall give the schematic expression of the physical Higgs states,
after EWSB, in terms of the gauge eigenstates, whose expressions contain only the vev of the various fields.
 In the following equations, ${\rm R}_{ij}^K\equiv{\rm R}_{ij}^K(m_1,m_2,m_3,\lambda_1,\lambda_2,\ldots)$ refers to the rotation matrix of each Higgs sector that depends on all the parameters of the potential in Eq.~(\ref{pot}). Starting from the scalar (CP-even) Higgs bosons we have
\bea
H_i = {\rm R}_{i1}^S {\rm Re}\, \rho^0 + {\rm R}_{i2}^S {\rm Re}\, \eta^0 +
{\rm R}_{i3}^S {\rm Re}\, \chi^0 +{\rm R}_{i4}^S {\rm Re}\, \sigma^0_1 +
{\rm R}_{i5}^S {\rm Re}\, \sigma_2^0,
\eea 
expressed in terms of the rotation matrix of the scalar components
${\rm R}^S$. There are similar expressions for the pseudoscalars 
\bea
Ah_i = {\rm R}_{i1}^P {\rm Im}\, \rho^0 + {\rm R}_{i2}^P {\rm Im}\,
\eta^0 +{\rm R}_{i3}^P{\rm Im}\, \chi^0 +{\rm R}_{i4}^P {\rm Im}\, \sigma^0_1 +{\rm R}_{i5}^P {\rm Im}\, \sigma_2^0
\eea 
in terms of the rotation matrix of the pseudoscalar components ${\rm R}^P$.
Here, however, we have two Goldstone bosons responsible for the generation of the masses of the neutral gauge bosons $Z$ and $Z^\prime$ given by
\begin{align}
A_0^1 &= \frac{1}{N_1}\left(v_\rho {\rm Im}\,\rho^0 -v_\eta {\rm Im}\,\eta^0 + v_\sigma {\rm Im}\,\sigma^0_1\right),\,\qquad N_1=\sqrt{v_\rho^2+v_\eta^2+v_\sigma^2}\ ;\\
A_0^2 &= \frac{1}{N_2}\left(-v_\rho {\rm Im}\,\rho^0+ v_\chi {\rm Im}\,\chi^0\right),\,\qquad N_2=\sqrt{v_\rho^2+v_\chi^2}.
\end{align}

For the charged Higgs bosons the interaction eigenstates are
\bea
H_i^+ = {\rm R}^{C}_{i1}\rho^{+} + {\rm R}^{C}_{i2}(\eta^{-})^* + {\rm R}^{C}_{i3}\eta^{+} + {\rm R}^{C}_{i4}(\chi^{-})^* + {\rm R}_{i5}^C \sigma_1^+ + {\rm R}_{i6}^C (\sigma_2^-)^*,
\eea 
with ${\rm R}^C$ being a rotation matrix of the charged sector. 
 We recall that even in this case two $H^+_i$ are massless  Goldstones bosons, because in the minimal 331 model there are the $W^\pm$ and the $V^\pm$ gauge bosons that both become massive after EWSB. The explicit expressions of the Goldstones are
\begin{align}
H_{W}^+ &= \frac{1}{N_W} \left(-v_\eta\eta^+ + v_\chi (\chi^-)^* + v_\sigma (\sigma_2^-)^*\right),\,\qquad N_W=\sqrt{v_\eta^2 + v_\chi^2 + v_\sigma^2}; \\
H_{V}^+ &= \frac{1}{N_V} \left(v_\rho \rho^+ - v_\eta (\eta^-)^* + v_\sigma \sigma_1^+\right),\,\qquad N_V=\sqrt{v_\rho^2 + v_\eta^2 + v_\sigma^2}.
\end{align}
In particular, we are interested in the doubly-charged Higgses, where the number of physical states, after EWSB, is three, whereas 
we would have had only
one physical doubly-charged Higgs, if we had not included the
sextet. The physical doubly-charged Higgs states are expressed in
terms of the gauge eigenstates and the elements of the
rotation matrix ${\rm R}^C$ as
\bea
H_i^{++} ={\rm R}^{2C}_{i1}\rho^{++} + {\rm R}^{2C}_{i2}(\chi^{--})^* + {\rm R}^{2C}_{i3}\sigma_1^{++} + {\rm R}^{2C}_{i4}(\sigma_2^{--})^*.
\eea
In particular, the structure of the corresponding Goldstone boson is
\bea
H_0^{++} =\frac{1}{N}\left(-v_\rho\rho^{++} + v_\chi(\chi^{--})^* - \sqrt2v_\sigma\sigma_1^{++} + \sqrt2v_\sigma(\sigma_2^{--})^*\right)
\eea
where $N=\sqrt{v_\rho^2+v_\chi^2+4v_\sigma^2}$ is a normalization factor.

\section{The Yukawa Sector}\label{331lep}
The model presented in the previous section exhibits
the interesting feature of having both scalar and vector doubly-charged
bosons, which is a peculiarity of the minimal version of the 331 model.
In fact it is possible to consider various versions of the $SU(3)_c\times SU(3)_L\times U(1)_X$ gauge symmetry, usually parametrized by $\beta$ \cite{cao_diphoton_2016,buras_anatomy_2013}. We discuss the case of $\beta=\sqrt3$, corresponding to the minimal version presented here \cite{frampton_chiral_1992,pisano_su3ensuremathbigotimesu1_1992}, leading to vector bosons with electric charge equal to $\pm 2$.\\
Doubly-charged states hold particular interest due to their potential for unique characteristics in terms of permissible decay channels, such as the production of same-sign lepton pairs \cite{corcella_exploring_2018,corcella_bilepton_2017,corcella_non-leptonic_2022,corcella_vector-like_2021,calabrese_331_2024}. Within the framework of the minimal 331 model, an even more intriguing prospect arises. It becomes possible to discern whether a same-sign lepton pair originates from either a scalar or a vector boson. As we will elaborate, this distinction also offers insights into the existence of a higher representation within the $SU(3)_c\times SU(3)_L\times U(1)_X$ gauge group, notably the sextet.

\subsection{The triplet sector}
In the previous section we have seen that the EWSB mechanism is realised in the 331 model by giving a vev to the neutral component of the triplets $\rho$, $\eta$ and $\chi$. The Yukawa interactions for SM and exotic quarks
are obtained by means of these scalar fields and are given by 
\begin{align}
    \mathcal{L}^{Y}_{q,\ triplet}=\ &- \overline{Q}_m\brak{Y^d_{m\a} \eta^*d_{\a R}+Y_{m\a}^u\chi^*u_{m\a}}
    -\overline{Q}_3( Y^d_{3\a}\chi d_{\a R}+Y^u_{3\a}\eta u_{m\a})+\nonumber\\
    &-\overline{Q}_{m}(Y^J_{mn}\chi J_{nR})-\overline{Q}_{3}Y^J_{3}\chi J_{3R} +\textnormal{h.c.}\ 
\end{align}
where $y^i_{d}$, $y^i_u$ and $y^i_E$ are the Yukawa couplings for down-,
up-type and exotic quarks, respectively. The masses of the exotic quarks are related to the vev of the neutral component of $\rho=(0,0,v_\rho)$ via the invariants
\begin{eqnarray}
 Q_1\,  \rho^*  D_R^*, Q_1\,  \rho^*  S_R^*&\sim & (3,3,-1/3)\times (1,\bar{3},-1)\times (\bar{3},1,4/3) \nonumber \\
 Q_3\,  \rho  T_R^* &\sim& (3,\bar{3},2/3)\times (1,{3},1)\times (\bar{3},1,-5/3), 
\end{eqnarray}
responsible of the breaking $SU(3)_c\times SU(3)_L\times U(1)_X \to SU(3)_c\times SU(2)_L\times U(1)_Y$. It is clear that, being
$v_\rho\gg v_{\eta,\chi}$, the masses of the exotic quarks are
${\cal O}(\rm{TeV})$ whenever the relation $Y^J\sim \mathbb{1}$ is satisfied. As we are going to see, the model also needs a sextet.

\subsection{The sextet Yukawa coupling}
The need for introducing a sextet sector can be summarised as follows. 
A typical Dirac mass term for the leptons in the SM
is associated with the operator $\bar{l}_LH e_R$, with $l_L=(v_{eL},e_L)$ being the $SU(2)_L $ doublet, with the representation content $(\bar{2},1/2)\times (2,1/2)\times(1,-1)$ (for $l, H$ and $e_R$, respectively) in $SU(2)_L\times U(1)_Y$. In the 331 model, the $L$ and $R$ components of the lepton $(e)$ belong to the same multiplet. Consequently, identifying an $SO(1,3)\times SU(3)_L$ singlet requires two leptons in the same representation. This can be achieved (at least partially) with the operator
\begin{eqnarray}
\mathcal{L}_{l,\, triplet}^{Yuk}&=& G^\eta_{a b}( l^i_{a \alpha}\epsilon^{\alpha \beta} l^j_{b \beta})\eta^{* k}\epsilon^{i j k} + \rm{h. c.}\nonumber\\
&=& G^\eta_{a b}\, l^i_{a}\cdot l^j_{b}\,\eta^{* k}\epsilon^{i j k} + \rm{h. c.} 
\end{eqnarray}
where the indices $a$ and $b$ run over the three generations of flavour, $\alpha$ and $\beta$ are Weyl indices contracted in order to generate  an $SO(1,3)$ invariant ($l^i_{a}\cdot l^j_{b}\equiv l^i_{a \alpha}\epsilon^{\alpha \beta} l^j_{b \beta}$) from two Weyl fermions, and $i,j,k=1,2,3$, are $SU(3)_L$ indices. \\
 The use of 
$\eta$ as a Higgs field is mandatory, since the components of the multiplet $l^j$ are $U(1)_X$ singlets. 
The representation content of the operator $l^i_a l^j_b$ according to $SU(3)_L$ is given by $3\times 3= 6 + \bar{3}$, with the 
$\bar{3}$ extracted by an anti-symmetrization over $i$ and $j$ via $\epsilon^{i j k}$. This allows to identify 
$l^i_a l^j_b \eta^{*k}\epsilon^{i j k}$ as an $SU(3)_L$ singlet. Considering that the two leptons are anticommuting Weyl spinors, and that the $\epsilon^{\alpha\beta}$ (Lorentz) and $\epsilon^{i j k}$ ($SU(3)_L$) contractions introduce two sign flips under the $a\leftrightarrow b$ exchange, the combination 
\begin{equation}
M_{a b}=( l^i_{a}\cdot l^j_{b })\eta^{* k}\epsilon^{i j k} 
\end{equation}
is therefore antisymmetric under the exchange of the two flavours, implying
that even $G_{a b}$ has to be antisymmetric.  However, an antisymmetric $G^\eta_{a b}$ is not sufficient to provide mass to all the leptons.\\
In fact, the diagonalization of $G^\eta$ by means a unitary matrix 
$U$, namely $G^\eta=U \Lambda U^\dagger$, with $G^\eta$ antisymmetric in
flavour space, implies that its 3 eigenvalues are given by
$\Lambda=(0,\lambda_{22}, \lambda_{33})$, with $\lambda_{22}=-\lambda_{33}$,
i.e. one eigenvalue is null and the other two are equal in magnitude.
At the minimim of $\eta$, i.e. $\eta=(0,v_\eta,0)$, one has
  \begin{equation} 
G^\eta_{a b}M^{a b}=-Tr (\Lambda\, U M U^\dagger)=
2 v_{\eta}\lambda_{22}\, U_{2 a}\, l^{1}_{a}\cdot l^{3}_{b}\,U_{2 b}^* + 2 v_{\eta}\lambda_{33}\, U_{3 a}\, l^{1}_{a}\cdot l^{3}_{b}\,U_{3 b}^*, 
 \end{equation}
with $ l^1_a=e_{a L}$ and  $l^3_b=e_{b R}^{c}$. Introducing the linear combinations 
\begin{equation} 
  E_{2 L}\equiv U_{2 a}\, l^1_a=U_{2 a} \, '\, e_{a L} \qquad U_{2 b}^*\,
  l^3_b=U_{2 b}^*\, e_{b R}^{c} = i\sigma_2(U_{2 b} \,e_{b R})^*\equiv
  E_{2 R}^{c},
\end{equation}
then the antisymmetric contribution in flavour space becomes 
\begin{equation}
\mathcal{L}_{l, \, triplet}^{Yuk}= 2 v_{\eta}\lambda_{22} \left( E_{2 L}E_{2 R}^{c}  - E_{3 L}E_{3 R}^{c}\right),
\end{equation}
which is clearly insufficient to generate the lepton masses of three
non-degenerate
lepton families. We shall solve this problem by introducing a second 
invariant operator, with the inclusion of a sextet $\sigma$
\begin{equation}
\sigma=\left(
\renewcommand*{\arraystretch}{1.5}
\begin{array}{ccc}
\sigma_1^{++}&\sigma_1^+/\sqrt2&\sigma^0/\sqrt2\\
\sigma_1^+/\sqrt2&\sigma_1^0&\sigma_2^-/\sqrt2\\
\sigma^0/\sqrt2&\sigma_2^-/\sqrt2&\sigma_2^{--}
\end{array}
\right)\in(1,6,0),
\end{equation}
leading to the Yukawa term
\begin{equation}\label{lag}
\mathcal{L}_{l, sextet}^{{Yuk.}}= G^\sigma_{a b} l^i_a\cdot l^j_b \sigma^*_{i,j},
\end{equation}
which allows to build a singlet out of the representation
$6$ of $SU(3)_L$, contained in $l^i_a\cdot l^j_b$, by combining it with
the flavour-symmetric
$\sigma^*$, i.e. $\bar{6}$.
Notice that $G^\sigma_{a b}$ is symmetric in flavour space. \\
It's interesting to note that without considering the sextet, a doubly-charged scalar wouldn't be able to decay into same-sign leptons. This is because, without the sextet, the interaction responsible for the leptons only involves the scalar triplet, denoted as $\eta$, which doesn't contain a doubly-charged state. So, if we detect a decay like $H^{\pm\pm}\to l^\pm l^\pm$, it would directly indicate the presence of the sextet in the 331 model.

\subsection{Lepton mass matrices}
Let's now come to discuss the lepton mass matrices in the model. They are related to the Yukawa interactions by the Lagrangian 
\begin{equation}
\mathcal{L}_{l}^{{Yuk.}}=\mathcal{L}_{l, sextet}^{{Yuk.}} + \mathcal{L}_{l, triplet}^{{Yuk.}} + \rm{h. c.}
\end{equation}
and are combinations of triplet and sextet contributions.
The structure of the mass matrix that emerges
from the vevs of the neutral components of  $\eta$ and $\sigma$ is thus
given by
\begin{equation} 
\mathcal{L}_{l}^{{Yuk.}}=\left(\sqrt{2} \sigma_0 G_{a,b}^\sigma  +2 v_\eta G^\eta_{a b}\right) (e_{a L}\cdot e_{b R}^{c}) 
+\sigma_1^0 G^\sigma_{a b} \left(\nu_L^T i \sigma_2 \nu_L\right)  +\rm{h. c.},
\end{equation}
which generates a Dirac mass matrix for the charged leptons $M_{ab}^l$
and a Majorana mass matrix for neutrinos $M_{a b}^{\nu_l}$
\begin{equation}\label{mlgen}
M^l_{a b} =\sqrt{2} \langle\sigma_0\rangle\, G_{a,b}^\sigma  +2 v_\eta\, G^\eta_{a b} \qquad , \qquad M_{a b}^{\nu_l}=\langle \sigma^0_1\rangle \, G^\sigma_{a b}. 
\end{equation}
In the expression above $\langle\sigma^0\rangle$ and $\langle\sigma_1^0\rangle$ are the vacuum expectation values of the neutral components of $\sigma$. For a vanishing $G^\sigma$, as we have already discussed, we will not be able to generate the lepton masses consistently, nor any mass for the neutrinos, i.e.   
\begin{equation}\label{mltip}
M^l_{a b}=2 v_\eta\, G^\eta_{a b} \qquad , \qquad M^{\nu_l}=0.
\end{equation} 
On the contrary, in the limit $G^\eta\to0$, Eq.~(\ref{mlgen}) becomes 
\begin{equation}\label{mlsext}
  M^l_{a b}=\sqrt{2} \langle\sigma_0\rangle\, G_{ab}^\sigma \qquad,
  \qquad M^{\nu_l}_{a,b}=\frac{\langle\sigma_1^0\rangle}{\sqrt2}G^\sigma_{a b},
\end{equation}
which has some interesting consequences.
Since the Yukawa couplings are the same for both leptons and neutrinos,
we have to require $\langle\sigma_1^0\rangle\ll\langle\sigma^0\rangle$, in order to obtain small neutrino masses. For the goal of our analysis, we will
assume that the vev of $\sigma_1^0$ vanishes, i.e.
$\langle\sigma_1^0\rangle\equiv0$. Clearly,
if the matrix $G^\sigma$ is diagonal in flavour space, from Eq.
(\ref{mlsext}) we will
immediately conclude that the Yukawa coupling $G^\sigma$ has
to be chosen to be proportional to the masses of the SM leptons.
An interesting consequence of this is that the decay $H^{\pm\pm}\to l^\pm l^\pm$, which is also proportional to $G^\sigma$, and therefore to the lepton masses,
will be enhanced for the heavier leptons, in particular for the $\tau$,
as thoroughly discussed in  \cite{tonasse_decay_2012}.
This is an almost unique situation which is not encountered in other models with doubly-charged scalars decaying into same-sign leptons \cite{muhlleitner_note_2003}.

\section{Flavor Physics in the minimal 331 Model}
One of the features of the minimal 331 Model lies in its arrangement of fermions within triplets of $SU(3)_L$. However, to maintain anomaly cancellation, it becomes necessary to assign one of the quark families to a different representation than the other two, ensuring an equal number of triplets and anti-triplets in the fermion sector.\\
This introduces several complexities, particularly concerning flavor physics within the model. Firstly, to achieve the spontaneous symmetry breaking of the 331 gauge symmetry to the $SU(3)_C\times U(1)_{em}$ symmetry, at least three scalar triplets of $SU(3)_L$ must be introduced. While these are sufficient to impart masses to quarks in the quark sector, the flavor structure becomes intricate due to the differing group representations of the three quark families.\\
Conversely, in the lepton sector, a fundamentally different situation arises. Realistic masses cannot be obtained using only three triplets. As previously argued, the introduction of a scalar sextet belonging to $(\textbf{1},\textbf{6},0)$ becomes necessary to generate appropriate masses for charged leptons.\\
A general feature of models of this kind, where mass terms arise from different scalar fields, is the introduction of flavor-changing neutral currents mediated by neutral scalars \cite{pleitez_challenges_2022}.

\subsection{Quark sector}

Let's revisit the presence of three scalar triplets within the model. The first triplet,

\begin{equation}
    \rho=\begin{pmatrix}
        \rho^{++}\\ \rho^+\\ \rho^0
    \end{pmatrix}\in (\textbf{1},\textbf{3},1)\ ,
\end{equation}

acquires a vacuum expectation value (vev) on the order of the spontaneous symmetry breaking of the 331 symmetry. The other two triplets,

\begin{equation}
    \eta=\begin{pmatrix}
        \eta^{+}_1\\ \eta^0\\ \eta^-_2
    \end{pmatrix}\in (\textbf{1},\textbf{3},0)\ ,
\end{equation}

\begin{equation}
    \chi=\begin{pmatrix}
        \chi^{0}\\ \chi^-\\ \chi^{--}
    \end{pmatrix}\in (\textbf{1},\textbf{3},-1)\ ,
\end{equation}

acquire non-zero vevs at the electroweak scale. In the quark sector, the Yukawa interactions are described by

\begin{align}\label{yuklagq}
    \mathcal{L}^{Y}_{q,\ triplet}=\ &- \overline{Q}_m\brak{Y^d_{m\a} \eta^*d_{\a R}+Y_{m\a}^u\chi^*u_{m\a}}
    -\overline{Q}_3( Y^d_{3\a}\chi d_{\a R}+Y^u_{3\a}\eta u_{m\a})+\nonumber\\
    &-\overline{Q}_{m}(Y^J_{mn}\chi J_{nR})-\overline{Q}_{3}Y^J_{3}\chi J_{3R} +\textnormal{h.c.}\ .
\end{align}

The mass matrices for the up-type and down-type quarks arise when all scalar triplets acquire real vevs: $\rho\rightarrow v_\rho/\sqrt{2}$, $\eta\rightarrow v_\eta/\sqrt{2}$, and $\chi\rightarrow v_\chi/\sqrt{2}$. Both matrices involve contributions from the triplets $\eta$ and $\chi$. Exotic quark mass terms emerge following the initial spontaneous symmetry breaking.

Since there are two quarks, namely $D$ and $S$ with electric charge $Q=-4/3\ e$, they undergo mixing via a Cabibbo-like $2\times2$ matrix
\begin{equation}
    \begin{pmatrix}
        D'_R\\S'_R
    \end{pmatrix}
    =\Tilde{V}_R^{-1}\begin{pmatrix}
        D_R\\S_R
    \end{pmatrix}\ ,\ \ \ \ \ \ \ \begin{pmatrix}
        D'_L\\S'_L
    \end{pmatrix}
    =\Tilde{V}_L^{-1}\begin{pmatrix}
        D_L\\S_L
    \end{pmatrix}\ .
\end{equation}

Notably, only the quark $T$ remains unmixed, being the sole quark with $Q=5/3\ e$. The resulting mass matrices for ordinary quarks from the Lagrangian in Equation \eqref{yuklagq} are as follows

\begin{equation}
    M_u=\begin{pmatrix}
        v_\chi Y^u_{11} & v_\chi Y^u_{12} &v_\chi Y^u_{13} \\
        v_\chi Y^u_{21} & v_\chi Y^u_{22} &v_\chi Y^u_{23}\\
        v_\eta Y^u_{31} & v_\eta Y^u_{32} &v_\eta Y^u_{33}\\
    \end{pmatrix}\ ,
\end{equation}

\begin{equation}
    M_d=\begin{pmatrix}
        v_\eta Y^d_{11} & v_\eta Y^d_{12} &v_\eta Y^d_{13} \\
        v_\eta Y^d_{21} & v_\eta Y^d_{22} &v_\eta Y^d_{23} \\
        v_\chi Y^d_{31} & v_\chi Y^d_{32} &v_\chi Y^d_{33} \\
    \end{pmatrix}\ .
\end{equation}

In the minimal version, there is typically no reason to initially place one type of quark in the diagonal basis, unless some specific additional symmetry is introduced. Therefore, similar to the Standard Model, we proceed by independently rotating down-type and up-type quarks into their mass eigenstates
\begin{equation}\label{rotu}
   \begin{pmatrix}
        d'_L\\s'_L\\b'_L
    \end{pmatrix}
    =V_L^{-1}\begin{pmatrix}
        d_L\\s_L\\b_L
    \end{pmatrix}\ \ \ \ \ \ \begin{pmatrix}
        d'_R\\s'_R\\b'_R
    \end{pmatrix}
    =V_R^{-1} \begin{pmatrix}
        d_R\\s_R\\b_R
    \end{pmatrix}
\end{equation}
\begin{equation}\label{rotd}
   \begin{pmatrix}
        u'_L\\c'_L\\t'_L
    \end{pmatrix}
    =U_L^{-1}\begin{pmatrix}
        u_L\\c_L\\t_L
    \end{pmatrix} \ \ \ \ \ \  \begin{pmatrix}
        u'_R\\c'_R\\t'_R
    \end{pmatrix}
    =U_R^{-1}\begin{pmatrix}
        u_R\\c_R\\t_R
    \end{pmatrix}
\end{equation}
Here, the primed fields represent flavor eigenstates, while unprimed fields denote mass eigenstates. Matrices $U_L$, $U_R$, $V_L$, and $V_R$ are unitary matrices satisfying
\begin{equation}
   V_L^\dagger V_L =V_R^\dagger V_R=U_L^\dagger U_L=U_R^\dagger U_R= \mathbb{1}\ ,
\end{equation}
which diagonalize $M_u$ and $M_d$ respectively via bi-unitary transformations
\begin{equation}
   U_L^\dagger M_u U_R=\hat{M}_u\ \ \ \ \  V_L^\dagger M_d V_R=\hat{M}_d\ ,
\end{equation}
where $\hat{M}_{u}$ and $\hat{M}_{d}$ are diagonal $3\times 3 $ matrices containing quark masses.\\
In the Standard Model all the three families are placed within the same group representation of the gauge symmetry, this results in no family distinction in the Lagrangian apart from the Yukawa sector. Here, non-diagonal terms between quark masses arise if the couplings are non-diagonal. The same field rotation procedure is implemented in the Standard Model to diagonalize Yukawa interactions while leaving the Lagrangian mostly unchanged, the only part sensitive to flavour rotation is the $W$ boson interactions, where the CKM matrix arises as
\begin{equation}
    V_{CKM}=U_L^\dagger V_L\ .
\end{equation}
Every other sector in the Lagrangian remains unaffected. Consequently, the GIM mechanism is naturally implemented, since there are no flavor-changing neutral currents (FCNCs) at tree-level, resulting in a natural suppression of these processes, which can only occur at loop-level. Up-type quark masses can be assumed diagonal from the outset, while down-type quark flavor and mass eigenstates can be related through the CKM matrices, as any other rotation effects cancel out in the Lagrangian.\\
However, in the context of the minimal 331 Model, the situation is fundamentally different. 
As previously emphasized, the appealing feature of constraining the number of families is that it causes every sector of the Lagrangian to be sensitive to flavor, owing to the different group representations to which the quarks are assigned. Consequently, flavor rotation matrices persist in various combinations in fermion interactions, either with scalars or gauge bosons.
From the Yukawa Lagrangian in Equation \eqref{yuklagq}, we can derive the fermion-scalar interactions after the field rotations discussed previously
\begin{equation}
    \mathcal{L}^{cc}_{scalar}=-\overline{\textbf{d}}_LV_L S_{du}^{cc}U_R^\dagger\textbf{u}_R-\overline{\textbf{u}}_LU_L S_{ud}^{cc}V_R^\dagger\textbf{d}_R+\textnormal{h.c.}\ ,
\end{equation}
where $\textbf{d}$ and $\textbf{u}$ represent the down-type and up-type flavor vectors in the mass basis, and
\begin{equation}\label{sdu}
    S_{du}^{cc}=\begin{pmatrix}
        \chi^- Y^u_{11} & \chi^- Y^u_{12} & \chi^- Y^u_{13} \\
        \chi^- Y^u_{21} & \chi^- Y^u_{22} & \chi^- Y^u_{23}\\
        \eta^-_1 Y^u_{31} & \eta^-_1 Y^u_{32} & \eta^-_1 Y^u_{33}\\
    \end{pmatrix}\ ,
\end{equation}

\begin{equation}\label{sud}
    S_{ud}^{cc}=\begin{pmatrix}
        \eta^+_1 Y^d_{11} &  \eta^+_1 Y^d_{12} & \eta^+_1 Y^d_{13} \\
        \eta^+_1 Y^d_{21} &  \eta^+_1 Y^d_{22} & \eta^+_1 Y^d_{23} \\
        \chi^+ Y^d_{31} &  \chi^+ Y^d_{32} & \chi^+ Y^d_{33} \\
    \end{pmatrix}\ .
\end{equation}
In the expression above, all the charged scalars are interaction egienstates, and we need to perform a rotation in \eqref{sud} and \eqref{sdu} in order to extract the corresponding mass eigenstates.
It's evident that in general, the combination of flavor rotation matrices cannot be further reduced, and they persist in a combination distinct from the CKM matrix. This behavior also manifests in the neutral currents mediated by scalar fields, where the Lagrangian becomes

\begin{equation}
    \mathcal{L}^{nc}_{scalar}=-\overline{\textbf{u}}_L U_L S_{u}^{nc}U_R^\dagger\textbf{u}_R-\overline{\textbf{d}}_LV_L S_{d}^{nc}V_R^\dagger\textbf{d}_R+\textnormal{h.c.}\ ,
\end{equation}
with
\begin{equation}\label{sdu}
    S_{u}^{nc}=\begin{pmatrix}
        \chi^0 Y^u_{11} & \chi^0 Y^u_{12} & \chi^0 Y^u_{13} \\
        \chi^0 Y^u_{21} & \chi^0 Y^u_{22} & \chi^0 Y^u_{23}\\
        \eta^0 Y^u_{31} & \eta^0 Y^u_{32} & \eta^0 Y^u_{33}\\
    \end{pmatrix}
\end{equation}
and
\begin{equation}\label{sud}
    S_{d}^{nc}=\begin{pmatrix}
        \eta^0 Y^d_{11} &  \eta^0 Y^d_{12} & \eta^0 Y^d_{13} \\
        \eta^0 Y^d_{21} &  \eta^0 Y^d_{22} & \eta^0 Y^d_{23} \\
        \chi^0 Y^d_{31} &  \chi^0 Y^d_{32} & \chi^0 Y^d_{33} \\
    \end{pmatrix}
\end{equation}

where, as in the case of charged scalars, also here we need a further rotation in order to extract the mass eigenstates of this sector.

Although the matrix combinations $U_L S_{u}^{nc}U_R^\dagger$ and $V_L S_{d}^{nc}V_R^\dagger$ resemble those in Equations \eqref{rotu} and \eqref{rotd}, they are insufficient to diagonalize the interaction. Hence, flavor-changing neutral currents in the scalar sector are a prediction of the minimal 331 Model. They can be controlled but not entirely avoided.\\
In the minimal 331 Model, eight bosons are present, including the neutral ones: the photon $A^\mu$, with straightforward interactions, $Z^\mu$ and $Z'^\mu$, and the charged ones, $W^\pm$, $V^\pm$, and the doubly charged bileptons $Y^{\pm\pm}$. However, rotations in flavor space do not leave the interactions unaffected. The d-type and u-type quarks interact with the $W^\pm$ boson as in the Standard Model
\begin{equation}
    \mathcal{L}_{duW}=\frac{g_2}{\sqrt{2}}\overline{\textbf{d}_L}\gamma^\mu V_{CKM}\textbf{u}_LW^-_\mu+\text{h.c.}
\end{equation}
Here, the CKM matrix appears as expected. Further interactions are possible between exotic quarks and ordinary quarks mediated by exotic gauge bosons. The interactions of u-type and d-type quarks with the exotic $T$ quark are given by

\begin{equation}
    \mathcal{L}_{uTV}=\frac{g_2}{\sqrt{2}}\overline{T_L}\gamma^\mu (U_{L})_{3j}\textbf{u}_{jL}V^+_\mu +\text{h.c.}
\end{equation}

\begin{equation}
    \mathcal{L}_{dTY}=\frac{g_2}{\sqrt{2}}\overline{T_L}\gamma^\mu ( V_{L})_{3j}\textbf{d}_{jL}Y^{++}_\mu +\text{h.c.}
\end{equation}

Here, only the third column of the rotation matrix appears due to the difference between the third quark generation and the other two. Similar interactions occur between up-type and down-type quarks with the other two exotic quarks $D$ and $S$
\begin{equation}\label{tt1}
    \mathcal{L}_{djY}=\frac{g_2}{\sqrt{2}}\overline{\textbf{j}_L}\gamma^\mu (O^\dagger V_{L})_{mj}\textbf{d}_{jL}V^{-}_\mu +\text{h.c.}\ \ \ \ \ \text{with } j_L=(D_L,\ S_L)
\end{equation}

\begin{equation}\label{tt2}
    \mathcal{L}_{ujY}=\frac{g_2}{\sqrt{2}}\overline{\textbf{j}_L}\gamma^\mu (O^\dagger U_{L})_{mj}\textbf{d}_{jL}Y^{++}_\mu +\text{h.c.}\ \ \ \ \ \text{with } j_L=(D_L,\ S_L)
\end{equation}
where the $O$ matrix in Equations \eqref{tt1} and \eqref{tt2} is a $2\times 2$ Cabibbo-like matrix which mixes the exotic quarks $D$ and $S$.\\
Finally, the interaction of ordinary quarks with the $Z'$ boson can be schematized as follows - omitting an overall coefficient - 
\begin{align}
\mathcal{L}_{qqZ'}=&\biggl(\overline{\textbf{u}_L}U_L^\dagger\gamma^\mu \Y^{u}_L U_L\textbf{u}_L+\overline{\textbf{d}_L}V_L^\dagger \gamma^\mu\Y^{d}_L V_L\textbf{d}_L\nonumber\\
    &+\overline{\textbf{u}_R}U_R^\dagger \gamma^\mu\Y^{u}_R U_R\textbf{u}_R +\overline{\textbf{d}_R}V_R^\dagger \gamma^\mu\Y^{d}_R V_R\textbf{d}_R \biggl)Z'_\mu
\end{align}
where we have defined the couplings with $\Y^u_L$ and $\Y^d_L$ which are proportional to the following matrix
\begin{equation}
    \Y_L^u=\Y_L^d\propto\begin{pmatrix}
        1-2\sin^2(\theta_W)&0&0\\
        0&1-2\sin^2(\theta_W)&0\\
        0&0&-1\\
    \end{pmatrix}
\end{equation}
In this interaction, there are flavor-changing neutral currents (FCNCs) in the left-handed interactions, but the right-handed neutral currents mediated by the gauge boson $Z'$ are diagonal in flavor space, as $\Y_R^u\propto\mathbb{1}$ and $\Y_R^d\propto\mathbb{1}$ \cite{machado_flavor-changing_2013}. \\
It's noteworthy that in the quark sector, the rotation matrices of the right-handed quarks cancel out from the Lagrangian, similar to the case of the Standard Model. Conversely, the left-handed $U_L, V_L$ matrices not only survive in a combination analogous to the CKM matrix of the Standard Model, but also independently. From a practical standpoint, it is possible to redefine the fields in the interactions - using the unitary condition $U_L^\dagger U_L=\mathbb{1}$ - to construct a Lagrangian for the quark sector in which only two matrices appear: the CKM matrix and $V_L$.\\
To obtain an appropriate parameterization for the matrix $V_L$, it is necessary to initially enumerate the additional parameters present within this matrix. Upon examining all conceivable interaction terms, it becomes apparent that, subsequent to employing phase transformations of the up and down-type quarks to simplify the CKM matrix, three more potential phases emerge from transformations in the $D$, $S$, and $T$ quarks. This results in a total of six supplementary parameters, encompassing three mixing angles and three phases. However, it is evident that only the $\tilde{V}_{3j}$ elements are essential when computing FCNCs, thus allowing for a parameterization that effectively diminishes the number of parameters involved. We obtain \cite{promberger_flavor-changing_2007}

\begin{equation}
    V_L=\begin{pmatrix}
        c_{12}c_{13}            & s_{12}c_{23}e^{i \d_3}-c_{12}s_{13}s_{23}e^{i (\d_1-\d_2)}& c_{12}c_{23}s_{13}e^{i \d_1}+s_{12}s_{23}e^{i (\d_2+\d_3)}\\
        -c_{13}s_{12}e^{-i \d_3}& c_{12}c_{23}+s_{12}s_{13}s_{23}e^{i (\d_1-\d_2-\d_3)}& -s_{12}c_{23}s_{13}e^{i (\d_1-\d_3)}-c_{12}s_{23}e^{i \d_2}\\
        -s_{13}e^{-i \d_1}      & -c_{13}s_{23}e^{-i \d_2}& c_{13}c_{23}\\
    \end{pmatrix}\ .
\end{equation}
where only two additional CP violating quantities $\d_1$ and $\d_2$ appear, that are responsible for the additional CP violating effects.

\subsection{Lepton Sector}
In the minimal 331 Model, similar to quarks, the leptonic sector exhibits a multitude of parameters. Despite the $Z$ and $Z'$ interactions being diagonal in flavor space (owing to the consistent transformation behavior of the three lepton generations under the electroweak symmetries $SU(3)_L\otimes U(1)_X$), Flavor-Changing Neutral Currents (FCNCs) occur in the scalar sector \cite{machado_lepton_2019}. These parameters result from the diverse contributions to the mass matrices of charged leptons, mirroring the scenario observed in quarks.

The Yukawa interactions in the lepton sector must incorporate the triplet $\eta$ and the sextet, whose components are detailed as follows
\begin{equation}
    \s=\begin{pmatrix}
      \s^{++}_1&\frac{\s_1^+}{\sqrt{2}}&\frac{\s^0_1}{\sqrt{2}}\\
      \frac{\s_1^+}{\sqrt{2}}&\s_2^0&\frac{\s^-_2}{\sqrt{2}}\\ 
      \frac{\s^0_1}{\sqrt{2}}&\frac{\s^-_2}{\sqrt{2}}&\s_2^{--}\\
    \end{pmatrix}\ .
\end{equation}
In the minimal 331 Model, including the sextet in the scalar sector is necessary to assign physical masses to charged leptons. This necessity arises from how the Yukawa interaction is constructed from group theory. When combining three triplets according to $\textbf{3} \otimes \textbf{3} \otimes \textbf{3}$, the resulting invariant structure demands antisymmetry among the triplets. As a consequence, the Yukawa matrix must exhibit antisymmetry in flavor indices to allow for vanishing interactions. However, this interaction pattern leads to eigenvalues of $(0, m, -m)$ of this matrix, which is evidently an unphysical solution.\\
Once spontaneous symmetry breaking occurs, the scalars acquire vacuum expectation values as follows

\begin{equation}
    \eta=\frac{1}{\sqrt{2}}\begin{pmatrix}
        \eta^{+}_1\\\frac{1}{\sqrt{2}} v_\eta +\frac{1}{\sqrt{2}}\Sigma_\eta+\frac{i}{\sqrt{2}}\z_\eta\\ \eta^-_2
    \end{pmatrix}
\end{equation}

\begin{equation}
    \s=\frac{1}{\sqrt{2}}\begin{pmatrix}
      \s^{++}_1&\frac{\s_1^+}{\sqrt{2}}&\frac{v_\s+\Sigma_\s+i\z_\s}{2}\\
      \frac{\s_1^+}{\sqrt{2}}&\s_2^0&\frac{\s^-_2}{\sqrt{2}}\\ 
      \frac{v_\s+\Sigma_\s+i\z_\s}{2}&\frac{\s^-_2}{\sqrt{2}}&\s_2^{--}\\
    \end{pmatrix}\ .
\end{equation}
In the context of the minimal 331 Model, as originally proposed, neutrinos are massless at tree level. However, a mechanism to generate massive neutrinos can be obtained through the scalar sextet. Indeed, the component $\s_2^0$ can also acquire a vev, which can be used in the Yukawa interactions to construct a Majorana mass term for the neutrinos
\begin{equation}
    \mathcal{L}^Y_\nu=-\overline{\nu_L^c}G^\s \nu_L\s+\text{h.c.}\ .
\end{equation}
This interaction is not invariant under flavor rotation, therefore an additional rotation matrix appears in this sector in order to diagonalize the neutrino mass terms
\begin{equation}
    V_\nu^T M_\nu V_\nu=\hat{M}_\nu\ ,
\end{equation}
where $\hat{M}_\nu$ is a diagonal matrix containing physical neutrino masses.\\
The inclusion of this additional vev carries further implications such as the mixing between the singly charged gauge bosons $W^\pm$ and $V^\pm$, which can be predicted to be a small value of $v_{\s_2}$ chosen to yield small neutrino masses. The interactions between charged leptons, neutrinos, and $W^\pm$ remain the same as in the Standard Model

\begin{equation}
    \mathcal{L}_{l\nu W}=\frac{ig_2}{2\sqrt{2}}\left(\overline{\nu_L}V_{PMNS}^\dagger\gamma^\mu l_L+\overline{(l^c)_R}V_{PMNS}\gamma^\mu(\nu^c)_R\right)W^+_\mu+\text{h.c.}
\end{equation}
where $V_{PMNS}$ is the Pontecorvo-Maki-Nakagawa-Sakata matrix given by

\begin{equation}
    V_{PMNS}=V_l^\dagger V^\nu\ .
\end{equation}
The model also predicts that the interaction between charged leptons and neutrinos must include additional contributions from $V^\pm$

\begin{equation}\label{aa1}
    \mathcal{L}_{l\nu V}=\frac{ig_2}{2\sqrt{2}}\left(\overline{l^c_L}V_{l\nu}^*\gamma^\mu \nu_L+\overline{\nu^c_R}V_{l\nu}\gamma^\mu(l)_R\right)V^+_\mu+\text{h.c.}
\end{equation}
where the definition
\begin{equation}\label{ab1}
    V_{l\nu}=(V^\nu_L)^\dagger (V_R^l)^*\ .
\end{equation}
In this model, there are also interactions between charged leptons and doubly charged vector bosons given by the Lagrangian
\begin{equation}
    \mathcal{L}_{llY}=\frac{ig_2}{2\sqrt{2}}\left(\overline{l^c}\gamma^\mu\left(\tilde{V}_{l\nu}-\tilde{V}_{l\nu}^T\right)-\gamma^5\gamma^\mu\left(\tilde{V}_{l\nu}+\tilde{V}_{l\nu}^T\right)\right)Y^{++}
\end{equation}
with
\begin{equation}\label{ab2}
    \tilde{V}_{l\nu}=(V^l_R)^T V_L^l\ .
\end{equation}
Finally, leptons couple universally to neutral vector bosons since no distinction has been made between generations of leptons, where the $Z'$ boson also has the property to be Leptophobic \cite{gomez_dumm_leptophobic_1997}.\\
Therefore, in the minimal 331 Model, $V_\nu$ is always not equal to $V_{PMNS}$, and $V^l_L$ and $V^l_R$ both appear separately in the combination shown in Equations \eqref{ab1} and \eqref{ab2}. Because the charged lepton mass matrix includes an anti-symmetric contribution, we cannot assume it is diagonal from the beginning. This means we cannot simply set $V^l_R$ equal to $\mathbb{1}$ in the interactions in Equation \eqref{aa1}. However, if interactions between leptons and the scalar triplet $\eta$ are forbidden by some discrete symmetry, then this simplification becomes possible.\\
In this scenario, the charged lepton mass matrix is diagonalized using the same unitary matrix as the neutrino mass matrix, making the PMNS matrix straightforward. Similarly, just like in the quark sector, determining the values of the entries in the matrices $V_\nu$, $V^l_L$, and $V^l_R$ is crucial for understanding the realistic behavior of this model.

\section{Conclusions}
The 331 model stands as a remarkable embodiment of the anomaly cancellation mechanism, encompassing all fermion generations. It represents a compelling avenue for extending the fermion sector of the Standard Model in an alternative direction. Within the framework of the 331 model, the significance of anomaly constraints cannot be overstated. They serve as pivotal guiding principles, steering us away from the conventional sequential structure of the Standard Model's fermion families. Unlike the traditional model, which lacks a clear rationale regarding the number of fermion families, the 331 model offers a fresh perspective, inviting exploration and deeper understanding of fundamental particle interactions.\\

\centerline{\bf Acknowledgements}
This work is dedicated to Paul Frampton, whose exceptional and insightful contributions to physics have greatly enriched the field as well as our perspective. We extend our heartfelt gratitude to him for the invaluable knowledge shared during our time together at the University of Salento, in Lecce and Martignano, and in numerous gatherings around the globe. Wishing him many more years filled with joy, success, and continued inspiration.\\
We thank Gennaro Corcella and Antonio Costantini for a long time collaborations and for sharing their insight on the analysis of the 331 model.
This work is partially supported by INFN within the Iniziativa Specifica QG-sky.  
The work of C. C. and D.M. is funded by the European Union, Next Generation EU, PNRR project "National Centre for HPC, Big Data and Quantum Computing", project code CN00000013. This work is partially supported by the the grant PRIN 2022BP52A MUR "The Holographic Universe for all Lambdas" Lecce-Naples.

\appendix

\section{Rotations for the determination of the mass eigenstates}\label{massmat} 
In this appendix we summarize some results concerning the scalar/psesudoscalar sectors of the model \cite{staub_automatic_2011,staub_sarah_2014}. 
 
\begin{itemize} 
\item {\bf Mass matrix of the scalar sector}, Basis: \( \left(\rho_1, \rho_2, \rho_3, {\rho}_{4}, {{\sigma}_{2}}^0\right), \left(\rho_1, \rho_2, \rho_3, {\rho}_{4}, {{\sigma}_{2}}^{0,*}\right) \) 
 
\begin{equation} 
m^2_{h} = \left( 
\begin{array}{ccccc}
m_{\rho_1\rho_1} &m_{\rho_2\rho_1} &m_{\rho_3\rho_1} &m_{{\rho}_{4}\rho_1} &0\\ 
m_{\rho_1\rho_2} &m_{\rho_2\rho_2} &m_{\rho_3\rho_2} &m_{{\rho}_{4}\rho_2} &0\\ 
m_{\rho_1\rho_3} &m_{\rho_2\rho_3} &m_{\rho_3\rho_3} &m_{{\rho}_{4}\rho_3} &0\\ 
m_{\rho_1{\rho}_{4}} &m_{\rho_2{\rho}_{4}} &m_{\rho_3{\rho}_{4}} &m_{{\rho}_{4}{\rho}_{4}} &0\\ 
0 &0 &0 &0 &m_{{{\sigma}_{2}}^0{{\sigma}_{2}}^{0,*}}\end{array} 
\right) 
 \end{equation} 
\begin{align} 
m_{\rho_1\rho_1} &= \frac{1}{4} \Big(12 \lambda_{1} v_{\rho}^{2}  + 2 \Big(\lambda_{12} v_{\eta}^{2}  + \lambda_{13} v_{\chi}^{2} \Big) -2 \sqrt{2} v_{\eta} v_{\sigma} {\xi}_{14}  + v_{\sigma}^{2} \Big(2 {\lambda}_{14}  + \zeta_{14}\Big)\Big) + {\mu}_{1}\\ 
m_{\rho_1\rho_2} &= f v_{\chi}  + v_{\rho} \Big(- \frac{1}{\sqrt{2}} v_{\sigma} {\xi}_{14}  + \lambda_{12} v_{\eta} \Big)\\ 
m_{\rho_2\rho_2} &= \frac{1}{2} \Big(6 \lambda_{2} v_{\eta}^{2}  + \lambda_{12} v_{\rho}^{2}  + \lambda_{23} v_{\chi}^{2}  + v_{\sigma}^{2} \Big(-2 {\xi}_{24}  + {\lambda}_{24}\Big)\Big) + {\mu}_{2}\\ 
m_{\rho_1\rho_3} &= \frac{1}{\sqrt{2}} {f}_{\sigma} v_{\sigma}  + f v_{\eta}  + \lambda_{13} v_{\rho} v_{\chi} \\ 
m_{\rho_2\rho_3} &= f v_{\rho}  + v_{\chi} \Big(\frac{1}{\sqrt{2}} v_{\sigma} {\xi}_{34}  + \lambda_{23} v_{\eta} \Big)\\ 
m_{\rho_3\rho_3} &= \frac{1}{4} \Big(12 \lambda_{3} v_{\chi}^{2}  + 2 \Big(\lambda_{13} v_{\rho}^{2}  + \lambda_{23} v_{\eta}^{2}  + \sqrt{2} v_{\eta} v_{\sigma} {\xi}_{34} \Big) + v_{\sigma}^{2} \Big(2 {\lambda}_{34}  + \zeta_{34}\Big)\Big) + {\mu}_{3}\\ 
m_{\rho_1{\rho}_{4}} &= \frac{1}{2} \Big(\sqrt{2} {f}_{\sigma} v_{\chi}  + v_{\rho} \Big(- \sqrt{2} v_{\eta} {\xi}_{14}  + v_{\sigma} \Big(2 {\lambda}_{14}  + \zeta_{14}\Big)\Big)\Big)\\ 
m_{\rho_2{\rho}_{4}} &= \frac{1}{2} \frac{1}{\sqrt{2}} \Big(v_{\chi}^{2} {\xi}_{34}  - v_{\rho}^{2} {\xi}_{14} \Big) + v_{\eta} v_{\sigma} \Big(-2 {\xi}_{24}  + {\lambda}_{24}\Big)\\ 
m_{\rho_3{\rho}_{4}} &= \frac{1}{2} \Big(\sqrt{2} {f}_{\sigma} v_{\rho}  + v_{\chi} \Big(\sqrt{2} v_{\eta} {\xi}_{34}  + v_{\sigma} \Big(2 {\lambda}_{34}  + \zeta_{34}\Big)\Big)\Big)\\ 
m_{{\rho}_{4}{\rho}_{4}} &= \frac{1}{4} \Big(2 v_{\eta}^{2} \Big(-2 {\xi}_{24}  + {\lambda}_{24}\Big) + 6 \Big(2 {\lambda}_4  + {\lambda}_{44}\Big)v_{\sigma}^{2}  + v_{\chi}^{2} \Big(2 {\lambda}_{34}  + \zeta_{34}\Big)\nonumber\\ &\ \ \ \ + v_{\rho}^{2} \Big(2 {\lambda}_{14}  + \zeta_{14}\Big)\Big) + {\mu}_{4}\\ 
m_{{{\sigma}_{2}}^0{{\sigma}_{2}}^{0,*}} &= \frac{1}{2} \Big(2 {\lambda}_4 v_{\sigma}^{2}  + {\lambda}_{14} v_{\rho}^{2}  + {\lambda}_{34} v_{\chi}^{2}  + v_{\eta}^{2} \Big({\lambda}_{24} + \zeta_{24}\Big)\Big) + {\mu}_{4}
\end{align} 
This matrix is diagonalized by \(R^{S}\)
\begin{equation} 
R^{S} m^2_{h} R^{S,\dagger} = m^{dia}_{2,h} 
\end{equation} 
with 
\begin{align} 
\rho_1 = \sum_{j}R^{{S},*}_{j 1}h_{{j}}\,, \hspace{1cm} 
\rho_2 = \sum_{j}R^{{S},*}_{j 2}h_{{j}}\,, \hspace{1cm} 
\rho_3 = \sum_{j}R^{{S},*}_{j 3}h_{{j}}\\ 
{\rho}_{4} = \sum_{j}R^{{S},*}_{j 4}h_{{j}}\,, \hspace{1cm} 
{{\sigma}_{2}}^0 = \sum_{j}R^{{S},*}_{j 5}h_{{j}}
\end{align} 
\item {\bf Mass matrix of the pseudoscalar sector}, Basis: \( \left(\sigma_1, \sigma_2, \sigma_3, {\sigma}_{4}\right), \left(\sigma_1, \sigma_2, \sigma_3, {\sigma}_{4}\right) \) 
 
\begin{equation} 
\label{ma0}
m^2_{A^0} = \left( 
\begin{array}{cccc}
m_{\sigma_1\sigma_1} &- f v_{\chi}  &- \frac{1}{\sqrt{2}} {f}_{\sigma} v_{\sigma}  - f v_{\eta}  &- \frac{1}{\sqrt{2}} {f}_{\sigma} v_{\chi} \\ 
- f v_{\chi}  &m_{\sigma_2\sigma_2} &- f v_{\rho}  &m_{{\sigma}_{4}\sigma_2}\\ 
- \frac{1}{\sqrt{2}} {f}_{\sigma} v_{\sigma}  - f v_{\eta}  &- f v_{\rho}  &m_{\sigma_3\sigma_3} &- \frac{1}{\sqrt{2}} {f}_{\sigma} v_{\rho} \\ 
- \frac{1}{\sqrt{2}} {f}_{\sigma} v_{\chi}  &m_{\sigma_2{\sigma}_{4}} &- \frac{1}{\sqrt{2}} {f}_{\sigma} v_{\rho}  &m_{{\sigma}_{4}{\sigma}_{4}}\end{array} 
\right) +  \xi_{Z}m^2(Z) +  \xi_{Z^\prime}m^2(Z^\prime) 
 \end{equation} 
\begin{align} 
m_{\sigma_1\sigma_1} &= \frac{1}{4} \Big(2 \Big(\lambda_{12} v_{\eta}^{2}  + \lambda_{13} v_{\chi}^{2} \Big) -2 \sqrt{2} v_{\eta} v_{\sigma} {\xi}_{14}  + 4 \lambda_{1} v_{\rho}^{2}  + v_{\sigma}^{2} \Big(2 {\lambda}_{14}  + \zeta_{14}\Big)\Big) + {\mu}_{1}\\ 
m_{\sigma_2\sigma_2} &= \frac{1}{2} \Big(2 \lambda_{2} v_{\eta}^{2}  + \lambda_{12} v_{\rho}^{2}  + \lambda_{23} v_{\chi}^{2}  + v_{\sigma}^{2} \Big(2 {\xi}_{24}  + {\lambda}_{24}\Big)\Big) + {\mu}_{2}\\ 
m_{\sigma_3\sigma_3} &= \frac{1}{4} \Big(2 \Big(\lambda_{13} v_{\rho}^{2}  + \lambda_{23} v_{\eta}^{2}  + \sqrt{2} v_{\eta} v_{\sigma} {\xi}_{34} \Big) + 4 \lambda_{3} v_{\chi}^{2}  + v_{\sigma}^{2} \Big(2 {\lambda}_{34}  + \zeta_{34}\Big)\Big) + {\mu}_{3}\\ 
m_{\sigma_2{\sigma}_{4}} &= \frac{1}{4} \Big(-8 v_{\eta} v_{\sigma} {\xi}_{24}  + \sqrt{2} \Big(v_{\chi}^{2} {\xi}_{34}  - v_{\rho}^{2} {\xi}_{14} \Big)\Big)\\ 
m_{{\sigma}_{4}{\sigma}_{4}} &= \frac{1}{4} \Big(2 \Big(\Big(2 {\lambda}_4  + {\lambda}_{44}\Big)v_{\sigma}^{2}  + v_{\eta}^{2} \Big(2 {\xi}_{24}  + {\lambda}_{24}\Big)\Big) + v_{\chi}^{2} \Big(2 {\lambda}_{34}  + \zeta_{34}\Big) \nonumber\\ &\ \ \ + v_{\rho}^{2} \Big(2 {\lambda}_{14}  + \zeta_{14}\Big)\Big) + {\mu}_{4}
\end{align} 
The gauge fixing contributions are
\begin{equation} 
m^2 (\xi_{Z}) = \left( 
\begin{array}{cccc}
m_{\sigma_1\sigma_1} &m_{\sigma_2\sigma_1} &m_{\sigma_3\sigma_1} &m_{{\sigma}_{4}\sigma_1}\\ 
m_{\sigma_1\sigma_2} &m_{\sigma_2\sigma_2} &m_{\sigma_3\sigma_2} &m_{{\sigma}_{4}\sigma_2}\\ 
m_{\sigma_1\sigma_3} &m_{\sigma_2\sigma_3} &m_{\sigma_3\sigma_3} &m_{{\sigma}_{4}\sigma_3}\\ 
m_{\sigma_1{\sigma}_{4}} &m_{\sigma_2{\sigma}_{4}} &m_{\sigma_3{\sigma}_{4}} &m_{{\sigma}_{4}{\sigma}_{4}}\end{array} 
\right) 
 \end{equation} 
\begin{align} 
m_{\sigma_1\sigma_1} &= \frac{1}{3} v_{\rho}^{2} \Big(-2 \sqrt{3} g_1 g_2 R_{{1 2}}^{Z} R_{{2 2}}^{Z}  + 3 g_{1}^{2} R_{{1 2}}^{Z,2}  + g_{2}^{2} R_{{2 2}}^{Z,2} \Big)\\ 
m_{\sigma_1\sigma_2} &= \frac{1}{6} g_2 v_{\rho} v_{\eta} \Big(g_1 R_{{1 2}}^{Z} \Big(-3 R_{{3 2}}^{Z}  + \sqrt{3} R_{{2 2}}^{Z} \Big) + g_2 R_{{2 2}}^{Z} \Big(- R_{{2 2}}^{Z}  + \sqrt{3} R_{{3 2}}^{Z} \Big)\Big)\\ 
m_{\sigma_2\sigma_2} &= \frac{1}{12} g_{2}^{2} v_{\eta}^{2} \Big(-2 \sqrt{3} R_{{2 2}}^{Z} R_{{3 2}}^{Z}  + 3 R_{{3 2}}^{Z,2}  + R_{{2 2}}^{Z,2}\Big)\\ 
m_{\sigma_1\sigma_3} &= -\frac{1}{6} v_{\rho} v_{\chi} \Big(-3 g_1 g_2 R_{{1 2}}^{Z} \Big(\sqrt{3} R_{{2 2}}^{Z}  + R_{{3 2}}^{Z}\Big) + 6 g_{1}^{2} R_{{1 2}}^{Z,2}  + g_{2}^{2} R_{{2 2}}^{Z} \Big(\sqrt{3} R_{{3 2}}^{Z}  + R_{{2 2}}^{Z}\Big)\Big)\\ 
m_{\sigma_2\sigma_3} &= \frac{1}{12} g_2 v_{\eta} v_{\chi} \Big(-2 g_1 R_{{1 2}}^{Z} \Big(-3 R_{{3 2}}^{Z}  + \sqrt{3} R_{{2 2}}^{Z} \Big) + g_2 \Big(-3 R_{{3 2}}^{Z,2}  + R_{{2 2}}^{Z,2}\Big)\Big)\\ 
m_{\sigma_3\sigma_3} &= \frac{1}{12} v_{\chi}^{2} \Big(12 g_{1}^{2} R_{{1 2}}^{Z,2}  -4 g_1 g_2 R_{{1 2}}^{Z} \Big(3 R_{{3 2}}^{Z}  + \sqrt{3} R_{{2 2}}^{Z} \Big) \nonumber\\ & \  \  \  + g_{2}^{2} \Big(2 \sqrt{3} R_{{2 2}}^{Z} R_{{3 2}}^{Z}  + 3 R_{{3 2}}^{Z,2}  + R_{{2 2}}^{Z,2}\Big)\Big)\\ 
m_{\sigma_1{\sigma}_{4}} &= \frac{1}{6} g_2 v_{\rho} v_{\sigma} \Big(g_1 R_{{1 2}}^{Z} \Big(-3 R_{{3 2}}^{Z}  + \sqrt{3} R_{{2 2}}^{Z} \Big) + g_2 R_{{2 2}}^{Z} \Big(- R_{{2 2}}^{Z}  + \sqrt{3} R_{{3 2}}^{Z} \Big)\Big)\\ 
m_{\sigma_2{\sigma}_{4}} &= \frac{1}{12} g_{2}^{2} v_{\eta} v_{\sigma} \Big(-2 \sqrt{3} R_{{2 2}}^{Z} R_{{3 2}}^{Z}  + 3 R_{{3 2}}^{Z,2}  + R_{{2 2}}^{Z,2}\Big)\\ 
m_{\sigma_3{\sigma}_{4}} &= \frac{1}{12} g_2 v_{\chi} v_{\sigma} \Big(-2 g_1 R_{{1 2}}^{Z} \Big(-3 R_{{3 2}}^{Z}  + \sqrt{3} R_{{2 2}}^{Z} \Big) + g_2 \Big(-3 R_{{3 2}}^{Z,2}  + R_{{2 2}}^{Z,2}\Big)\Big)\\ 
m_{{\sigma}_{4}{\sigma}_{4}} &= \frac{1}{12} g_{2}^{2} v_{\sigma}^{2} \Big(-2 \sqrt{3} R_{{2 2}}^{Z} R_{{3 2}}^{Z}  + 3 R_{{3 2}}^{Z,2}  + R_{{2 2}}^{Z,2}\Big)
\end{align} 

\begin{equation} 
m^2 (\xi_{Z^\prime}) = \left( 
\begin{array}{cccc}
m_{\sigma_1\sigma_1} &m_{\sigma_2\sigma_1} &m_{\sigma_3\sigma_1} &m_{{\sigma}_{4}\sigma_1}\\ 
m_{\sigma_1\sigma_2} &m_{\sigma_2\sigma_2} &m_{\sigma_3\sigma_2} &m_{{\sigma}_{4}\sigma_2}\\ 
m_{\sigma_1\sigma_3} &m_{\sigma_2\sigma_3} &m_{\sigma_3\sigma_3} &m_{{\sigma}_{4}\sigma_3}\\ 
m_{\sigma_1{\sigma}_{4}} &m_{\sigma_2{\sigma}_{4}} &m_{\sigma_3{\sigma}_{4}} &m_{{\sigma}_{4}{\sigma}_{4}}\end{array} 
\right) 
 \end{equation} 
\begin{align} 
m_{\sigma_1\sigma_1} &= \frac{1}{3} v_{\rho}^{2} \Big(-2 \sqrt{3} g_1 g_2 R_{{1 3}}^{Z} R_{{2 3}}^{Z}  + 3 g_{1}^{2} R_{{1 3}}^{Z,2}  + g_{2}^{2} R_{{2 3}}^{Z,2} \Big)\\ 
m_{\sigma_1\sigma_2} &= \frac{1}{6} g_2 v_{\rho} v_{\eta} \Big(g_1 R_{{1 3}}^{Z} \Big(-3 R_{{3 3}}^{Z}  + \sqrt{3} R_{{2 3}}^{Z} \Big) + g_2 R_{{2 3}}^{Z} \Big(- R_{{2 3}}^{Z}  + \sqrt{3} R_{{3 3}}^{Z} \Big)\Big)\\ 
m_{\sigma_2\sigma_2} &= \frac{1}{12} g_{2}^{2} v_{\eta}^{2} \Big(-2 \sqrt{3} R_{{2 3}}^{Z} R_{{3 3}}^{Z}  + 3 R_{{3 3}}^{Z,2}  + R_{{2 3}}^{Z,2}\Big)\\ 
m_{\sigma_1\sigma_3} &= -\frac{1}{6} v_{\rho} v_{\chi} \Big(-3 g_1 g_2 R_{{1 3}}^{Z} \Big(\sqrt{3} R_{{2 3}}^{Z}  + R_{{3 3}}^{Z}\Big) + 6 g_{1}^{2} R_{{1 3}}^{Z,2}  + g_{2}^{2} R_{{2 3}}^{Z} \Big(\sqrt{3} R_{{3 3}}^{Z}  + R_{{2 3}}^{Z}\Big)\Big)\\ 
m_{\sigma_2\sigma_3} &= \frac{1}{12} g_2 v_{\eta} v_{\chi} \Big(-2 g_1 R_{{1 3}}^{Z} \Big(-3 R_{{3 3}}^{Z}  + \sqrt{3} R_{{2 3}}^{Z} \Big) + g_2 \Big(-3 R_{{3 3}}^{Z,2}  + R_{{2 3}}^{Z,2}\Big)\Big)\\ 
m_{\sigma_3\sigma_3} &= \frac{1}{12} v_{\chi}^{2} \Big(12 g_{1}^{2} R_{{1 3}}^{Z,2}  -4 g_1 g_2 R_{{1 3}}^{Z} \Big(3 R_{{3 3}}^{Z}  + \sqrt{3} R_{{2 3}}^{Z} \Big)\nonumber\\ &\ \  \  + g_{2}^{2} \Big(2 \sqrt{3} R_{{2 3}}^{Z} R_{{3 3}}^{Z}  + 3 R_{{3 3}}^{Z,2}  + R_{{2 3}}^{Z,2}\Big)\Big)\\ 
m_{\sigma_1{\sigma}_{4}} &= \frac{1}{6} g_2 v_{\rho} v_{\sigma} \Big(g_1 R_{{1 3}}^{Z} \Big(-3 R_{{3 3}}^{Z}  + \sqrt{3} R_{{2 3}}^{Z} \Big) + g_2 R_{{2 3}}^{Z} \Big(- R_{{2 3}}^{Z}  + \sqrt{3} R_{{3 3}}^{Z} \Big)\Big)\\ 
m_{\sigma_2{\sigma}_{4}} &= \frac{1}{12} g_{2}^{2} v_{\eta} v_{\sigma} \Big(-2 \sqrt{3} R_{{2 3}}^{Z} R_{{3 3}}^{Z}  + 3 R_{{3 3}}^{Z,2}  + R_{{2 3}}^{Z,2}\Big)\\ 
m_{\sigma_3{\sigma}_{4}} &= \frac{1}{12} g_2 v_{\chi} v_{\sigma} \Big(-2 g_1 R_{{1 3}}^{Z} \Big(-3 R_{{3 3}}^{Z}  + \sqrt{3} R_{{2 3}}^{Z} \Big) + g_2 \Big(-3 R_{{3 3}}^{Z,2}  + R_{{2 3}}^{Z,2}\Big)\Big)\\ 
m_{{\sigma}_{4}{\sigma}_{4}} &= \frac{1}{12} g_{2}^{2} v_{\sigma}^{2} \Big(-2 \sqrt{3} R_{{2 3}}^{Z} R_{{3 3}}^{Z}  + 3 R_{{3 3}}^{Z,2}  + R_{{2 3}}^{Z,2}\Big)
\end{align} 
where $R^Z$ is the rotation matrix that diagonalize the mass of the neutral gauge boson components in the $\{W^3,W^8,X\}$ basis which is given by the following matrix
\begin{equation}
    \left(
\begin{array}{ccc}
 g_1^2 v_\r^2+g_1^2 v_\chi^2 & -\frac{g_1 g_2 v_\r^2}{\sqrt{3}}-\frac{g_1 g_2 v_\chi^2}{2 \sqrt{3}} & -\frac{1}{2} g_1 g_2 v_\chi^2 \\
 -\frac{g_1 g_2 v_\r^2}{\sqrt{3}}-\frac{g_1 g_2 v_\chi^2}{2 \sqrt{3}} & \frac{g_2^2 v_\r^2}{3}+\frac{g_2^2 v_\eta^2}{12}+\frac{g_2^2 v_\chi^2}{12}+\frac{g_2^2 v_\sigma^2}{12} & -\frac{g_2^2 v_\eta^2}{4 \sqrt{3}}+\frac{g_2^2 v_\chi^2}{4 \sqrt{3}}-\frac{g_2^2 v_\sigma^2}{4 \sqrt{3}} \\
 -\frac{1}{2} g_1 g_2 v_\chi^2 & -\frac{g_2^2 v_\eta^2}{4 \sqrt{3}}+\frac{g_2^2 v_\chi^2}{4 \sqrt{3}}-\frac{g_2^2 v_\sigma^2}{4 \sqrt{3}} & \frac{g_2^2 v_\eta^2}{4}+\frac{g_2^2 v_\chi^2}{4}+\frac{g_2^2 v_\sigma^2}{4} \\
\end{array}
\right)
\end{equation}
The matrix \eqref{ma0} is diagonalized by the matrix \(R^{P}\) 
\begin{equation} 
R^{P} m^2_{A^0} R^{P,\dagger} = m^{dia}_{2,A^0} 
\end{equation} 
with 
\begin{align} 
\sigma_1 = \sum_{j}R^{{P},*}_{j 1}A^0_{{j}}\,, \hspace{1cm} 
\sigma_2 = \sum_{j}R^{{P},*}_{j 2}A^0_{{j}}\,, \hspace{1cm} 
\sigma_3 = \sum_{j}R^{{P},*}_{j 3}A^0_{{j}}\\ 
{\sigma}_{4} = \sum_{j}R^{{P},*}_{j 4}A^0_{{j}}
\end{align} 
\item {\bf Mass matrix for Charged Higgs}, Basis: \( \left({\rho}_{+}^*, {\eta}_1^{{+},*}, {\sigma}_2^{{p},*}, {\eta}_2^{-}, {\chi}_{-}, {\sigma}_1^{-}\right), \left({\rho}_{+}, {\eta}_1^{+}, {\sigma}_2^{p}, {\eta}_2^{{-},*}, {\chi}_{-}^*, {\sigma}_1^{{-},*}\right) \) 
 
\begin{equation} 
m^2_{H^-} = \left( 
\begin{array}{cccccc}
m_{{\rho}_{+}^*{\rho}_{+}} &0 &0 &m^*_{{\eta}_2^{-}{\rho}_{+}} &0 &m^*_{{\sigma}_1^{-}{\rho}_{+}}\\ 
0 &m_{{\eta}_1^{{+},*}{\eta}_1^{+}} &m^*_{{\sigma}_2^{{p},*}{\eta}_1^{+}} &0 &m^*_{{\chi}_{-}{\eta}_1^{+}} &0\\ 
0 &m_{{\eta}_1^{{+},*}{\sigma}_2^{p}} &m_{{\sigma}_2^{{p},*}{\sigma}_2^{p}} &0 &m^*_{{\chi}_{-}{\sigma}_2^{p}} &0\\ 
m_{{\rho}_{+}^*{\eta}_2^{{-},*}} &0 &0 &m_{{\eta}_2^{-}{\eta}_2^{{-},*}} &0 &m^*_{{\sigma}_1^{-}{\eta}_2^{{-},*}}\\ 
0 &m_{{\eta}_1^{{+},*}{\chi}_{-}^*} &m_{{\sigma}_2^{{p},*}{\chi}_{-}^*} &0 &m_{{\chi}_{-}{\chi}_{-}^*} &0\\ 
m_{{\rho}_{+}^*{\sigma}_1^{{-},*}} &0 &0 &m_{{\eta}_2^{-}{\sigma}_1^{{-},*}} &0 &m_{{\sigma}_1^{-}{\sigma}_1^{{-},*}}\end{array} 
\right) +  \xi_{W^+}m^2(W^+) +  \xi_{{W^\prime}^+}m^2({W^\prime}^+) 
 \end{equation} 
\begin{align} 
m_{{\rho}_{+}^*{\rho}_{+}} &= \frac{1}{2} \Big(2 \lambda_{1} v_{\rho}^{2}  + \Big(\lambda_{12} + \zeta_{12}\Big)v_{\eta}^{2}  + \lambda_{13} v_{\chi}^{2}  + {\lambda}_{14} v_{\sigma}^{2} \Big) + {\mu}_{1}\\ 
m_{{\eta}_1^{{+},*}{\eta}_1^{+}} &= \frac{1}{4} \Big(2 \Big(\lambda_{12} v_{\rho}^{2}  + \Big(\lambda_{23} + \zeta_{23}\Big)v_{\chi}^{2} \Big) + 4 \lambda_{2} v_{\eta}^{2}  + v_{\sigma}^{2} \Big(2 {\lambda}_{24}  + \zeta_{24}\Big)\Big) + {\mu}_{2}\\ 
m_{{\eta}_1^{{+},*}{\sigma}_2^{p}} &= \frac{1}{4} \Big(\sqrt{2} v_{\rho}^{2} {\xi}_{14}  + v_{\eta} v_{\sigma} \Big(4 {\xi}_{24}  + \zeta_{24}\Big)\Big)\\ 
m_{{\sigma}_2^{{p},*}{\sigma}_2^{p}} &= \frac{1}{4} \Big(2 \Big(\Big(2 {\lambda}_4  + {\lambda}_{44}\Big)v_{\sigma}^{2}  + {\lambda}_{34} v_{\chi}^{2} \Big) + v_{\eta}^{2} \Big(2 {\lambda}_{24}  + \zeta_{24}\Big) + v_{\rho}^{2} \Big(2 {\lambda}_{14}  + \zeta_{14}\Big)\Big) + {\mu}_{4}\\ 
m_{{\rho}_{+}^*{\eta}_2^{{-},*}} &= \frac{1}{4} \Big(-4 f v_{\chi}  + v_{\rho} \Big(2 \zeta_{12} v_{\eta}  + \sqrt{2} v_{\sigma} {\xi}_{14} \Big)\Big)\\ 
m_{{\eta}_2^{-}{\eta}_2^{{-},*}} &= \frac{1}{4} \Big(2 \Big(\Big(\lambda_{12} + \zeta_{12}\Big)v_{\rho}^{2}  + \lambda_{23} v_{\chi}^{2} \Big) + 4 \lambda_{2} v_{\eta}^{2}  + v_{\sigma}^{2} \Big(2 {\lambda}_{24}  + \zeta_{24}\Big)\Big) + {\mu}_{2}\\ 
m_{{\eta}_1^{{+},*}{\chi}_{-}^*} &= \frac{1}{4} \Big(-4 f v_{\rho}  + v_{\chi} \Big(2 \zeta_{23} v_{\eta}  - \sqrt{2} v_{\sigma} {\xi}_{34} \Big)\Big)\\ 
m_{{\sigma}_2^{{p},*}{\chi}_{-}^*} &= \frac{1}{4} \Big(\sqrt{2} \Big(2 {f}_{\sigma} v_{\rho}  + v_{\eta} v_{\chi} {\xi}_{34} \Big) + v_{\chi} v_{\sigma} \zeta_{34} \Big)\\ 
m_{{\chi}_{-}{\chi}_{-}^*} &= \frac{1}{2} \Big(2 \lambda_{3} v_{\chi}^{2}  + \lambda_{13} v_{\rho}^{2}  + \Big(\lambda_{23} + \zeta_{23}\Big)v_{\eta}^{2}  + {\lambda}_{34} v_{\sigma}^{2} \Big) + {\mu}_{3}\\ 
m_{{\rho}_{+}^*{\sigma}_1^{{-},*}} &= \frac{1}{4} \Big(\sqrt{2} \Big(2 {f}_{\sigma} v_{\chi}  - v_{\rho} v_{\eta} {\xi}_{14} \Big) + v_{\rho} v_{\sigma} \zeta_{14} \Big)\\ 
m_{{\eta}_2^{-}{\sigma}_1^{{-},*}} &= \frac{1}{4} \Big(- \sqrt{2} v_{\chi}^{2} {\xi}_{34}  + v_{\eta} v_{\sigma} \Big(4 {\xi}_{24}  + \zeta_{24}\Big)\Big)\\ 
m_{{\sigma}_1^{-}{\sigma}_1^{{-},*}} &= \frac{1}{4} \Big(2 \Big(\Big(2 {\lambda}_4  + {\lambda}_{44}\Big)v_{\sigma}^{2}  + {\lambda}_{14} v_{\rho}^{2} \Big) + v_{\chi}^{2} \Big(2 {\lambda}_{34}  + \zeta_{34}\Big) + v_{\eta}^{2} \Big(2 {\lambda}_{24}  + \zeta_{24}\Big)\Big) + {\mu}_{4}
\end{align} 
The gauge fixing contributions are
\begin{equation} 
m^2 (\xi_{W^+}) = \left( 
\begin{array}{cccccc}
0 &0 &0 &0 &0 &0\\ 
0 &\frac{1}{4} g_{2}^{2} v_{\eta}^{2}  &-\frac{1}{4} g_{2}^{2} v_{\eta} v_{\sigma}  &0 &-\frac{1}{4} g_{2}^{2} v_{\eta} v_{\chi}  &0\\ 
0 &-\frac{1}{4} g_{2}^{2} v_{\eta} v_{\sigma}  &\frac{1}{4} g_{2}^{2} v_{\sigma}^{2}  &0 &\frac{1}{4} g_{2}^{2} v_{\chi} v_{\sigma}  &0\\ 
0 &0 &0 &0 &0 &0\\ 
0 &-\frac{1}{4} g_{2}^{2} v_{\eta} v_{\chi}  &\frac{1}{4} g_{2}^{2} v_{\chi} v_{\sigma}  &0 &\frac{1}{4} g_{2}^{2} v_{\chi}^{2}  &0\\ 
0 &0 &0 &0 &0 &0\end{array} 
\right) 
 \end{equation} 
\begin{equation} 
m^2 (\xi_{{W^\prime}^+}) = \left( 
\begin{array}{cccccc}
\frac{1}{4} g_{2}^{2} v_{\rho}^{2}  &0 &0 &-\frac{1}{4} g_{2}^{2} v_{\rho} v_{\eta}  &0 &\frac{1}{4} g_{2}^{2} v_{\rho} v_{\sigma} \\ 
0 &0 &0 &0 &0 &0\\ 
0 &0 &0 &0 &0 &0\\ 
-\frac{1}{4} g_{2}^{2} v_{\rho} v_{\eta}  &0 &0 &\frac{1}{4} g_{2}^{2} v_{\eta}^{2}  &0 &-\frac{1}{4} g_{2}^{2} v_{\eta} v_{\sigma} \\ 
0 &0 &0 &0 &0 &0\\ 
\frac{1}{4} g_{2}^{2} v_{\rho} v_{\sigma}  &0 &0 &-\frac{1}{4} g_{2}^{2} v_{\eta} v_{\sigma}  &0 &\frac{1}{4} g_{2}^{2} v_{\sigma}^{2} \end{array} 
\right) 
 \end{equation} 
This matrix is diagonalized by \(R^{C}\)
\begin{equation} 
R^{C} m^2_{H^-} R^{C,\dagger} = m^{dia}_{2,H^-} 
\end{equation} 
with 
\begin{align} 
{\rho}_{+} = \sum_{j}R_{{j 1}}^{C}H^+_{{j}}\,, \hspace{1cm} 
{\eta}_1^{+} = \sum_{j}R_{{j 2}}^{C}H^+_{{j}}\,, \hspace{1cm} 
{\sigma}_2^{p} = \sum_{j}R_{{j 3}}^{C}H^+_{{j}}\\ 
{\eta}_2^{-} = \sum_{j}R^{{C},*}_{j 4}H^-_{{j}}\,, \hspace{1cm} 
{\chi}_{-} = \sum_{j}R^{{C},*}_{j 5}H^-_{{j}}\,, \hspace{1cm} 
{\sigma}_1^{-} = \sum_{j}R^{{C},*}_{j 6}H^-_{{j}}
\end{align} 
\item {\bf Mass matrix for Doubly Charged Higgs} \\Basis: \( \left({\rho}^{{++},*}, {\chi}^{--}, {{\sigma}_{1}}^{--}, {{\sigma}_{2}}^{{++},*}\right), \left({\rho}^{++}, {\chi}^{{--},*}, {{\sigma}_{1}}^{{--},*}, {{\sigma}_{2}}^{++}\right) \) 
 
\begin{equation} 
m^2_{H^{--}} = \left( 
\begin{array}{cccc}
m_{{\rho}^{{++},*}{\rho}^{++}} &m^*_{{\chi}^{--}{\rho}^{++}} &m^*_{{{\sigma}_{1}}^{--}{\rho}^{++}} &m^*_{{{\sigma}_{2}}^{{++},*}{\rho}^{++}}\\ 
m_{{\rho}^{{++},*}{\chi}^{{--},*}} &m_{{\chi}^{--}{\chi}^{{--},*}} &m^*_{{{\sigma}_{1}}^{--}{\chi}^{{--},*}} &m^*_{{{\sigma}_{2}}^{{++},*}{\chi}^{{--},*}}\\ 
m_{{\rho}^{{++},*}{{\sigma}_{1}}^{{--},*}} &m_{{\chi}^{--}{{\sigma}_{1}}^{{--},*}} &m_{{{\sigma}_{1}}^{--}{{\sigma}_{1}}^{{--},*}} &\frac{1}{2} {\lambda}_{44} v_{\sigma}^{2}  + v_{\eta}^{2} {\xi}_{24} \\ 
m_{{\rho}^{{++},*}{{\sigma}_{2}}^{++}} &m_{{\chi}^{--}{{\sigma}_{2}}^{++}} &\frac{1}{2} {\lambda}_{44} v_{\sigma}^{2}  + v_{\eta}^{2} {\xi}_{24}  &m_{{{\sigma}_{2}}^{{++},*}{{\sigma}_{2}}^{++}}\end{array} 
\right) +  \xi_{Y^{++}}m^2(Y^{++}) 
 \end{equation} 
\begin{align} 
m_{{\rho}^{{++},*}{\rho}^{++}} &= \frac{1}{4} \Big(2 \Big(\lambda_{12} v_{\eta}^{2}  + \Big(\lambda_{13} + \zeta_{13}\Big)v_{\chi}^{2}  + \sqrt{2} v_{\eta} v_{\sigma} {\xi}_{14} \Big) + 4 \lambda_{1} v_{\rho}^{2}  \nonumber\\ &\ \ \ + v_{\sigma}^{2} \Big(2 {\lambda}_{14}  + \zeta_{14}\Big)\Big) + {\mu}_{1}\\ 
m_{{\rho}^{{++},*}{\chi}^{{--},*}} &= \frac{1}{2} \Big(-2 f v_{\eta}  + \sqrt{2} {f}_{\sigma} v_{\sigma}  + \zeta_{13} v_{\rho} v_{\chi} \Big)\\ 
m_{{\chi}^{--}{\chi}^{{--},*}} &= \frac{1}{4} \Big(2 \Big(\Big(\lambda_{13} + \zeta_{13}\Big)v_{\rho}^{2}  + \lambda_{23} v_{\eta}^{2} \Big) -2 \sqrt{2} v_{\eta} v_{\sigma} {\xi}_{34}  + 4 \lambda_{3} v_{\chi}^{2} \nonumber\\ &\ \ \  + v_{\sigma}^{2} \Big(2 {\lambda}_{34}  + \zeta_{34}\Big)\Big) + {\mu}_{3}\\ 
m_{{\rho}^{{++},*}{{\sigma}_{1}}^{{--},*}} &= \frac{1}{4} \Big(4 {f}_{\sigma} v_{\chi}  + v_{\rho} \Big(-2 v_{\eta} {\xi}_{14}  + \sqrt{2} v_{\sigma} \zeta_{14} \Big)\Big)\\ 
m_{{\chi}^{--}{{\sigma}_{1}}^{{--},*}} &= \frac{1}{4} v_{\chi} \Big(-2 v_{\eta} {\xi}_{34}  + \sqrt{2} v_{\sigma} \zeta_{34} \Big)\\ 
m_{{{\sigma}_{1}}^{--}{{\sigma}_{1}}^{{--},*}} &= \frac{1}{2} \Big(2 \Big({\lambda}_4 + {\lambda}_{44}\Big)v_{\sigma}^{2}  + {\lambda}_{14} v_{\rho}^{2}  + {\lambda}_{24} v_{\eta}^{2}  + v_{\chi}^{2} \Big({\lambda}_{34} + \zeta_{34}\Big)\Big) + {\mu}_{4}\\ 
m_{{\rho}^{{++},*}{{\sigma}_{2}}^{++}} &= \frac{1}{4} v_{\rho} \Big(2 v_{\eta} {\xi}_{14}  + \sqrt{2} v_{\sigma} \zeta_{14} \Big)\\ 
m_{{\chi}^{--}{{\sigma}_{2}}^{++}} &= \frac{1}{4} \Big(4 {f}_{\sigma} v_{\rho}  + v_{\chi} \Big(2 v_{\eta} {\xi}_{34}  + \sqrt{2} v_{\sigma} \zeta_{34} \Big)\Big)\\ 
m_{{{\sigma}_{2}}^{{++},*}{{\sigma}_{2}}^{++}} &= \frac{1}{2} \Big(2 \Big({\lambda}_4 + {\lambda}_{44}\Big)v_{\sigma}^{2}  + {\lambda}_{24} v_{\eta}^{2}  + {\lambda}_{34} v_{\chi}^{2}  + v_{\rho}^{2} \Big({\lambda}_{14} + \zeta_{14}\Big)\Big) + {\mu}_{4}
\end{align} 
The gauge fixing contributions are
\begin{equation} 
m^2 (\xi_{Y^{++}}) = \left( 
\begin{array}{cccc}
\frac{1}{4} g_{2}^{2} v_{\rho}^{2}  &-\frac{1}{4} g_{2}^{2} v_{\rho} v_{\chi}  &\frac{1}{2} \frac{1}{\sqrt{2}} g_{2}^{2} v_{\rho} v_{\sigma}  &-\frac{1}{2} \frac{1}{\sqrt{2}} g_{2}^{2} v_{\rho} v_{\sigma} \\ 
-\frac{1}{4} g_{2}^{2} v_{\rho} v_{\chi}  &\frac{1}{4} g_{2}^{2} v_{\chi}^{2}  &-\frac{1}{2} \frac{1}{\sqrt{2}} g_{2}^{2} v_{\chi} v_{\sigma}  &\frac{1}{2} \frac{1}{\sqrt{2}} g_{2}^{2} v_{\chi} v_{\sigma} \\ 
\frac{1}{2} \frac{1}{\sqrt{2}} g_{2}^{2} v_{\rho} v_{\sigma}  &-\frac{1}{2} \frac{1}{\sqrt{2}} g_{2}^{2} v_{\chi} v_{\sigma}  &\frac{1}{2} g_{2}^{2} v_{\sigma}^{2}  &-\frac{1}{2} g_{2}^{2} v_{\sigma}^{2} \\ 
-\frac{1}{2} \frac{1}{\sqrt{2}} g_{2}^{2} v_{\rho} v_{\sigma}  &\frac{1}{2} \frac{1}{\sqrt{2}} g_{2}^{2} v_{\chi} v_{\sigma}  &-\frac{1}{2} g_{2}^{2} v_{\sigma}^{2}  &\frac{1}{2} g_{2}^{2} v_{\sigma}^{2} \end{array} 
\right) 
 \end{equation} 
This matrix is diagonalized by \(R_{2C}\)
\begin{equation} 
R_{2C} m^2_{H^{--}} R_{2C}^{\dagger} = m^{dia}_{2,H^{--}} 
\end{equation} 
with 
\begin{align} 
{\rho}^{++} = \sum_{j}R_{2C,{j 1}}H^{++}_{{j}}\,, \hspace{1cm} 
{\chi}^{--} = \sum_{j}R^*_{{2C},{j 2}}H^{--}_{{j}}\,, \hspace{1cm} 
{{\sigma}_{1}}^{--} = \sum_{j}R^*_{{2C},{j 3}}H^{--}_{{j}}\\ 
{{\sigma}_{2}}^{++} = \sum_{j}R_{2C,{j 4}}H^{++}_{{j}}
\end{align} 
\end{itemize}

\section{Renormalization group equations}
In this Appendix we present the Beta fuction of all the parameter of the minimal 331 Model in presence of the scalar sextet \cite{staub_automatic_2011,staub_sarah_2014,thomsen_rgbeta_2021}.
Here we recall the form of the scalar potential
\begin{align}
    V=&m_1\r^\dagger\r+m_2\eta^\dagger\eta+m_3\chi^\dagger\chi+\l_1(\rho^\dagger\r)^2+\l_2(\eta^\dagger\eta)^2+\l_3(\chi^\dagger\chi)^2\nonumber\\
    &+\l_{12}\r^\dagger\r\eta^\dagger\eta+\l_{13}\r^\dagger\r\chi^\dagger\chi+\l_{23}\chi^\dagger\chi\eta^\dagger\eta+\z_{12}\r^\dagger\eta\eta^\dagger\r+\z_{13}\r^\dagger\chi\chi^\dagger\r+\z_{23}\eta^\dagger\chi\chi^\dagger\eta\nonumber\\
    &+m_4\Tr(\s^\dagger\s)+\l_4(\Tr(\s^\dagger\s))^2+\l_{14}\r^\dagger\r\Tr(\s^\dagger\s)+\l_{24}\eta^\dagger\eta\Tr(\s^\dagger\s)+\l_{34}\chi^\dagger\chi\Tr(\s^\dagger\s)\nonumber\\
    &+\l_{44}\Tr(\s^\dagger\s\s^\dagger\s)+\z_{14}\r^\dagger\s\s^\dagger\rho+\z_{24}\eta^\dagger\s\s^\dagger\eta+\z_{34}\chi^\dagger\s\s^\dagger\chi\nonumber\\
    &+(\sqrt{2}f_{\r\eta\chi}\epsilon^{ijk}\r_i\eta_j\chi_k+\sqrt{2}f_{\r\s\chi}\r^T\s^\dagger\chi\nonumber\\
    &+\xi_{14}\epsilon^{ijk}\r^{*l}\s_{li}\r_j\eta_k+\xi_{24}\epsilon^{ijk}\epsilon^{lmn}\eta_i\eta_l\s_{jm}\s_{kn}+\xi_{34}\epsilon^{ijk}\chi^{*l}\s_{li}\chi_j\eta_k )+\textnormal{h.c.}\ ,
\end{align}
and the Yukawa interactions 
\begin{align}
    \mathcal{L}^{Y}_{q,\ triplet}=\ &- \overline{Q}_m\brak{Y^d_{m\a} \eta^*d_{\a R}+Y_{m\a}^u\chi^*u_{m\a}}
    -\overline{Q}_3( Y^d_{3\a}\chi d_{\a R}+Y^u_{3\a}\eta u_{m\a})+\nonumber\\
    &-\overline{Q}_{m}(Y^J_{mn}\chi J_{nR})-\overline{Q}_{3}Y^J_{3}\chi J_{3R} +\textnormal{h.c.}\ .
\end{align}

\begin{equation}
    \mathcal{L}^Y_{l,\ triplet}=G^\eta_{ab}l^i_a\cdot l^j_b\eta^{*k}\epsilon^{ijk}+h.c.
\end{equation}
considering also the sextet contribution to the leptonic Yukawa
\begin{equation}
    \mathcal{L}^Y_{l,\ sextet}=G^\s_{ab}l^i_a\cdot l^j_b\s^{*}_{ij}+\textnormal{h.c.}
\end{equation}

\subsection{Gauge Couplings}
{\allowdisplaybreaks  \begin{align} 
\beta_{g_1}^{(1)} & =  
22 g_{1}^{3} \\  
\beta_{g_2}^{(1)} & =  
-\frac{17}{3} g_{2}^{3} \\ 
\beta_{g_3}^{(1)} & =  
-5 g_{3}^{3} \\ 
\end{align}} 
\subsection{Quartic scalar couplings}
{\allowdisplaybreaks  \begin{align} 
\beta_{{\xi}_{34}}^{(1)} & =  
+2 \zeta_{13} {\xi}_{14} -6 g_{1}^{2} {\xi}_{34} -22 g_{2}^{2} {\xi}_{34} +4 \lambda_{23} {\xi}_{34} -2 \zeta_{23} {\xi}_{34} +2 {\lambda}_{24} {\xi}_{34} +4 \lambda_{3} {\xi}_{34} +4 {\lambda}_{34} {\xi}_{34} -12 {\xi}_{24} {\xi}_{34}\nonumber \\ 
 & - {\xi}_{34} \zeta_{24} +3 {\xi}_{34} \zeta_{34} +3 {\xi}_{34} \Big({Y^d_1  Y^{d *}_1}\Big) +3 {\xi}_{34} \Big({Y^d_2  Y^{d *}_2}\Big) +6 {\xi}_{34} \Big({Y^d_3  Y^{d *}_3}\Big) +6 {\xi}_{34} \Big({Y^u_1  Y^{u *}_1}\Big) \nonumber \\ 
 &+6 {\xi}_{34} \Big({Y^u_2  Y^{u *}_2}\Big) +3 {\xi}_{34} \Big({Y^u_3  Y^{u *}_3}\Big) +2 {\xi}_{34} \mbox{Tr}\Big({G^\sigma  G^{\sigma *}}\Big) \\ 
\beta_{{\xi}_{14}}^{(1)} & =  
-6 g_{1}^{2} {\xi}_{14} -22 g_{2}^{2} {\xi}_{14} +4 \lambda_{1} {\xi}_{14} +4 \lambda_{12} {\xi}_{14} -2 \zeta_{12} {\xi}_{14} +4 {\lambda}_{14} {\xi}_{14} +2 {\lambda}_{24} {\xi}_{14} \nonumber \\ 
 &-12 {\xi}_{14} {\xi}_{24} +2 \zeta_{13} {\xi}_{34} +3 {\xi}_{14} \zeta_{14} - {\xi}_{14} \zeta_{24} +6 {\xi}_{14} |Y^J_1|^2 +3 {\xi}_{14} \Big({Y^d_1  Y^{d *}_1}\Big) +3 {\xi}_{14} \Big({Y^d_2  Y^{d *}_2}\Big)\nonumber \\ 
 & +6 {\xi}_{14} \Big({Y^J_1  Y^{J *}_1}\Big) +6 {\xi}_{14} \Big({Y^J_2  Y^{J *}_2}\Big) +3 {\xi}_{14} \Big({Y^u_3  Y^{u *}_3}\Big) +2 {\xi}_{14} \mbox{Tr}\Big({G^\sigma  G^{\sigma *}}\Big) \\ 
\beta_{{\xi}_{24}}^{(1)} & =  
- {\xi}_{14}^{2} -28 g_{2}^{2} {\xi}_{24} +4 \lambda_{2} {\xi}_{24} +8 {\lambda}_{24} {\xi}_{24} +4 {\lambda}_4 {\xi}_{24} -2 {\lambda}_{44} {\xi}_{24} - {\xi}_{34}^{2} -4 {\xi}_{24} \zeta_{24} +6 {\xi}_{24} \Big({Y^d_1  Y^{d *}_1}\Big) \nonumber \\ 
 &+6 {\xi}_{24} \Big({Y^d_2  Y^{d *}_2}\Big) +6 {\xi}_{24} \Big({Y^u_3  Y^{u *}_3}\Big) +4 {\xi}_{24} \mbox{Tr}\Big({G^\sigma  G^{\sigma *}}\Big) \\ 
\beta_{\lambda_{3}}^{(1)} & =  
+6 g_{1}^{4} +4 g_{1}^{2} g_{2}^{2} +\frac{13}{6} g_{2}^{4} +3 \lambda_{13}^{2} +2 \lambda_{13} \zeta_{13} +\zeta_{13}^{2}+3 \lambda_{23}^{2} +2 \lambda_{23} \zeta_{23} +\zeta_{23}^{2}-12 g_{1}^{2} \lambda_{3}\nonumber \\ 
 & -16 g_{2}^{2} \lambda_{3} +28 \lambda_{3}^{2} +6 {\lambda}_{34}^{2} +2 {\xi}_{34}^{2} +4 {\lambda}_{34} \zeta_{34} +\frac{3}{2} \zeta_{34}^{2} +12 \lambda_{3} \Big({Y^d_3  Y^{d *}_3}\Big) -6 \Big(\Big({Y^d_3  Y^{d *}_3}\Big)\Big)^{2} \nonumber \\ 
 &+12 \lambda_{3} \Big({Y^u_1  Y^{u *}_1}\Big) -6 \Big(\Big({Y^u_1  Y^{u *}_1}\Big)\Big)^{2} -12 \Big({Y^u_1  Y^{u *}_2}\Big) \Big({Y^u_2  Y^{u *}_1}\Big) +12 \lambda_{3} \Big({Y^u_2  Y^{u *}_2}\Big) \nonumber \\ 
 &-6 \Big(\Big({Y^u_2  Y^{u *}_2}\Big)\Big)^{2} \\  
\beta_{{\lambda}_{34}}^{(1)} & =  
+\frac{13}{3} g_{2}^{4} +6 \lambda_{13} {\lambda}_{14} +2 \zeta_{13} {\lambda}_{14} +6 \lambda_{23} {\lambda}_{24} +2 \zeta_{23} {\lambda}_{24} -6 g_{1}^{2} {\lambda}_{34} -28 g_{2}^{2} {\lambda}_{34} +16 \lambda_{3} {\lambda}_{34}\nonumber \\ 
 & +4 {\lambda}_{34}^{2} +28 {\lambda}_{34} {\lambda}_4 +16 {\lambda}_{34} {\lambda}_{44} +2 {\xi}_{34}^{2} +2 \lambda_{13} \zeta_{14} +2 \lambda_{23} \zeta_{24} +4 \lambda_{3} \zeta_{34} +8 {\lambda}_4 \zeta_{34} +2 {\lambda}_{44} \zeta_{34} \nonumber \\ 
 &+\zeta_{34}^{2}+6 {\lambda}_{34} \Big({Y^d_3  Y^{d *}_3}\Big) +6 {\lambda}_{34} \Big({Y^u_1  Y^{u *}_1}\Big) +6 {\lambda}_{34} \Big({Y^u_2  Y^{u *}_2}\Big) +4 {\lambda}_{34} \mbox{Tr}\Big({G^\sigma  G^{\sigma *}}\Big) \\ 
\beta_{\zeta_{34}}^{(1)} & =  
+7 g_{2}^{4} +2 {\xi}_{34}^{2} +2 \zeta_{13} \zeta_{14} +2 \zeta_{23} \zeta_{24} -6 g_{1}^{2} \zeta_{34} -28 g_{2}^{2} \zeta_{34} +4 \lambda_{3} \zeta_{34} +8 {\lambda}_{34} \zeta_{34} \nonumber \\ 
 & +4 {\lambda}_4 \zeta_{34} +10 {\lambda}_{44} \zeta_{34} +5 \zeta_{34}^{2}+6 \zeta_{34} \Big({Y^d_3  Y^{d *}_3}\Big) +6 \zeta_{34} \Big({Y^u_1  Y^{u *}_1}\Big) +6 \zeta_{34} \Big({Y^u_2  Y^{u *}_2}\Big)  \nonumber \\ 
 &+4 \zeta_{34} \mbox{Tr}\Big({G^\sigma  G^{\sigma *}}\Big) \\ 
\beta_{\zeta_{23}}^{(1)} & =  
+\frac{5}{2} g_{2}^{4} +2 \zeta_{12} \zeta_{13} -6 g_{1}^{2} \zeta_{23} -16 g_{2}^{2} \zeta_{23} +4 \lambda_{2} \zeta_{23} +8 \lambda_{23} \zeta_{23} +6 \zeta_{23}^{2} +4 \zeta_{23} \lambda_{3} -5 {\xi}_{34}^{2} \nonumber \\ 
 &+\frac{5}{2} \zeta_{24} \zeta_{34} +6 \zeta_{23} \Big({Y^d_3  Y^{d *}_3}\Big) +6 \Big({Y^d_1  Y^{d *}_1}\Big) \Big(-2 \Big({Y^u_1  Y^{u *}_1}\Big)  + \zeta_{23}\Big)+6 \zeta_{23} \Big({Y^u_1  Y^{u *}_1}\Big)\nonumber \\ 
 & -12 \Big({Y^d_2  Y^{d *}_1}\Big) \Big({Y^u_1  Y^{u *}_2}\Big) -12 \Big({Y^d_3  Y^{d *}_1}\Big) \Big({Y^u_1  Y^{u *}_3}\Big) -12 \Big({Y^d_1  Y^{d *}_2}\Big) \Big({Y^u_2  Y^{u *}_1}\Big) \nonumber \\ 
 &+6 \Big({Y^d_2  Y^{d *}_2}\Big) \Big(-2 \Big({Y^u_2  Y^{u *}_2}\Big)  + \zeta_{23}\Big)+6 \zeta_{23} \Big({Y^u_2  Y^{u *}_2}\Big) -12 \Big({Y^d_3  Y^{d *}_2}\Big) \Big({Y^u_2  Y^{u *}_3}\Big) \nonumber \\ 
 &-12 \Big({Y^d_1  Y^{d *}_3}\Big) \Big({Y^u_3  Y^{u *}_1}\Big) -12 \Big({Y^d_2  Y^{d *}_3}\Big) \Big({Y^u_3  Y^{u *}_2}\Big) +6 \zeta_{23} \Big({Y^u_3  Y^{u *}_3}\Big) \nonumber \\ 
 &-12 \Big({Y^d_3  Y^{d *}_3}\Big) \Big({Y^u_3  Y^{u *}_3}\Big) \\ 
\beta_{\lambda_{23}}^{(1)} & =  
+\frac{11}{6} g_{2}^{4} +6 \lambda_{12} \lambda_{13} +2 \zeta_{12} \lambda_{13} +2 \lambda_{12} \zeta_{13} -6 g_{1}^{2} \lambda_{23} -16 g_{2}^{2} \lambda_{23} +16 \lambda_{2} \lambda_{23} +4 \lambda_{23}^{2}  \nonumber \\ 
 &+4 \lambda_{2} \zeta_{23} +2 \zeta_{23}^{2} +16 \lambda_{23} \lambda_{3}+4 \zeta_{23} \lambda_{3} +12 {\lambda}_{24} {\lambda}_{34} +7 {\xi}_{34}^{2} +4 {\lambda}_{34} \zeta_{24} +4 {\lambda}_{24} \zeta_{34} \nonumber \\ 
 &+\frac{1}{2} \zeta_{24} \zeta_{34} +6 \lambda_{23} \Big({Y^d_1  Y^{d *}_1}\Big) +6 \lambda_{23} \Big({Y^d_2  Y^{d *}_2}\Big) -12 \Big({Y^d_1  Y^{d *}_3}\Big) \Big({Y^d_3  Y^{d *}_1}\Big) \nonumber \\ 
 &+6 \lambda_{23} \Big({Y^d_3  Y^{d *}_3}\Big) +6 \lambda_{23} \Big({Y^u_1  Y^{u *}_1}\Big) +6 \lambda_{23} \Big({Y^u_2  Y^{u *}_2}\Big) -12 \Big({Y^u_1  Y^{u *}_3}\Big) \Big({Y^u_3  Y^{u *}_1}\Big)\nonumber \\ 
 & -12 \Big({Y^u_2  Y^{u *}_3}\Big) \Big({Y^u_3  Y^{u *}_2}\Big) +6 \lambda_{23} \Big({Y^u_3  Y^{u *}_3}\Big) -12 \Big({Y^d_2  Y^{d *}_3}\Big) \Big({Y^d_3  Y^{d *}_2}\Big) \\ 
\beta_{\lambda_{2}}^{(1)} & =  
+\frac{13}{6} g_{2}^{4} +3 \lambda_{12}^{2} +2 \lambda_{12} \zeta_{12} +\zeta_{12}^{2}-16 g_{2}^{2} \lambda_{2} +28 \lambda_{2}^{2} +3 \lambda_{23}^{2} +2 \lambda_{23} \zeta_{23} \nonumber \\ 
 &+\zeta_{23}^{2}+6 {\lambda}_{24}^{2} +12 {\xi}_{24}^{2}  +4 {\lambda}_{24} \zeta_{24} +\frac{3}{2} \zeta_{24}^{2}+12 \lambda_{2} \Big({Y^d_1  Y^{d *}_1}\Big) -6 \Big(\Big({Y^d_1  Y^{d *}_1}\Big)\Big)^{2} \nonumber \\ 
 &-12 \Big({Y^d_1  Y^{d *}_2}\Big) \Big({Y^d_2  Y^{d *}_1}\Big) +12 \lambda_{2} \Big({Y^d_2  Y^{d *}_2}\Big) -6 \Big(\Big({Y^d_2  Y^{d *}_2}\Big)\Big)^{2} +12 \lambda_{2} \Big({Y^u_3  Y^{u *}_3}\Big)\nonumber \\ 
 & -6 \Big(\Big({Y^u_3  Y^{u *}_3}\Big)\Big)^{2} \\ 
\beta_{{\lambda}_{24}}^{(1)} & =  
+\frac{13}{3} g_{2}^{4} +6 \lambda_{12} {\lambda}_{14} +2 \zeta_{12} {\lambda}_{14} -28 g_{2}^{2} {\lambda}_{24} +16 \lambda_{2} {\lambda}_{24} +4 {\lambda}_{24}^{2} +6 \lambda_{23} {\lambda}_{34} +2 \zeta_{23} {\lambda}_{34} \nonumber \\ 
 &+28 {\lambda}_{24} {\lambda}_4 +16 {\lambda}_{24} {\lambda}_{44} +2 {\xi}_{14}^{2} +48 {\xi}_{24}^{2} +2 {\xi}_{34}^{2} +2 \lambda_{12} \zeta_{14} +4 \lambda_{2} \zeta_{24} +8 {\lambda}_4 \zeta_{24} +2 {\lambda}_{44} \zeta_{24}\nonumber \\ 
 & +\zeta_{24}^{2}+2 \lambda_{23} \zeta_{34}  +6 {\lambda}_{24} \Big({Y^d_1  Y^{d *}_1}\Big) +6 {\lambda}_{24} \Big({Y^d_2  Y^{d *}_2}\Big)+6 {\lambda}_{24} \Big({Y^u_3  Y^{u *}_3}\Big) +4 {\lambda}_{24} \mbox{Tr}\Big({G^\sigma  G^{\sigma *}}\Big) \\ 
\beta_{\zeta_{24}}^{(1)} & =  
+7 g_{2}^{4} -2 {\xi}_{14}^{2} -48 {\xi}_{24}^{2} -2 {\xi}_{34}^{2} +2 \zeta_{12} \zeta_{14} -28 g_{2}^{2} \zeta_{24} +4 \lambda_{2} \zeta_{24} +8 {\lambda}_{24} \zeta_{24} +4 {\lambda}_4 \zeta_{24} \nonumber \\ 
 & +10 {\lambda}_{44} \zeta_{24} +5 \zeta_{24}^{2} +2 \zeta_{23} \zeta_{34} +6 \zeta_{24} \Big({Y^d_1  Y^{d *}_1}\Big) +6 \zeta_{24} \Big({Y^d_2  Y^{d *}_2}\Big) +6 \zeta_{24} \Big({Y^u_3  Y^{u *}_3}\Big) \nonumber \\ 
 & +4 \zeta_{24} \mbox{Tr}\Big({G^\sigma  G^{\sigma *}}\Big) \\ 
\beta_{\zeta_{13}}^{(1)} & =  
-12 g_{1}^{2} g_{2}^{2} +\frac{5}{2} g_{2}^{4} -12 g_{1}^{2} \zeta_{13} -16 g_{2}^{2} \zeta_{13} +4 \lambda_{1} \zeta_{13} +8 \lambda_{13} \zeta_{13} +6 \zeta_{13}^{2} +2 \zeta_{12} \zeta_{23} +4 \zeta_{13} \lambda_{3}\nonumber \\ 
 & +6 {\xi}_{14} {\xi}_{34} +\frac{5}{2} \zeta_{14} \zeta_{34} +6 |Y^J_1|^2 \Big(-2 \Big({Y^d_3  Y^{d *}_3}\Big)  + \zeta_{13}\Big)+6 \zeta_{13} \Big({Y^d_3  Y^{d *}_3}\Big) +6 \zeta_{13} \Big({Y^J_1  Y^{J *}_1}\Big)\nonumber \\ 
 & +6 \zeta_{13} \Big({Y^J_2  Y^{J *}_2}\Big)  +6 \zeta_{13} \Big({Y^u_1  Y^{u *}_1}\Big)-12 \Big({Y^J_1  Y^{J *}_1}\Big) \Big({Y^u_1  Y^{u *}_1}\Big) -12 \Big({Y^J_2  Y^{J *}_1}\Big) \Big({Y^u_1  Y^{u *}_2}\Big)  \nonumber \\ 
 &-12 \Big({Y^J_1  Y^{J *}_2}\Big) \Big({Y^u_2  Y^{u *}_1}\Big)+6 \zeta_{13} \Big({Y^u_2  Y^{u *}_2}\Big) -12 \Big({Y^J_2  Y^{J *}_2}\Big) \Big({Y^u_2  Y^{u *}_2}\Big) \\ 
\beta_{\zeta_{12}}^{(1)} & =  
+\frac{5}{2} g_{2}^{4} -6 g_{1}^{2} \zeta_{12} -16 g_{2}^{2} \zeta_{12} +4 \lambda_{1} \zeta_{12} +8 \lambda_{12} \zeta_{12} +6 \zeta_{12}^{2} +4 \zeta_{12} \lambda_{2} +2 \zeta_{13} \zeta_{23} -5 {\xi}_{14}^{2} \nonumber \\ 
 &+\frac{5}{2} \zeta_{14} \zeta_{24} +6 \zeta_{12} \Big({Y^d_2  Y^{d *}_2}\Big) +6 \Big({Y^d_1  Y^{d *}_1}\Big) \Big(-2 \Big({Y^J_1  Y^{J *}_1}\Big)  + \zeta_{12}\Big)+6 \zeta_{12} \Big({Y^J_1  Y^{J *}_1}\Big) \nonumber \\ 
 &-12 \Big({Y^d_2  Y^{d *}_1}\Big) \Big({Y^J_1  Y^{J *}_2}\Big) -12 \Big({Y^d_1  Y^{d *}_2}\Big) \Big({Y^J_2  Y^{J *}_1}\Big) +6 \zeta_{12} \Big({Y^J_2  Y^{J *}_2}\Big)\nonumber \\ 
 &  -12 \Big({Y^d_2  Y^{d *}_2}\Big) \Big({Y^J_2  Y^{J *}_2}\Big) +6 |Y^J_1|^2 \Big(-2 \Big({Y^u_3  Y^{u *}_3}\Big) + \zeta_{12}\Big)+6 \zeta_{12} \Big({Y^u_3  Y^{u *}_3}\Big) \\ 
\beta_{\lambda_{13}}^{(1)} & =  
+12 g_{1}^{4} +4 g_{1}^{2} g_{2}^{2} +\frac{11}{6} g_{2}^{4} -12 g_{1}^{2} \lambda_{13} -16 g_{2}^{2} \lambda_{13} +16 \lambda_{1} \lambda_{13} +4 \lambda_{13}^{2} +4 \lambda_{1} \zeta_{13} +2 \zeta_{13}^{2}\nonumber \\ 
 & +6 \lambda_{12} \lambda_{23}  +2 \zeta_{12} \lambda_{23} +2 \lambda_{12} \zeta_{23} +16 \lambda_{13} \lambda_{3} +4 \zeta_{13} \lambda_{3} +12 {\lambda}_{14} {\lambda}_{34} -2 {\xi}_{14} {\xi}_{34} \nonumber \\ 
 &+4 {\lambda}_{34} \zeta_{14} +4 {\lambda}_{14} \zeta_{34} +\frac{1}{2} \zeta_{14} \zeta_{34} +6 \lambda_{13} |Y^J_1|^2 +6 \lambda_{13} \Big({Y^d_3  Y^{d *}_3}\Big) +6 \lambda_{13} \Big({Y^J_1  Y^{J *}_1}\Big)\nonumber \\ 
 & +6 \lambda_{13} \Big({Y^J_2  Y^{J *}_2}\Big) +6 \lambda_{13} \Big({Y^u_1  Y^{u *}_1}\Big) +6 \lambda_{13} \Big({Y^u_2  Y^{u *}_2}\Big) \\ 
\beta_{\lambda_{12}}^{(1)} & =  
+\frac{11}{6} g_{2}^{4} -6 g_{1}^{2} \lambda_{12} -16 g_{2}^{2} \lambda_{12} +16 \lambda_{1} \lambda_{12} +4 \lambda_{12}^{2} +4 \lambda_{1} \zeta_{12} +2 \zeta_{12}^{2} +16 \lambda_{12} \lambda_{2} +4 \zeta_{12} \lambda_{2} \nonumber \\ 
 &+6 \lambda_{13} \lambda_{23} +2 \zeta_{13} \lambda_{23} +2 \lambda_{13} \zeta_{23} +12 {\lambda}_{14} {\lambda}_{24} +7 {\xi}_{14}^{2} +4 {\lambda}_{24} \zeta_{14} +4 {\lambda}_{14} \zeta_{24} +\frac{1}{2} \zeta_{14} \zeta_{24} \nonumber \\ 
 &+6 \lambda_{12} |Y^J_1|^2 +6 \lambda_{12} \Big({Y^d_1  Y^{d *}_1}\Big) +6 \lambda_{12} \Big({Y^d_2  Y^{d *}_2}\Big) +6 \lambda_{12} \Big({Y^J_1  Y^{J *}_1}\Big) +6 \lambda_{12} \Big({Y^J_2  Y^{J *}_2}\Big)\nonumber \\ 
 & +6 \lambda_{12} \Big({Y^u_3  Y^{u *}_3}\Big) \\ 
\beta_{\lambda_{1}}^{(1)} & =  
+6 g_{1}^{4} +4 g_{1}^{2} g_{2}^{2} +\frac{13}{6} g_{2}^{4} -12 g_{1}^{2} \lambda_{1} -16 g_{2}^{2} \lambda_{1} +28 \lambda_{1}^{2} +3 \lambda_{12}^{2} +2 \lambda_{12} \zeta_{12} +\zeta_{12}^{2}+3 \lambda_{13}^{2} \nonumber \\ 
 &+2 \lambda_{13} \zeta_{13} +\zeta_{13}^{2}+6 {\lambda}_{14}^{2} +2 {\xi}_{14}^{2} +4 {\lambda}_{14} \zeta_{14} +\frac{3}{2} \zeta_{14}^{2} +12 \lambda_{1} |Y^J_1|^2 -6 |Y^J_1|^4 \nonumber \\ 
 &+12 \lambda_{1} \Big({Y^J_1  Y^{J *}_1}\Big) -6 \Big(\Big({Y^J_1  Y^{J *}_1}\Big)\Big)^{2} -12 \Big({Y^J_1  Y^{J *}_2}\Big) \Big({Y^J_2  Y^{J *}_1}\Big) +12 \lambda_{1} \Big({Y^J_2  Y^{J *}_2}\Big)\nonumber \\ 
 & -6 \Big(\Big({Y^J_2  Y^{J *}_2}\Big)\Big)^{2} \\ 
\beta_{{\lambda}_{14}}^{(1)} & =  
+\frac{13}{3} g_{2}^{4} -6 g_{1}^{2} {\lambda}_{14} -28 g_{2}^{2} {\lambda}_{14} +16 \lambda_{1} {\lambda}_{14} +4 {\lambda}_{14}^{2} +6 \lambda_{12} {\lambda}_{24} +2 \zeta_{12} {\lambda}_{24}\nonumber \\ 
 &  +6 \lambda_{13} {\lambda}_{34} +2 \zeta_{13} {\lambda}_{34}+28 {\lambda}_{14} {\lambda}_4 +16 {\lambda}_{14} {\lambda}_{44} +2 {\xi}_{14}^{2} +4 \lambda_{1} \zeta_{14} +8 {\lambda}_4 \zeta_{14} +2 {\lambda}_{44} \zeta_{14}\nonumber \\ 
 & +\zeta_{14}^{2}+2 \lambda_{12} \zeta_{24} +2 \lambda_{13} \zeta_{34} +6 {\lambda}_{14} |Y^J_1|^2 +6 {\lambda}_{14} \Big({Y^J_1  Y^{J *}_1}\Big) +6 {\lambda}_{14} \Big({Y^J_2  Y^{J *}_2}\Big)\nonumber \\ 
 & +4 {\lambda}_{14} \mbox{Tr}\Big({G^\sigma  G^{\sigma *}}\Big) \\ 
\beta_{\zeta_{14}}^{(1)} & =  
+7 g_{2}^{4} +2 {\xi}_{14}^{2} -6 g_{1}^{2} \zeta_{14} -28 g_{2}^{2} \zeta_{14} +4 \lambda_{1} \zeta_{14} +8 {\lambda}_{14} \zeta_{14} +4 {\lambda}_4 \zeta_{14} +10 {\lambda}_{44} \zeta_{14} +5 \zeta_{14}^{2}\nonumber \\ 
 & +2 \zeta_{12} \zeta_{24} +2 \zeta_{13} \zeta_{34} +6 \zeta_{14} |Y^J_1|^2 +6 \zeta_{14} \Big({Y^J_1  Y^{J *}_1}\Big) +6 \zeta_{14} \Big({Y^J_2  Y^{J *}_2}\Big) \nonumber \\ 
 &+4 \zeta_{14} \mbox{Tr}\Big({G^\sigma  G^{\sigma *}}\Big) \\ 
\beta_{{\lambda}_4}^{(1)} & =  
+\frac{35}{3} g_{2}^{4} +3 {\lambda}_{14}^{2} +3 {\lambda}_{24}^{2} +3 {\lambda}_{34}^{2} -40 g_{2}^{2} {\lambda}_4 +40 {\lambda}_{4}^{2} +32 {\lambda}_4 {\lambda}_{44} +6 {\lambda}_{44}^{2} +8 {\xi}_{24}^{2} +2 {\lambda}_{14} \zeta_{14}\nonumber \\ 
 & +2 {\lambda}_{24} \zeta_{24} +2 {\lambda}_{34} \zeta_{34} +8 {\lambda}_4 \mbox{Tr}\Big({G^\sigma  G^{\sigma *}}\Big) \\ 
\beta_{{\lambda}_{44}}^{(1)} & =  
-16 \mbox{Tr}\Big({G^\sigma  G^{\sigma *}  G^\sigma  G^{\sigma *}}\Big)  + 22 {\lambda}_{44}^{2}  + 24 {\lambda}_4 {\lambda}_{44}  -40 g_{2}^{2} {\lambda}_{44}  + 5 g_{2}^{4}  + 8 {\lambda}_{44} \mbox{Tr}\Big({G^\sigma  G^{\sigma *}}\Big) \nonumber \\ 
 &   -8 {\xi}_{24}^{2}+ \zeta_{14}^{2} + \zeta_{24}^{2} + \zeta_{34}^{2}
\end{align}}

\subsection{Yukawa Couplings}
{\allowdisplaybreaks  \begin{align} 
\beta_{G^\eta}^{(1)} & =  
-8 {G^\sigma  G^{\sigma *}  G^\sigma}  + G^\sigma \Big(3 \Big(\Big({Y^d_1  Y^{d *}_1}\Big) \Big({Y^d_2  Y^{d *}_2}\Big)  + \Big({Y^u_3  Y^{u *}_3}\Big)\Big) -4 \mbox{Tr}\Big({G^\eta  G^{\eta *}}\Big)  -8 g_{2}^{2} \Big)\\ 
\beta_{G^\sigma}^{(1)} & =  
8 {G^\sigma  G^{\sigma *}  G^\sigma}  + G^\sigma \Big(2 \mbox{Tr}\Big({G^\sigma  G^{\sigma *}}\Big)  -8 g_{2}^{2} \Big)\\ 
\beta_{Y_{1,{{i_1}}}^{d}}^{(1)} & =  
\frac{1}{6} \Big(\Big(18 \Big({Y^d_2  Y^{d *}_2}\Big)  + 18 \Big({Y^u_3  Y^{u *}_3}\Big)  -24 g_{2}^{2}  + 30 \Big({Y^d_1  Y^{d *}_1}\Big)  + 3 \Big({Y^J_1  Y^{J *}_1}\Big)  + 3 \Big({Y^u_1  Y^{u *}_1}\Big) \nonumber \\ 
 & -48 g_{3}^{2}  -4 g_{1}^{2} \Big)Y_{1,{{i_1}}}^{d} +3 \Big(3 \Big({Y^d_1  Y^{d *}_3}\Big) Y_{3,{{i_1}}}^{d}  + 4 \Big({Y^d_1  Y^{d *}_2}\Big) Y_{2,{{i_1}}}^{d}  + 4 \Big({Y^u_1  Y^{u *}_3}\Big) Y_{3,{{i_1}}}^{d} \nonumber \\ 
 &  + \Big({Y^J_1  Y^{J *}_2}\Big) Y_{2,{{i_1}}}^{d} + \Big({Y^u_1  Y^{u *}_2}\Big) Y_{2,{{i_1}}}^{d} \Big)\Big)\\ 
\beta_{Y_{1,{{i_1}}}^{J}}^{(1)} & =  
\frac{1}{6} \Big(\Big(18 |Y^J_1|^2  + 18 \Big({Y^J_2  Y^{J *}_2}\Big)  -24 g_{2}^{2}  + 30 \Big({Y^J_1  Y^{J *}_1}\Big)  -34 g_{1}^{2}  + 3 \Big({Y^d_1  Y^{d *}_1}\Big)  \nonumber \\ 
 & + 3 \Big({Y^u_1  Y^{u *}_1}\Big) -48 g_{3}^{2} \Big)Y_{1,{{i_1}}}^{J} +3 \Big(4 \Big({Y^J_1  Y^{J *}_2}\Big)  + \Big({Y^d_1  Y^{d *}_2}\Big) + \Big({Y^u_1  Y^{u *}_2}\Big)\Big)Y_{2,{{i_1}}}^{J} \Big)\\ 
\beta_{Y_{1,{{i_1}}}^{u}}^{(1)} & =  
\frac{1}{6} \Big(\Big(-10 g_{1}^{2}  + 18 \Big({Y^d_3  Y^{d *}_3}\Big)  + 18 \Big({Y^u_2  Y^{u *}_2}\Big)  -24 g_{2}^{2}  + 30 \Big({Y^u_1  Y^{u *}_1}\Big)  + 3 \Big({Y^d_1  Y^{d *}_1}\Big)  \nonumber \\ 
 &+ 3 \Big({Y^J_1  Y^{J *}_1}\Big)  -48 g_{3}^{2} \Big)Y_{1,{{i_1}}}^{u} +3 \Big(3 \Big({Y^u_1  Y^{u *}_3}\Big) Y_{3,{{i_1}}}^{u}  + 4 \Big({Y^d_1  Y^{d *}_3}\Big) Y_{3,{{i_1}}}^{u}  + 4 \Big({Y^u_1  Y^{u *}_2}\Big) Y_{2,{{i_1}}}^{u}  \nonumber \\ 
 &+ \Big({Y^d_1  Y^{d *}_2}\Big) Y_{2,{{i_1}}}^{u}  + \Big({Y^J_1  Y^{J *}_2}\Big) Y_{2,{{i_1}}}^{u} \Big)\Big)\\ 
\beta_{Y_{2,{{i_1}}}^{d}}^{(1)} & =  
\frac{1}{6} \Big(12 \Big({Y^d_2  Y^{d *}_1}\Big) Y_{1,{{i_1}}}^{d} +3 \Big({Y^J_2  Y^{J *}_1}\Big) Y_{1,{{i_1}}}^{d} +3 \Big({Y^u_2  Y^{u *}_1}\Big) Y_{1,{{i_1}}}^{d} -4 g_{1}^{2} Y_{2,{{i_1}}}^{d} -24 g_{2}^{2} Y_{2,{{i_1}}}^{d} \nonumber \\ 
 &-48 g_{3}^{2} Y_{2,{{i_1}}}^{d} +18 \Big({Y^d_1  Y^{d *}_1}\Big) Y_{2,{{i_1}}}^{d} +30 \Big({Y^d_2  Y^{d *}_2}\Big) Y_{2,{{i_1}}}^{d} +3 \Big({Y^J_2  Y^{J *}_2}\Big) Y_{2,{{i_1}}}^{d} \nonumber \\ 
 &+3 \Big({Y^u_2  Y^{u *}_2}\Big) Y_{2,{{i_1}}}^{d} +18 \Big({Y^u_3  Y^{u *}_3}\Big) Y_{2,{{i_1}}}^{d} +9 \Big({Y^d_2  Y^{d *}_3}\Big) Y_{3,{{i_1}}}^{d} +12 \Big({Y^u_2  Y^{u *}_3}\Big) Y_{3,{{i_1}}}^{d} \Big)\\ 
\beta_{Y_{2,{{i_1}}}^{J}}^{(1)} & =  
\frac{1}{6} \Big(3 \Big({Y^d_2  Y^{d *}_1}\Big) Y_{1,{{i_1}}}^{J} +12 \Big({Y^J_2  Y^{J *}_1}\Big) Y_{1,{{i_1}}}^{J} +3 \Big({Y^u_2  Y^{u *}_1}\Big) Y_{1,{{i_1}}}^{J} -34 g_{1}^{2} Y_{2,{{i_1}}}^{J} -24 g_{2}^{2} Y_{2,{{i_1}}}^{J} \nonumber \\ 
 &-48 g_{3}^{2} Y_{2,{{i_1}}}^{J} +18 |Y^J_1|^2 Y_{2,{{i_1}}}^{J} +3 \Big({Y^d_2  Y^{d *}_2}\Big) Y_{2,{{i_1}}}^{J} +18 \Big({Y^J_1  Y^{J *}_1}\Big) Y_{2,{{i_1}}}^{J} +30 \Big({Y^J_2  Y^{J *}_2}\Big) Y_{2,{{i_1}}}^{J} \nonumber \\ 
 &+3 \Big({Y^u_2  Y^{u *}_2}\Big) Y_{2,{{i_1}}}^{J} \Big)\\ 
\beta_{Y_{2,{{i_1}}}^{u}}^{(1)} & =  
\frac{1}{6} \Big(3 \Big({Y^d_2  Y^{d *}_1}\Big) Y_{1,{{i_1}}}^{u} +3 \Big({Y^J_2  Y^{J *}_1}\Big) Y_{1,{{i_1}}}^{u} +12 \Big({Y^u_2  Y^{u *}_1}\Big) Y_{1,{{i_1}}}^{u} -10 g_{1}^{2} Y_{2,{{i_1}}}^{u} -24 g_{2}^{2} Y_{2,{{i_1}}}^{u} \nonumber \\ 
 & -48 g_{3}^{2} Y_{2,{{i_1}}}^{u}+3 \Big({Y^d_2  Y^{d *}_2}\Big) Y_{2,{{i_1}}}^{u} +18 \Big({Y^d_3  Y^{d *}_3}\Big) Y_{2,{{i_1}}}^{u} +3 \Big({Y^J_2  Y^{J *}_2}\Big) Y_{2,{{i_1}}}^{u}\nonumber \\ 
 & +18 \Big({Y^u_1  Y^{u *}_1}\Big) Y_{2,{{i_1}}}^{u} +30 \Big({Y^u_2  Y^{u *}_2}\Big) Y_{2,{{i_1}}}^{u} +12 \Big({Y^d_2  Y^{d *}_3}\Big) Y_{3,{{i_1}}}^{u} +9 \Big({Y^u_2  Y^{u *}_3}\Big) Y_{3,{{i_1}}}^{u} \Big)\\ 
\beta_{Y_{3,{{i_1}}}^{d}}^{(1)} & =  
+\frac{3}{2} \Big({Y^d_3  Y^{d *}_1}\Big) Y_{1,{{i_1}}}^{d} +2 \Big({Y^u_3  Y^{u *}_1}\Big) Y_{1,{{i_1}}}^{d} +\frac{3}{2} \Big({Y^d_3  Y^{d *}_2}\Big) Y_{2,{{i_1}}}^{d} +2 \Big({Y^u_3  Y^{u *}_2}\Big) Y_{2,{{i_1}}}^{d}\nonumber \\ 
 & -\frac{5}{3} g_{1}^{2} Y_{3,{{i_1}}}^{d} -4 g_{2}^{2} Y_{3,{{i_1}}}^{d} -8 g_{3}^{2} Y_{3,{{i_1}}}^{d} +\frac{1}{2} |Y^J_1|^2 Y_{3,{{i_1}}}^{d} +5 \Big({Y^d_3  Y^{d *}_3}\Big) Y_{3,{{i_1}}}^{d} +3 \Big({Y^u_1  Y^{u *}_1}\Big) Y_{3,{{i_1}}}^{d} \nonumber \\ 
 &+3 \Big({Y^u_2  Y^{u *}_2}\Big) Y_{3,{{i_1}}}^{d} +\frac{1}{2} \Big({Y^u_3  Y^{u *}_3}\Big) Y_{3,{{i_1}}}^{d} \\ 
\beta_{Y^J_1}^{(1)} & =  
\frac{1}{6} Y^J_1 \Big(18 \Big({Y^J_1  Y^{J *}_1}\Big)  + 18 \Big({Y^J_2  Y^{J *}_2}\Big)  -24 g_{2}^{2}  + 30 |Y^J_1|^2  + 3 \Big({Y^d_3  Y^{d *}_3}\Big) \nonumber \\ 
 & + 3 \Big({Y^u_3  Y^{u *}_3}\Big)  -48 g_{3}^{2}  -58 g_{1}^{2} \Big)\\ 
\beta_{Y_{3,{{i_1}}}^{u}}^{(1)} & =  
+2 \Big({Y^d_3  Y^{d *}_1}\Big) Y_{1,{{i_1}}}^{u} +\frac{3}{2} \Big({Y^u_3  Y^{u *}_1}\Big) Y_{1,{{i_1}}}^{u} +2 \Big({Y^d_3  Y^{d *}_2}\Big) Y_{2,{{i_1}}}^{u} +\frac{3}{2} \Big({Y^u_3  Y^{u *}_2}\Big) Y_{2,{{i_1}}}^{u} \nonumber \\ 
 &-\frac{8}{3} g_{1}^{2} Y_{3,{{i_1}}}^{u} -4 g_{2}^{2} Y_{3,{{i_1}}}^{u} -8 g_{3}^{2} Y_{3,{{i_1}}}^{u} +\frac{1}{2} |Y^J_1|^2 Y_{3,{{i_1}}}^{u} +3 \Big({Y^d_1  Y^{d *}_1}\Big) Y_{3,{{i_1}}}^{u} +3 \Big({Y^d_2  Y^{d *}_2}\Big) Y_{3,{{i_1}}}^{u} \nonumber \\ 
 &+\frac{1}{2} \Big({Y^d_3  Y^{d *}_3}\Big) Y_{3,{{i_1}}}^{u} +5 \Big({Y^u_3  Y^{u *}_3}\Big) Y_{3,{{i_1}}}^{u} 
\end{align}}

\subsection{Trilinear Scalar couplings}
{\allowdisplaybreaks  \begin{align} 
\beta_{{f}_{\sigma}}^{(1)} & =  
-6 {f}_{\sigma} g_{1}^{2} -18 {f}_{\sigma} g_{2}^{2} +2 {f}_{\sigma} \lambda_{13} +2 {f}_{\sigma} \zeta_{13} +2 {f}_{\sigma} {\lambda}_{14} +2 {f}_{\sigma} {\lambda}_{34} -4 f {\xi}_{14} \nonumber \\ 
 &+4 f {\xi}_{34} +4 {f}_{\sigma} \zeta_{14} +4 {f}_{\sigma} \zeta_{34} +3 {f}_{\sigma} |Y^J_1|^2 +3 {f}_{\sigma} \Big({Y^d_3  Y^{d *}_3}\Big) +3 {f}_{\sigma} \Big({Y^J_1  Y^{J *}_1}\Big) \nonumber \\ 
 &+3 {f}_{\sigma} \Big({Y^J_2  Y^{J *}_2}\Big) +3 {f}_{\sigma} \Big({Y^u_1  Y^{u *}_1}\Big) +3 {f}_{\sigma} \Big({Y^u_2  Y^{u *}_2}\Big) +2 {f}_{\sigma} \mbox{Tr}\Big({G^\sigma  G^{\sigma *}}\Big) \\ 
\beta_{f}^{(1)} & =  
-6 f g_{1}^{2} -12 f g_{2}^{2} +2 f \lambda_{12} -2 f \zeta_{12} +2 f \lambda_{13} -2 f \zeta_{13} +2 f \lambda_{23} -2 f \zeta_{23} \nonumber \\ 
 &-4 {f}_{\sigma} {\xi}_{14}  +4 {f}_{\sigma} {\xi}_{34} +3 f |Y^J_1|^2+3 f \Big({Y^d_1  Y^{d *}_1}\Big) +3 f \Big({Y^d_2  Y^{d *}_2}\Big) +3 f \Big({Y^d_3  Y^{d *}_3}\Big) \nonumber \\ 
 &+3 f \Big({Y^J_1  Y^{J *}_1}\Big) +3 f \Big({Y^J_2  Y^{J *}_2}\Big) +3 f \Big({Y^u_1  Y^{u *}_1}\Big) +3 f \Big({Y^u_2  Y^{u *}_2}\Big) +3 f \Big({Y^u_3  Y^{u *}_3}\Big) 
\end{align}}

\subsection{Scalar Mass Terms}
{\allowdisplaybreaks  \begin{align} 
\beta_{{\mu}_{3}}^{(1)} & =  
2 \Big(4 f^{2} +4 {f}_{\sigma}^{2} +3 \lambda_{13} {\mu}_{1} +\zeta_{13} {\mu}_{1} +3 \lambda_{23} {\mu}_{2} +\zeta_{23} {\mu}_{2} -3 g_{1}^{2} {\mu}_{3} -4 g_{2}^{2} {\mu}_{3} +8 \lambda_{3} {\mu}_{3}\nonumber \\ 
 & +6 {\lambda}_{34} {\mu}_{4}  +2 {\mu}_{4} \zeta_{34}+3 {\mu}_{3} \Big({Y^d_3  Y^{d *}_3}\Big) +3 {\mu}_{3} \Big({Y^u_1  Y^{u *}_1}\Big) +3 {\mu}_{3} \Big({Y^u_2  Y^{u *}_2}\Big) \Big)\\ 
\beta_{{\mu}_{2}}^{(1)} & =  
2 \Big(4 f^{2} +3 \lambda_{12} {\mu}_{1} +\zeta_{12} {\mu}_{1} -4 g_{2}^{2} {\mu}_{2} +8 \lambda_{2} {\mu}_{2} +3 \lambda_{23} {\mu}_{3} +\zeta_{23} {\mu}_{3} +6 {\lambda}_{24} {\mu}_{4} \nonumber \\ 
 &+2 {\mu}_{4} \zeta_{24} +3 {\mu}_{2} \Big({Y^d_1  Y^{d *}_1}\Big) +3 {\mu}_{2} \Big({Y^d_2  Y^{d *}_2}\Big) +3 {\mu}_{2} \Big({Y^u_3  Y^{u *}_3}\Big) \Big)\\ 
\beta_{{\mu}_{1}}^{(1)} & =  
2 \Big(4 f^{2} +4 {f}_{\sigma}^{2} -3 g_{1}^{2} {\mu}_{1} -4 g_{2}^{2} {\mu}_{1} +8 \lambda_{1} {\mu}_{1} +3 \lambda_{12} {\mu}_{2} +\zeta_{12} {\mu}_{2} +3 \lambda_{13} {\mu}_{3} \nonumber \\ 
 &+\zeta_{13} {\mu}_{3} +6 {\lambda}_{14} {\mu}_{4}  +2 {\mu}_{4} \zeta_{14}+3 {\mu}_{1} |Y^J_1|^2 +3 {\mu}_{1} \Big({Y^J_1  Y^{J *}_1}\Big) +3 {\mu}_{1} \Big({Y^J_2  Y^{J *}_2}\Big) \Big)\\ 
\beta_{{\mu}_{4}}^{(1)} & =  
2 \Big(-10 g_{2}^{2} {\mu}_{4}  + 14 {\lambda}_4 {\mu}_{4}  + 2 {f}_{\sigma}^{2}  + 2 {\mu}_{4} \mbox{Tr}\Big({G^\sigma  G^{\sigma *}}\Big)  + 3 {\lambda}_{14} {\mu}_{1}  + 3 {\lambda}_{24} {\mu}_{2}  + 3 {\lambda}_{34} {\mu}_{3}  \nonumber \\ 
 & + 8 {\lambda}_{44} {\mu}_{4}  + {\mu}_{1} \zeta_{14}  + {\mu}_{2} \zeta_{24}  + {\mu}_{3} \zeta_{34} \Big)
\end{align}}

\subsection{Vacuum expectation values}
{\allowdisplaybreaks  \begin{align} 
\beta_{v_{\rho}}^{(1)} & =  
\frac{1}{3} v_{\rho} \Big(12 g_{2}^{2}  + 3 g_{1}^{2} \xi  + 4 g_{2}^{2} \xi  + 9 g_{1}^{2}  -9 |Y^J_1|^2  -9 \Big({Y^J_1  Y^{J *}_1}\Big)  -9 \Big({Y^J_2  Y^{J *}_2}\Big) \Big)\\ 
\beta_{v_{\eta}}^{(1)} & =  
\frac{1}{3} v_{\eta} \Big(12 g_{2}^{2}  + 4 g_{2}^{2} \xi  -9 \Big({Y^d_1  Y^{d *}_1}\Big)  -9 \Big({Y^d_2  Y^{d *}_2}\Big)  -9 \Big({Y^u_3  Y^{u *}_3}\Big) \Big)\\ 
\beta_{v_{\chi}}^{(1)} & =  
\frac{1}{3} v_{\chi} \Big(12 g_{2}^{2}  + 3 g_{1}^{2} \xi  + 4 g_{2}^{2} \xi  + 9 g_{1}^{2}  -9 \Big({Y^d_3  Y^{d *}_3}\Big)  -9 \Big({Y^u_1  Y^{u *}_1}\Big)  -9 \Big({Y^u_2  Y^{u *}_2}\Big) \Big)\\ 
\beta_{v_{\sigma}}^{(1)} & =  
\frac{2}{3} v_{\sigma} \Big(-3 \mbox{Tr}\Big({G^\sigma  G^{\sigma *}}\Big)  + 5 g_{2}^{2} \Big(3 + \xi\Big)\Big)
\end{align}}

\providecommand{\href}[2]{#2}\begingroup\raggedright\endgroup


\end{document}